\documentclass[
aps,
prb,
reprint,
superscriptaddress,
floatfix,
notitlepage,
]{revtex4-1}
\usepackage{amsfonts,amsmath,amssymb}
\usepackage{graphicx}%
\usepackage{xcolor}%
\usepackage{comment}
\usepackage{cleveref}
\usepackage{natbib}

\DeclareMathOperator{\Tr}{Tr}

\def\br{{\mathbf r}}
\def\bs{{\mathbf s}}

\def\bp{{\mathbf p}}
\def\bq{{\mathbf q}}
\def\bP{{\mathbf P}}

\def\bu{{\mathbf u}}
\def\bw{{\mathbf w}}

\newcommand*\diff{\mathop{}\!\mathrm{d}}
\def\dd{{\diff}}
\def\dr{{\!\!\dd\br\,}}

\def\ext{{\mathrm{ext}}}
\def\H{{\mathrm{H}}}
\def\s{{\mathrm{s}}}
\def\xc{{\mathrm{xc}}}
\def\TF{{\mathrm{TF}}}

\def\calL{{\mathcal{L}}}

\def\nbar{\overline{n}}
\def\ebar{\overline{e}}
\def\bpbar{\overline{\bp}}

\renewcommand{\vec}[1]{\boldsymbol{\mathbf{#1}}}

\begin{document}
%\preprint{ }

\title{Generalized Hydrodynamics Revisited}

\author{James Dufty}
%\email{dufty@ufl.edu}
%\email{dufty@ufl.edu}
\affiliation{Department of Physics, University of Florida, Gainesville, FL 32611}

\author{Kai Luo}
\email{kluo@carnegiescience.edu}
%\affiliation{Quantum Theory Project, Department of Physics, University of Florida, Gainesville, FL 32611}
\affiliation{Extreme Materials Initiative, Geophysical Laboratory, Carnegie Institution for Science, Washington, DC 20015-1305, USA}

\author{Jeffrey Wrighton}
%\email{jeffwrighton@gmail.com}
\affiliation{Department of Physics, University of Florida, Gainesville, FL 32611}

\date{\today}

\begin{abstract}
During the past decade a number of attempts to formulate a continuum
description of complex states of matter have been proposed to circumvent more cumbersome
many-body and simulation methods. Typically these have been quantum systems
(e.g., electrons) and the resulting phenomenologies collectively often called ``quantum hydrodynamics".
However, there is extensive work from the past based
in non-equilibrium statistical mechanics on the microscopic origins of
macroscopic continuum dynamics that has not been exploited in this context.
Although formally exact, its original target was the derivation of
Navier-Stokes hydrodynamics for slowly varying states in space and time. The
objective here is to revisit that work for the present interest in complex
quantum states - possible strong degeneracy, strong coupling, and all
space-time scales. The result is an exact representation of generalized
hydrodynamics suitable for introducing controlled approximations for diverse
specific cases, and for critiquing existing work.
\end{abstract}
\maketitle

\section{Introduction}
\label{sec:intro}
Traditional hydrodynamics for a simple fluid describes the dynamics of the
average local conserved fields associated with symmetries of the Galilean
group (number density $\nbar(\br,t)$, energy
density $\ebar(\br,t)$, momentum density
$\bpbar(\br,t)  $ 
%\cite{Landau59,DeGroot62}%
\cite{Landau1987,DeGroot2013}%
. These fields are chosen since they dominate the behavior on large space and
time scales, leading to a closed dynamics simpler than that of the full
many-body degrees of freedom. 
The governing equations have their origin in
the  \textit{macroscopic} conservation laws that follow from averages of the
corresponding exact \textit{microscopic} conservation laws for the operators
representing these fields. 
The averages are defined with respect to a chosen
state (ensemble) whose dynamics is governed by the Liouville-von Neumann equation. 
The generic forms of the macroscopic conservation laws, classical or
quantum, do not depend on the specific state of interest except through the
explicit forms for the average energy and momentum fluxes. 
The latter are
``constitutive equations" relating those fluxes to the hydrodynamic variables
for a closed description of their dynamics. This is analogous to
thermodynamics whose formulation is general, but which requires an equation of
state for each specific system. 
For variations on small space and time scales
the Liouville-von Neumann equation can be solved to obtain the average
fluxes in the form of Fourier's law and Newton's viscosity law
\cite{McLennan1989,Zubarev1974}. The resulting hydrodynamic equations are the
non-linear Navier-Stokes equations, characterized by a thermodynamic pressure
and transport coefficients given formally by Green-Kubo time correlation functions \cite {Zwanzig1965}. 

The objective here is to extend the continuum description to general complex
states being studied now in various 
evolving fields of condensed matter physics and
materials sciences. Examples of recent reviews include warm, dense matter \cite{IPAMreview}, high energy density physics \cite{Murillo2017,Murillo2019}, thermo-electric transport \cite{Eich2016}, quantum plasmas \cite{Bonitz2019,Bonitz2018}, and electrons in graphene \cite{LucasFong2018}. 
This entails accommodating conditions of strong
coupling, strong quantum degeneracy, and/or shorter space and time scales.
This idea has been explored extensively some years ago for calculating equilibrium time
correlation functions 
%\cite{Mountain1977,Boon13}.
\cite{Mountain1977,Boon1991}.
 Such functions obey the
\emph{linear} Navier-Stokes equations at low frequencies and long wavelengths
\cite{Onsager1931,KadanoffMartin1963}. These conditions are met for the differential cross section
measured by laser light scattering, and the latter is a useful tool for
measuring the hydrodynamic transport coefficients. However, that approach does
not apply to the neutron scattering cross section which samples much smaller
wavelengths and larger frequencies. To address this problem the hydrodynamic
description is modified (generalized) to include these extended domains.
Typically this is done by making thermodynamic derivatives wave vector
dependent and transport coefficients frequency and wave vector dependent. An
early review of the methods with references is given by Mountain
\cite{Mountain1966}.

The scope of the objective here is far greater, to provide the basis for a
continuum description of macroscopic dynamics under the most general
conditions. It is a formally exact resolution of the macroscopic conservation
laws into physically transparent components suitable for constructing
approximate models in specific cases of interest. In this way much of the
present phenomenology about ``new" hydrodynamics can be critiqued and
controlled. The analysis uses an application of non-equilibrium statistical
mechanics developed long ago for linear response, but not limited to that
case. The two earliest and most complete formulations are those of McLennan
and collaborators \cite{McLennan1963,Wong1978,Dufty1979,McLennan1989} and of Zubarev
and collaborators \cite{Zubarev1974,Zubarev1996}. %\cite{Zubarev1974,Zubarev2,Zubarev3}. 
The present work can be
viewed as a ``revisiting" of that approach assuring its exact treatment of
extreme quantum and non-local effects. It is hoped in this way an
important largely forgotten tool will become available for new developments in
the formulation of practical continuum dynamics models.

One motivation for this revisiting of generalized hydrodynamics is a growing
current interest in a continuum description for the electron density in
quantum devices (e.g.,nanomaterials) and other charged particle quantum
systems (e.g., plasmonics).
%; see Moldabekov et al.\cite{Bonitz2018} for extensive references.
A very recent review of current activity in quantum hydrodynamics for plasmas has just appeared \cite{Bonitz2019}.
 There is a much earlier history of
quantum hydrodynamics associated with attempts to reinterpret the Schrodinger
equation as an equivalent continuum description \cite{Madelung1927,Bohm1952}, time
dependent Thomas - Fermi models \cite{Palade2015}, or from a macroscopic
Hamiltonian \cite{Bloch1934}, each leading to different quantum potentials
occuring in related quantum continuum equations. See Stanton and Murillo
\cite{Stanton2015} for some resolution of these different quantum potentials.
%and detailed comments below in section \ref{sec:perfect_fluid_hydrodynamics}. 
Previous quantum
hydrodynamics have often focused on nearly ideal plasmas (weak coupling, weak
spatial inhomogeneities). One interest here is for the quite different states
of warm dense matter \cite{IPAMreview}, comprised of strongly coupled ions and electrons at
solid state densities and a wide range of temperatures and
spatial inhomogeneity. %(clearly, electron densities near ions have very strong gradients). 
A direct approach by standard many-body methods is very
challenging. Density functional theory (DFT) for equilibrium properties of
such states has met with some significant success \cite{Karasiev2018}.
Applications of time dependent density functional theory (TDDFT) are still at
an early stage in this context \cite{Tokatly2005a, Tokatly2005b, Gao2010}. In principle the density and momentum
conservation laws provide the basis for a closed description of the density if
the momentum flux can be expressed as a functional of the density via TDDFT
\cite{Tokatly2005a,Yan2015}. Extensions of this idea including energy
conservation are also being considered \cite{Eich2016}. Here, the closure of all
conservation laws is provided by a representation of the fluxes in terms of a
formal solution to the Liouville-von Neumann equation as a functional of the
fields, non-local in both space and time.

In the next few sections the basis for an exact continuum description for the
average density, internal energy density, and momentum density (equivalently,
flow velocity) is reviewed. The corresponding conservation laws are not closed
since the energy and momentum fluxes are not specified as functionals of the
dependent variables. Instead they are defined as averages of specific
operators over a solution to the Liouville-von Neumann equation. That
solution is represented as a reference state plus its deviation. The reference
state is taken to be local equilibrium (also known as the ``relevant state"\cite{Zubarev1996}).
This is a composite state representing an equilibrium system at each space and
time point, constrained to have the same values for the average local
conserved densities as those for the actual non-equilibrium state. If only the hydrodynamic
fields were measured, the local equilibrium state would be indistinguishable
from that for the actual system - it is the best choice within information
entropy theory \cite{Jaynes1957a,Jaynes1957b,Robertson1993,Zubarev1996}. This local equilibrium
contribution to the fluxes gives results that are specific functionals of the
hydrodynamic fields. If deviations from the local equilibrium fluxes are
neglected, a closed set of equations for the fields is obtained. This is
referred to as ``perfect fluid hydrodynamics". The terminology is chosen since
these equations have no entropy production. The perfect fluid hydrodynamics is
local in time, but non-local in space. Hence there are no inherent limitations
on space and time scales. Calculation of the local equilibrium energy and
momentum fluxes is a problem closely related to that of DFT for the
thermodynamics of a non-uniform system. Specific limits of these perfect fluid
equations subsume all previous ``quantum hydrodynamics", and
important differences are noted. In the ``local density approximation"
where spatial non-locality is neglected these become the usual Euler equations
of hydrodynamics extended to quantum states.

In Section \ref{sec:dissipative_fluxes} the remainder of the energy and momentum fluxes, additions to the
local equilibrium contributions, are considered. These are defined by averages
over a special solution to the Liouville-von Neumann equation that is the
deviation from the local equilibrium ensemble. It is not actually a solution
but rather a representation of a solution in terms of the unknown macroscopic
fields of interest. In this way averages of the fluxes are given as
functionals of these fields, allowing the desired closure of the macroscopic
conservation laws. The results are similar to those known for linear
constitutive equations, with the equilibrium time correlation functions
replaced by corresponding non-linear local equilibrium time correlation
functions. They apply for both classical and quantum conditions, and no
limitations are imposed on length and time scales. With the energy and momentum fluxes determined in this way 
as functionals of the hydrodynamic fields, the macroscopic conservation laws become the
desired exact generalized hydrodynamic equations.

The derivation of the macroscopic local conservation laws in the next section
is based on the fundamental microscopic Heisenberg dynamics for the system.
This is contrasted with some earlier treatments starting from a macroscopic
mean field Hamiltonian \cite{Bloch1934,Yan2015,Bonitz2018}. The associated
approximate macroscopic Hamiltonian dynamics of the latter entails several
limitations, most importantly an irrotational velocity field, a mean field
free energy functional, and no dissipative contributions to the fluxes. The
details of the exact microscopic derivation here are relegated to the Supplemental Material, so the next section is devoted to their averages, notational
conventions, and their transformation from energy and momentum fields to the
more usual internal energy density and flow velocity fields. Also, the energy
and momentum fluxes are divided into a local equilibrium contribution and
dissipative contribution. The interpretation of each is discussed. Finally,
the objectives and results are reviewed in the last section with an outlook
for future developments.

As noted above, the approach to macroscopic transport given here follows closely the earlier applications of non-equilibrium statistical mechanics \cite{McLennan1989,Zubarev1974,McLennan1963,Wong1978,Dufty1979,Zubarev1996}. The present work updates this, and provides easier access to that complex work in the context of complex states not experimentally accessible until recently. It retains validity for quantum effects, non-linearity, strong coupling and spatial inhomogeneity. Being a formally exact representation of a continuum description, it subsumes the phenomenological versions developed in recent years, providing a structure for their critique and extension. Some examples of this connection to earlier work are given in the text and conclusions section.

On conclusion of this work a new review of quantum hydrodynamics appeared \cite{Bonitz2019},
with extensive critique of various approximations in the context of plasmas
and related electronic systems. Consequently, no attempt will be made here to
make contact with the rather complete discussion there of existing work. A
primary premise there is that phenomenological quantum hydrodynamics has often
been misused due to a lack of understanding of its validity conditions. The
need for strong theoretical underpinnings of any approximation or application
is stressed. The objective here is to provide an exact hydrodynamic
formulation as the basis for introduction of controlled applications.

The presentation given here is all within the context of a one component fluid with simple constituents (e.g., charges, atoms, molecules) governed by non-relativistic quantum mechanics. However, the philosophy of constructing an exact formulation for the macroscopic fields prior to the introduction of approximations extends to diverse other systems $\it{mutatis}$   $\it{mutandis}$  with different fields, (e.g. polymers, mixtures, broken symmetry states) or different dynamics (e.g., relativistic, granular/active). 

\section{Macroscopic Local Conservation Laws}
\label{sec:macro_local_conservation_laws}

\subsection{Microscopic local conservation laws}
\label{sub:microscopic_local_conservation_laws}
For simplicity of notation only a one component system is considered here and
below. A system of $N$ identical particles (Bosons or Fermions) contained in
a volume $\Omega$ enclosed by a surface $\sum$ has a Hamiltonian operator
given by%
\begin{equation}
H_{N}=\sum_{\alpha=1}^{N}(\frac{\bp_{\alpha}^{2}}{2m}+v^{\ext}(
\bq_{\alpha},t)) +\frac{1}{2}\sum_{\alpha \neq \beta=1}%
^{N}U(\left\vert \bq_{\alpha}-\bq_{\beta}\right\vert
) . 
\label{eq:hamiltonian}%
\end{equation}
Here $U(\left\vert \bq_{\alpha}-\bq_{\beta}\right\vert)$ is the interaction potential for the pair of particles at positions
$\bq_{\alpha},\bq_{\beta}$, and $v^{\ext}(\bq_{\alpha}) $ is an external single particle potential. The boundary
conditions are contained in the external potential by the condition
$v^{\ext}(\bq_{\alpha}) \to \infty$ for $\bq%
_{\alpha}\in\sum$. The state of the system is specified by a statistical
density operator $\rho$, defined in terms of its components for each $N$
particle Hilbert space, $\rho_{N}$, and normalized according to
\begin{equation}
%\rho_{N}=\overline{\rho}_{N}\calS_{N}\,, \quad 
\sum_{N>0}
\Tr^{(N)}\rho_{N}=1 \,. 
\label{eq:densitymatrix}%
\end{equation}
%\kl{what is $\bar{\rho}$ here?}
The definition of $\rho_{N}$ is such that it includes a restriction to the
appropriate subspace of symmetrized or anti-symmetrized elements of the $N$
particle Hilbert space.
% via the symmetrization operator $\calS_{N}$. 
For a given observable represented by the operator $A_{N}$ in the $N$ particle
Hilbert space its expectation value in the state $\rho$ is defined by%
\begin{equation}
  \overline{A}= \langle A;\rho \rangle \equiv\sum_{N>0}\Tr^{(N)}\rho_{N}A_{N} \,.
\label{eq:ensemble_avg}%
\end{equation}
The notation $\langle A;\rho \rangle$ instead of the simplified $\overline{A}$  will be used should any confusion arise.

The Hamiltonian without the external potential has all the continuous
symmetries of the Galilei group (time translations, spatial translation, rotations, boosts) as well as discrete symmetries of parity and time reversal.
Associated with the continuous symmetries are the usual conservation laws for
number, energy, linear momentum, and angular momentum. Here only point
particles (point sources of force) are considered so the relevant conservation
laws are those of number density, momentum density, and energy density. The
operators representing these densities are, respectively
\begin{subequations}
\begin{align}
  n(\br) &= \sum_{\alpha=1}^{N}\Delta(\br -\bq_{\alpha}) \,,
  \label{eq:def_density}%
  \\
  \bp (\br) &= \sum_{\alpha=1}^{N} \frac{1}{2}\left[ \bp_{\alpha},\Delta(
  \br-\bq_{\alpha}) \right]_{+} \,,
  \label{eq:def_momentum}%
  \\
  e(\br) &= \sum_{\alpha=1}^{N}\frac{1}{2}\left[ 
    \frac{\bp_{\alpha}^{2}}{2m}%+v^{\ext}(\bq_{\alpha})
 , \Delta( \br-\bq_{\alpha}) \right] _{+} \nonumber
 \\ 
 &\qquad+\frac{1}{2}\sum_{\alpha\neq\beta=1}^{N}U(\left\vert \bq%
_{\alpha}-\bq_{\beta}\right\vert) \Delta(\br-\bq_{\alpha}) \,,
\label{eq:def_energy}%
\end{align}%
\end{subequations}
where $[,]_{+}$ denotes the anticommutator, $\left[A,B\right]_{+} = AB+BA$.
Here $\Delta(\br) $ is a function localized about some
small domain centered at $\br$ and normalized to unity%
\begin{equation}
\int \dr \Delta(\br) =1. 
\label{eq:normalization}%
\end{equation}
Often in the literature the ideal case of a delta function is considered,
$\Delta(\br) \to\delta(\br)$. Exact microscopic local conservation laws follow directly from the
Heisenberg dynamics for the densities in ~\cref{eq:def_density,eq:def_energy,eq:def_momentum}. The analysis is straightforward but
lengthy. For completeness, the details are repeated in the Supplemental
Material, with the results%
\begin{subequations}
\begin{align}
  \partial_{t}n(\br,t) + m^{-1}%\frac{1}{m}
  \nabla\cdot \bp(\br,t) &= 0 \,,
  \label{eq:eom_density}
  \\
  \partial_{t}e(\br,t) + \nabla\cdot\bs( \br,t)&=-\frac{1}{m}\vec p(\vec r)\cdot\nabla v^{\rm ext}(\vec r,t) 
  \label{eq:eom_energy}
  \\
  \partial_{t}p_{i}(\br,t)+\partial_{j}t_{ij}(
  \br,t)&= -n(\br,t)\partial_{i}v^{\ext}(\br,t) \,.
  \label{eq:eom_momentum}
\end{align}
\end{subequations}
These operator equations are exact, with explicit expressions for the energy
flux $\bs(\br,t) $ and momentum flux $t_{ij}(\br,t) $ given in the Supplemental Material. A summation convention is adopted for repeated indices for Cartesian coordinates.

\subsection{Macroscopic local conservation laws}
\label{sub:macroscopic_local_conservation_laws}
The corresponding macroscopic averages of these microscopic equations (defined
as in %\cref{eq:eom_density,eq:eom_energy,eq:eom_momentum})
(\ref{eq:ensemble_avg}))
 give the corresponding macroscopic local conservation laws.
\begin{subequations}
\begin{align}
 & \partial_{t}\nbar(\br,t)+m^{-1} %\frac{1}{m}
  \nabla\cdot \bpbar(\br,t) = 0 \,,
  \label{eq:macro_eom_density}%
  \\
&  \partial_{t}\ebar(\br,t)+\nabla\cdot \overline{\bs}(\br,t) = -\frac{1}{m}\overline{\vec p}(\vec r)\cdot\nabla v^{\rm ext}(\vec r,t) \,, %\overline{w}(\br,t) \,,
  \label{eq:macro_eom_energy}
  \\
 & \partial_{t}\overline{p}_{i}(\br,t) +\partial_{j}
  \overline{t}_{ij}(\br,t)=-\nbar(\br,t) \partial_{i} v^{\ext}(\br,t) \,.
\label{eq:macro_eom_momentum}
\end{align}
\end{subequations}
%(\kl{work till here, 2019May06,22.24})
The sources in (\ref{eq:macro_eom_energy}) and (\ref{eq:macro_eom_momentum}) represent the average work done by the external force
and the average external force density, respectively. The microscopic conservation laws are independent of the state of the system
considered, while the macroscopic conservation laws are specific to the state.
Still, they are quite general and apply for both mixed and pure states,
classical and quantum. It remains to determine the fluxes such as
to provide a closed set of equations for the average fields
$\nbar(\br,t),\ebar(\br,t)$, and $\bpbar(\br,t)$.

Instead of the average momentum density, it is traditional to consider the
flow velocity field $\bu(\br,t) $ defined by%
\begin{equation}
  \bpbar(\br,t) \equiv m\nbar(\br,t) \bu(\br,t)\,.
  \label{eq:flowvelocity}
\end{equation}
Also, it is useful to extract the purely convective parts of the energy
density and fluxes of energy and momentum \cite{McLennan1989}
\begin{subequations}
\begin{align}
	\ebar(\br,t) &= \ebar_{0}(\br,t) +\frac{1}{2}m\nbar(\br,t) u^{2}(\br,
	t) \label{eq:density}%
	\\
	\overline{s}_{i}(\br,t)
	&= \overline{s}_{0i}(\br,t) +\ebar(\br,t) u_{i}(\br,t) + \overline{t}_{0ij}(\br,t) u_{j}(\br,t) \label{2.16}%
	\\
	\overline{t}_{ij}(\br,t)
	&=  \overline{t}_{0ij}(\br,t) 
	+m\nbar(\br,t) u_{i}(\br,t) u_{j}(\br,t). \label{2.17}%
\end{align}%
\end{subequations}
The operators with a subscript $0$ are the same as those without the
subscript, except with all particle momentum operators replaced by
$\bp_{\alpha} \to\bp_{\alpha} -m\,\bu(\br, t)$. Hence the corresponding averages are those quantities in the
local rest frame of the fluid at point $\br$. For example,
$ \overline e_{0}(\br,t)  $ is
the average local internal energy density. 

%To simplify the notation in the
%following, define the average number density and internal energy density by%
%\begin{equation}
%\nbar(\br,t) \equiv n(\br,t) ,\hspace{0.25in}\ebar_{0}(\br,t) \equiv e_{0}(\br,t)  . \label{2.17a}%
%\end{equation}
The local conservation laws 
%for $\nbar(\br,t),\ebar_{0}(\br,t)$, and $\bu(\br,t)$
become
\begin{widetext}
\begin{subequations}
\begin{align}
	D_{t}\nbar(\br,t) +\nbar(
	\br,t) \nabla\cdot\bu(\br,t) &= 0 \,,
	\label{eq:eom_nbar}%
	\\
	D_{t}\ebar_{0}(\br,t) +\ebar_{0}(
	\br,t) \nabla\cdot\bu(\br,t)
	+ \overline{t}_{0ij}(\br,t) \partial_{i} u_{j}(\br,t) +\nabla\cdot
	\overline{\bs}_{0}(\br,t) &=- \nbar(
	\br,t)  \bu(\br,t)\cdot\nabla v^{\rm ext}(\vec r,t) \,, 
	\label{eq:eom_e0}%
	\\
	m\nbar(\br,t) D_{t} u_{i}(\br,t) +\partial_{j} \overline{t}_{0ij}(\br,t) &= -\overline{n}(\br,t) \partial_{i} v^{\ext}(\br,t) \,, 
	\label{eq:eom_u}%
\end{align}%
\end{subequations}
\end{widetext}
where $D_{t}=\partial_{t}+\bu \cdot\nabla$
is the material derivative. The dependent fields are now $\nbar(
\br,t),\ebar_{0}(\br,t),$ and
$\bu(\br,t) .$  The utility of these exact equations depends on determination of the energy and
momentum fluxes, $ \overline{\bs}_{0}(\br,t)$ and $\overline{t}_{0ij}(\br,t)$, in the local rest frame. 

\subsection{Analysis of energy and momentum fluxes}

For a closed set of equations,
the required local rest frame fluxes should be expressed as functionals of the fields so that the macroscopic
conservation laws would become a closed, deterministic set of equations for
the fields. That is the final objective of the presentation here. It is
accomplished in two steps. The first is to identify the result for a ``perfect
fluid" without dissipation. Next the remainder responsible for dissipation is
obtained from a formal solution to the Liouville-von Neumann equation.

Note that the time dependence can be shifted to the state $\rho$ using the
cyclic invariance of the trace
\begin{align}
\overline{\bs}_{0} (\br,t)
%= \langle \bs_{0}(\br,t); \rho \rangle
&= \langle \bs_{0}(\br);\rho(t) \rangle,
\\
\overline{t}_{0ij} (\br,t)
%= \langle t_{0ij}(\br,t); \rho \rangle 
&= \langle t_{0ij}(\br); \rho(t) \rangle\,, \label{2.22}%
\end{align}
which in turn requires the solution to the Liouville-von Neumann equation for 
$\rho_N(t) $%
\begin{equation}
\partial_{t}\rho_{N}\left(  t\right)  +\mathcal{L}_{N}\left(  t\right)
\rho_{N}\left(  t\right)  =0 \label{eq13}%
\end{equation}
where $\mathcal{L}_N$ is the Liouville operator defined by%
\begin{equation}
\mathcal{L}_{N}\left(  t\right)  X\equiv i\left[  H_{N}\left(  t\right)  ,X\right],
\label{eq13a}%
\end{equation}
for each $N$, for any operator $X$.

The solution is written as the sum of a reference state $\rho^{\ell}_N\left[
y(t)\right] $ plus its deviation $\Delta_N(t) $
\begin{equation}
\rho_{N}(t)=\rho_{N}^{\ell}\left[ y\left(  t\right)  \right]  +\Delta
_{N}\left(  t\right).
\label{2.24}%
\end{equation}
The reference state is entirely\ determined by a set of conjugate fields
$\left\{y(t)\right\}$  in one-to-one correspondence with the
macroscopic conserved fields. This correspondence is defined by the
requirements%
\begin{subequations}
\begin{align}
	\nbar^{\ell} (\br| y(t)) &\equiv\nbar(\br,t) \,,
	\label{eq:nequiv}
	\\
	\ebar_{0}^{\ell}(\br| y(t)) &\equiv\ebar_{0}(\br,t) \,,
	\label{eq:e0equiv}%
	\\
	\bpbar^{\ell}(\br| y(t))
	&\equiv m\nbar(\br,t)\bu(\br,t) \,,
	\label{eq:pequiv}
\end{align}
\end{subequations}
where the superscript $\ell$ denotes a reference ensemble average, $\overline{A}^{\ell} = \langle A; \rho^{\ell} \rangle$.
The left sides of these equations are functionals of the conjugate fields
while the right sides are the fields of the local conservation laws. In this
way the conjugate fields $\left\{y(t)\right\} $ are functionals of the average
conserved fields $\nbar(\br,t),\ebar%
_{0}(\br,t),\bu(\br,t)$, and vice versa.
The reference state therefore has the exact average values for the conserved
fields by construction.

A choice for $\rho^{\ell}$ with these properties is the local equilibrium
ensemble \cite{McLennan1989,Zubarev1996}. To motivate it, note that the grand ensemble for a system at rest is%
\begin{equation}
\rho_{eN}\left[ \beta,\nu\right] =e^{-Q\left[ \beta,\nu\right] }e^{-\beta
H+\nu N}  \,, %\calS_{N} 
\label{eq:eqensemble}
\end{equation}
where by normalization
\begin{equation}
Q\left[\beta,\nu\right] =\ln \sum_{N>0}
\Tr^{(N)}e^{-\beta H+\nu N} \,. %\calS_{N} \,, 
\label{eq:Qnormalization}%
\end{equation}
%with the symmetrization operator $\calS_{N}$.
The corresponding result for a system moving with velocity $\bw$ is
obtained by the replacement (transformation to the moving frame) $H \to H-\bw \cdot \bP + \frac{1}{2}m N w^{2}$, where $\bP$ is the total
momentum operator. Now consider an assembly of equilibrium systems where
$\beta,\nu,\bw$ vary at each point in space and time
\begin{widetext}
	\begin{equation}
	\rho^{\ell}_N\left[ \beta(t),\nu(t),\bw(t)
	\right] \equiv e^{-Q^\ell\left[ \beta(t),\nu(t),\bw(
	t) \right] }e^{-\int \dr\left[ \beta(\br,t)(
	e(\br)-\bw(\br,t)
	\cdot \bp (\br) +\frac{1}{2}m w^{2}(
	\br,t) n(\br))-\nu(
	\br,t) n(\br) \right] }\,, %\calS_{N},
	\label{eq:localensemble}%
	\end{equation}
	where
\begin{equation}
Q^\ell\left[ \beta(t),\nu(t),\bw(t) \right]
= \ln\sum_{N>0}
\Tr^{(N)}e^{-\int \dr\left[ \beta(\br,t)(e(
\br)-\bw(\br,t) \cdot
\bp(\br) +\frac{1}{2}mw^{2}(\br,t) n(\br))-\nu(\br,t) n(\br%
) \right] }\,.  %\calS_{N} \,. 
\label{eq:Qnorm2}%
\end{equation}
\end{widetext}
The operators $N, H,\bP$ have been replaced accordingly by the
associated operators for their densities 
\begin{equation}
 \left\{ \psi_\kappa(\br) \right\} \equiv\left\{n(\br), e(\br),\bp(\br) \right\}.
 \label{densitieslist}
\end{equation}
 In this notation the local equilibrium ensemble becomes 
 \begin{equation}
 \rho^{\ell}_N [\beta(t),\nu(t),\vec w(t)] = e^{-\eta}, \; \eta = Q^\ell +\!\! \int \dr \psi_\kappa (\br) y_\kappa (\br) ,
\end{equation}
where summation over repeated indices is implied.
%above is referred to in supplementary material
The ``conjugate fields" of this local equilibrium ensemble are the coefficients of
the corresponding conserved field operators %$n(\br), e(\br),\bp(\br) $
\begin{align}
\left\{ y(\br, t) \right\} \leftrightarrow\left\{-
\nu(\br,t)+ \beta(\br,t)
m w^{2}(\br,t)/2, \right. \nonumber
\\
\left.\beta(\br,
t),-\beta(\br,t)\bw(\br,t)
\right\} \,.
\label{eq:localensemblec}%
\end{align}
It is interesting to note that this local equilibrium ensemble is also the ``best choice" in the sense that it maximizes the information entropy for the
given values of the conservative fields \cite{Jaynes1957a,Jaynes1957b,Robertson1993}. 
Furthermore, the local equilibrium state describes a ``perfect fluid" in the
sense that there is no dissipation, as discussed in the next section.

With $\rho^{\ell}_N$ specified, the left sides of Eqs.  (\ref{eq:nequiv})--(\ref{eq:pequiv})
%(\ref{eq:localensemble}) and (\ref{eq:localensembleb}) \kl{which is eq:localensembleb?}
can be calculated as functionals of $\beta(t),\nu(t)
,\bw(t) $. Inverting these equations 
%Eqs. (\ref{eq:nequiv}),(\ref{eq:e0equiv}),  and (\ref{eq:pequiv}) %(\ref{2.24b})
%\kl{which two equations?}
then gives these auxiliary fields as functionals of the average conserved fields of interest, and consequently%
\begin{equation}
\rho^{\ell}_N\left[ \beta(t),\nu(t),\bw(t)
\right] \to\rho^{\prime \ell}_N\left[\nbar(t)
,\ebar_{0}(t),\bu(t) \right] \,.
\label{eq:localensemblea}%
\end{equation}
 To simplify and clarify the notation the fields
$\overline{n}\left(  \mathbf{r},t\right)  ,\overline{e}_{0}\left(
\mathbf{r},t\right)  ,$ and $\overline{\mathbf{u}}\left(  \mathbf{r},t\right)
$ will be denoted%
\begin{equation}
\left\{  \zeta\left(  \mathbf{r},t\right)  \right\} \equiv \left\{
\overline{n}\left(  \mathbf{r},t\right)  ,\overline{e}_{0}\left(
\mathbf{r},t\right)  ,\overline{\mathbf{u}}\left(  \mathbf{r},t\right)
\right\}
\end{equation}
so that (\ref{eq:localensemblea}) becomes
\begin{equation}
\rho_{N}^{\ell}\left[  y(t)\right]  =\rho_{N}^{\prime\ell}\left[
\zeta(t)\right]  .
\end{equation}
All local equilibrium averages therefore become functionals of the conjugate
variables, or equivalently of the average conserved fields. To make these two
representation clear the notation in the following will be
\begin{equation}
\left\langle X\left(  \mathbf{r}\right)  ;\rho^{\ell}\left[  y(t)\right]
\right\rangle \equiv\overline{X}^{\ell}\left(  \mathbf{r\mid}y(t)\right)
\equiv\overline{X}\left(  \mathbf{r\mid}\zeta(t)\right)  ,
\label{eq27}
\end{equation}
where the one-to-one relationship of (\ref{eq:nequiv}) - (\ref{eq:pequiv}) can be expressed as
\begin{equation}
y(\mathbf{r}\text{,}t)=y(\mathbf{r\mid}\zeta(t))\mbox{ or its inverse }%
\zeta(\mathbf{r}\text{,}t)=\zeta(\mathbf{r\mid}y(t))  .%
\end{equation}
This functional relationship is local in time. As a specific example the local
equilibrium fluxes are%
\begin{equation}
\left\langle \mathbf{s}_{0}\left(  \mathbf{r}\right)  ;\rho^{\ell}\left[
y(t)\right]  \right\rangle =\overline{\mathbf{s}}_{0}^{\ell}\left(
\mathbf{r\mid}y(t)\right)  =\overline{\mathbf{s}}_{0}\left(  \mathbf{r\mid
}\zeta(t)\right) \label{eq:s0local} %
\end{equation}%
\begin{equation}
\left\langle t_{0ij}\left(  \mathbf{r}\right)  ;\rho^{\ell}\left[
y(t)\right]  \right\rangle =\overline{t}_{0ij}^{\ell}\left(  \mathbf{r\mid
}y(t)\right)  =\overline{t}_{0ij}\left(  \mathbf{r\mid}\zeta(t)\right)
\label{eq:t0local} %
\end{equation}

Since the operators $\bs_{0}(\br) $ and
$t_{0 ij}(\br) $ are given by the derivation in the Supplemental Material,
 the functionals %$\overline{\bs}_{0}(
%\br|\nbar(t),\ebar_{0}(t),\bu%
%(t)) $ and $\overline{t}_{0ij}(\br |\nbar(t),\ebar_{0}(t),\bu(t)
%) $ 
$\overline{\bs}_{0}(
\br|\zeta(t)) $ and $\overline{t}_{0ij}(\br |\zeta(t)) $ 
are given as explicit local equilibrium averages, and can be taken
as formally known. In particular, 
%in the classical limit 
it can be shown (see Appendix \ref{ap:A}) that
%\begin{align}
%\overline{\bs}_{0}(\br |\nbar(t)
%,\ebar_{0}(t),\bu(t))
%&\to 0,
%\\
%\overline{t}_{0,ij}(\br |%
%\nbar(t),\ebar_{0}(t)
%,\bu(t)) &\to\delta_{ij} \mathrm{p}(
%\br |\nbar(t),\ebar_{0}(t)
%) . \label{2.27}%
%\end{align}

\begin{align}
\overline{\bs}_{0}(\br |\zeta(t))
&= 0,
\\
\overline{t}_{0ij}(\br |\zeta(t)) &=
%\delta_{ij} 
\overline{ {\pi}}_{ij}(
\br |\nbar(t),\ebar_{0}(t)
) . \label{2.27}%
\end{align}
Here, $\overline{{\pi}}_{ij}(\br |\nbar(t),\ebar%
_{0}(t)) $ is the equilibrium pressure tensor for a non-uniform
system as a functional of the local density and internal energy density, independent of the flow velocity. With these results the macroscopic conservation laws become%
\begin{widetext}
	\begin{subequations}
\begin{align}
 D_{t}\nbar(\br,t) +\nbar(
\br,t) \nabla\cdot\bu(\br,t) &=0\,,
\label{2.29}%
\\
%& D_{t}\ebar_{0}(\br,t) +\ebar_{0}(
%\br,t) \nabla\cdot\bu(\br,t)
%+\overline{t}_{0,ij}(\br |\nbar(t)
%,\ebar_{0}(t),\bu(t)) \partial_{i}%
%u_{j}(\br,t) +\nabla\cdot\overline{\bs}_{0}(
%\br |\nbar(t),\ebar_{0}(t)
%,\bu(t)) \nonumber
%\\
%&\qquad +t_{0,ij}^{\ast}(\br,t) \partial_{i}u_{j}(\br,t) +\nabla\cdot\bs_{0}^{\ast}(\br,t) = 0 
%\label{2.30}%
 D_{t}\ebar_{0}(\br,t) +\ebar_{0}(
\br,t) \nabla\cdot\bu(\br,t)
%+\overline{t}^{\ell}_{0,ij}(\br | \zeta(t))
+\overline{{\pi}}_{ij}(\br | \zeta(t)) \partial_{i}%
u_{j}(\br,t) 
%+\nabla\cdot\overline{\bs}^{\ell}_{0}(\br | \zeta(t)) 
\nonumber
\\
\qquad +\overline{t}_{0ij}^{\ast}(\br,t) \partial_{i}u_{j}(\br,t) +\nabla\cdot \overline{\bs}_{0}^{\ast}(\br,
t) &= -\overline n(\vec r,t)\vec u(\vec r,t)  \cdot\mathbf{\nabla}%
v^{\rm ext}\left(  \mathbf{r},t\right) \,,
\label{2.30}%
\\
 m\nbar(\br,t) D_{t}u_{i}(\br,
t) +\partial_{j}\overline{{\pi}}_{ij}(\br | \zeta(t)) + \partial_{j} \overline{t}_{0ij}^{\ast}(\br,t) 
&= -\nbar(
\br,t) \partial_{i}v^{\ext}(\br,t) \,.
\label{2.31}%
\end{align}
\end{subequations}
\end{widetext}
The fluxes have been separated explicitly%
%\begin{align}
%	\overline{\bs}_{0}(\br,t)
%	 &=\overline{\bs}_{0}(\br |\overline
%	{n}(t),\ebar_{0}(t),\bu(
%	t)) +\bs_{0}^{\ast}(\br,t)
%	\label{2.31a}%
%	\\
%	 \overline{t}_{0,ij}(\br,t)
%	 &=\overline{t}_{0,ij}(\br |\nbar(
%	t),\ebar_{0}(t),\bu(t)) +t_{0,ij}^{\ast}(\br,t) \,.
%\label{2.31b}%
%\end{align}
%
\begin{align}
\overline{\bs}_{0}(\br,t)
&=
%\overline{\bs}^{\ell}_{0}(\br |\zeta(t)) 
%+
\overline{\bs}_{0}^{\ast}(\br,t)
\label{2.31a}%
\\
\overline{t}_{0ij}(\br,t)
&=\overline{{\pi}}_{ij}(\br | \zeta(t))
%\overline{t}^{\ell}_{0ij}(\br |\zeta(t)) 
+\overline{t}_{0ij}^{\ast}(\br,t) \,.
\label{2.31b}%
\end{align}
The contributions to the fluxes $\overline{\bs}_{0}^{\ast}(\br,
t) $ and $\overline{t}_{0ij}^{\ast}(\br,t) $ are those from
$\Delta_N(t)$ in (\ref{2.24})
\begin{equation}
\overline{\bs}_{0}^{\ast}(\br,t) \equiv
\langle \bs_{0}(\br) ;\Delta(t) \rangle \,,\; \overline{t}_{0ij}^{\ast}(\br,t)
\equiv \langle t_{0ij}(\br) ;\Delta(t) \rangle \,, \label{2.32}%
\end{equation}
and are discussed in Section \ref{sec:dissipative_fluxes}.

Equations (\ref{2.29})-(\ref{2.31}) are the basis for all analysis of
continuum descriptions described here. They are still exact and general, with
the constitutive equations for $\overline{\bs}_{0}^{\ast}$ and $\overline{t}_{0ij}^{\ast}$ to be determined for specific states of interest. They remain to be
specified from a solution to the Liouville-von Neumann equation.
Explicit expressions are well-known for states with small space and time
variations. In that case, the fluxes are obtained to first order in the
spatial gradients of the fields, with transport coefficients given by
Green-Kubo time correlation functions. For the fluid phase these are Fourier's
law for the energy flux and Newton's viscosity law for the momentum flux. The
resulting continuum equations are the Navier-Stokes hydrodynamic equations.
The exact general form for arbitrary states is discussed in Section \ref{sec:dissipative_fluxes}.

\section{Perfect fluid hydrodynamics}
\label{sec:perfect_fluid_hydrodynamics}
In this section it is first shown that the contributions to
dissipation (entropy production) are entirely due to the components of the
fluxes $\overline{\bs}_{0}^{\ast}$ and $\overline{t}_{0ij}^{\ast}$. Hence, their neglect
results in a ``perfect fluid hydrodynamics." In the local density approximation
(see below) they are the usual Euler-level hydrodynamics. More generally they
extend the Euler equations to strong spatial non-locality.

%\kl{remove the entropy production section?}
\subsection{Entropy production}
\label{sub:entropyproduction}
The entropy and local entropy are defined here from the information entropy,
maximized for the given average conserved fields \cite{Jaynes1957a,Jaynes1957b,Robertson1993}
%\kl{which citation?}
\begin{equation}
S=-\max \ln P\left[ y(t) \right] 
_{P}= \overline{\eta}^{\ell}[y(t)]
%\langle \eta\left[ y \right] ;t \rangle ^{\ell}
\label{3.01}%
\end{equation}
where
\begin{equation}
\eta^\ell[y] \equiv  -\ln \rho^{\ell} = Q^{\ell}\left[ y(t) \right] + %\sum_{\kappa}
\int \dr y_{\kappa}(\br,t)\psi_{\kappa}(\br)
\end{equation}
is the index function for the local equilibrium
ensemble in (\ref{eq:localensemble}), so (\ref{3.01}) becomes 
%\kl{what is $\eta$ here? I am confused...}
%\kl{check the changes here.}
\begin{equation}
S\left[ y(t) \right] =Q^{\ell} \left[ y(t) \right]
+ %\sum_{\kappa} 
\int \dr y_{\kappa}(\br,t) \overline{\psi}_{\kappa}^{\ell}
(\br|y(t)). \label{3.02}%
\end{equation}
From Eq. (\ref{densitieslist}) the operators for the conserved fields are denoted by $\psi_{\kappa}(
\br) $ ($\kappa=1,2,3$)
and the conjugate average fields are those of Eq. (\ref{eq:localensemblec}).
\begin{comment}
\begin{equation}
\left\{y_{\kappa}(\br)\right\}=\left\{ \left[-\nu(\br)+\frac{\beta(\br,t)}{2}%
m w^{2}(\vec r) \right],\beta(\br),-\beta(\br%
) \bw(\br)\right\} \,. 
\label{3.04}%
\end{equation}
\end{comment}
They are defined in terms of the average conserved fields by the conditions \cref{eq:nequiv,eq:e0equiv,eq:pequiv}.
%(\ref{2.24a}) and (\ref{2.24b}) 
%\kl{We need to rename all the labels of the equations wherever a question mark appears!}
%\begin{equation}
% \psi_{\kappa}(\br) ;t^{\ell}=\overline{\psi}_{\kappa}(\br,t),\psi_{\kappa}(\br) ;t ^{\ell}=\overline{\psi}_{\kappa}(\br,t) \label{3.05}%
%\end{equation}
%Normalization requires%
In this notation, Eq. (\ref{eq:Qnorm2}) becomes
\begin{equation}
Q^{\ell}\left[ y(t) \right] =
\ln \sum_{N}
\Tr^{(N)}e^{-%\sum_{\kappa} 
\int \dr y_{\kappa}(\br,t)\psi_{\kappa}(\br%
)}. \label{3.06}%
\end{equation}

The entropy production is defined by%
\begin{equation}
\sigma (t)  =\partial_{t}S\left[ y(t) \right] = \int \dr \, y_{\kappa}(\br,t)\partial
_{t} \overline{\psi}_{\kappa}^{\ell}(\br\,|y(t)). \label{3.07}%
\end{equation}
Next use (\ref{eq:nequiv}) -
(\ref{eq:pequiv}),  which define the conjugate fields in terms of the average conserved
fields, to write%
\begin{equation}
\partial_{t}\overline{\psi}_{\kappa}^{\ell}\left(  \mathbf{r}|y(t)\right)
=\partial_{t}\overline{\psi}_{\kappa}\left(  \mathbf{r}, t\right)
\end{equation}
Then using  %(\ref{b.4})
% and 
the macroscopic conservation laws,
\begin{equation}
%\partial_{t} \overline{\psi}_{\kappa}^{\ell}(\br,t) =
\partial_{t}\overline{\psi}_{\kappa}(
\br,t) =-\partial_{j} \overline{\gamma}_{j\kappa}%
(\br,t) + \overline{f}_{\kappa}(\br,t)\,, \label{3.08}%
\end{equation}%
%\kl{definition of $\gamma_{i\kappa}$}
where
\begin{align}
 \overline{\gamma}_{j\kappa}(\br,t)
&= \left\{\overline n(\vec r,t)u_j(\vec r,t), %\frac{\overline{p}_{i}(\br,t)}{m},
 \overline{s}_{j}(\br,t), \overline{t}_{ij}(\br,t)\right\}
 \,,
\\
 \overline{f}_{\kappa}(\br,t)
&=\left\{ 0, -\overline n(\vec r,t)\vec u(\vec r)  \cdot\mathbf{\nabla}%
v^{{\rm ext}}\left(  \mathbf{r},t\right),-\nbar(\br,t)\partial_{i}v^{\ext}(
\br,t) \right\} \,.
\label{3.09}%
\end{align}
The entropy production becomes%
\begin{align}
\sigma (t)    
%= & -\int \dr y_{\kappa}(\br,t)\partial_{i} \overline{\gamma}_{i\kappa}(\br,t) \nonumber
%\\
%& -\int \dr\beta(\br)\nbar(\br,t)\bw(\br)\cdot \nabla v^{\ext}(\br)
%\nonumber\\
=&  \int \dr \, \overline{\gamma}_{i \kappa}(\br,t)
\partial_{i}y_{\kappa}(\br,t)+\sigma_{\ext}(t),
\label{3.010}%
\end{align}
where $\sigma_{\ext}(t) $ is the entropy production due to the
external force%
\begin{equation}
\sigma_{\ext}(t) = \int \dr y_{\kappa}(\br,t) \overline{f}_{\kappa}(\br,t)
%-\int \dr\beta(\br)\overline{n}(\br,t)\bw(\br)\cdot \nabla v^{\ext}(\br) 
\,.
\label{3.010a}%
\end{equation}
%\kl{will work out the missing term due to energy source}

To interpret this further write $ \overline{\gamma}_{i \kappa}(\br,t) = \overline{\gamma}^{\ell}_{i \kappa}(\br|y(t))
+\overline{\gamma}_{i \kappa}^{\ast}(\br,t)$ and use the cyclic
invariance of the trace to show that the contributions from $
\overline{\gamma}_{i \kappa}^{\ell}(\br| y(t)) $ and the external force
cancel (see Eq. (\ref{a.17}) of Appendix \ref{ap:A})%
\begin{align}
0&=\sum_{N}
\Tr^{(N)}\calL \rho^{\ell}[y(t)] \nonumber
\\
&=\int \dr
\overline{\gamma}_{i \kappa} ^{\ell}(\br|y(t)) \partial_{i}y_{\kappa
}(\br,t)+\sigma_{\ext}(t) . \label{3.011}%
\end{align}
Consequently all of the entropy production is due to $\overline{\gamma
}_{i \kappa}^{\ast}(\br,t)$%

\begin{equation}
\sigma (t)   =\int \dr \, \overline{\gamma
}_{i \kappa}^{\ast}(\br,t)\partial_{i}y_{\kappa}(\br,t).
\label{3.012}%
\end{equation}
All dissipation is associated with $\overline{\gamma}_{i \kappa}^{\ast
}(\br,t)$.

\subsection{Hydrodynamics without dissipation}

A special case of interest is conditions for which the irreversible energy and
momentum fluxes can be neglected, $\overline{s}_{0i}^{\ast}\to 0$ and
$\overline{t}_{0ij}^{\ast}\to 0$. This is never strictly true but can be
a reasonable approximation for certain flows.%
%\begin{subequations}
%\begin{align}
%D_{t}\nbar(\br,t) +\nbar(
%\br,t) \nabla\cdot\bu(\br,t) &=0 \,,
%\label{3.1}%
%\\
%D_{t}\ebar_{0}(\br,t) +\ebar_{0}(
%\br,t) \nabla\cdot\bu(\br,t)
%%+\overline{p}_{ij}(\br |\nbar(t),\ebar_{0},\bu(t))
%+\overline{t}_{ij}(\br,t)
% \partial_{i} u_{j}(\br,t) &= 0 \,, %w(\br,t)  \,.
%\label{3.2}%
%\\
%m\nbar(\br,t) D_{t}u_{i}(\br,t) %+\partial_{j}{p}_{ij}(\br|\nbar(t),\ebar_{0},\bu(t))
%+\partial_{j}\overline{t}_{ij}(\br,t)
% &= -\nbar(
% \br,t) \partial_{i}v^{\ext}(\br) %\overline{f}_{i}(\br,t)  
% \,. 
%\label{3.3}%
%\end{align}
%\end{subequations}
\begin{widetext}
	\begin{subequations}
		\begin{align}
		& D_{t}\nbar(\br,t) +\nbar(
		\br,t) \nabla\cdot\bu(\br,t) =0,
		\label{eq:eom_density_nodissipation}%
		\\
		& D_{t}\ebar_{0}(\br,t) +\ebar_{0}(
		\br,t) \nabla\cdot\bu(\br,t)
		%+\overline{t}^{\ell}_{0ij}(\br | \zeta(t)) \partial_{i}%
		+\overline{{\pi}}_{ij}(\br|\zeta(t)) \partial_{i} u_{j}(\br,t) 
		= -\overline n(\vec r,t)\vec u(\vec r,t)  \cdot\mathbf{\nabla}%
v^{\rm ext}\left(  \mathbf{r},t\right)
		\label{eq:eom_e0_nodissipation}%
		\\
		& m\nbar(\br,t) D_{t}u_{i}(\br,
		t) +\partial_{j}\overline{{\pi}}_{ij}(\br|
		\zeta(t)) 
%		+ \partial_{j} \overline{t}_{0ij}^{\ast}(\br,t) 
        = -\nbar(
		\br,t) \partial_{i}v^{\ext}(\br,t) \,.
		\label{eq:eom_u_nodissipation}%
		\end{align}
	\end{subequations}
\end{widetext}
%\kl{define $\mathrm{p}_{ij}$ here?}
The resulting continuum equations are referred to as the perfect fluid equations. These equations have no unknown components (except the external
forces to be chosen) and hence comprise a closed set of equations for the
fields, justifying the terminology ``hydrodynamics."
%\kl{May07 13:23, work on it later}1
The explicit functional dependence of 
%the average heat flux 
%$\overline{\bs}^{\ell}_{0}(\br |\zeta(t)) $ and
${\pi}_{ij}(\br |\zeta(t) ) $ 
%$\overline{\bs}_{0}(\br |\nbar(t),\ebar_{0}(t),\bu(t)) $ and
%$\overline{t}_{ij}(\br |\nbar(t),\ebar_{0}(t),\bu(t)) $
on the fields
requires evaluation of the local equilibrium average, Eq. (\ref{2.27}).

%\begin{align}
%\overline{\bs}_{0}(\br |\zeta(t)
%%\nbar(t),\ebar_{0}(t),\bu(t)
%)
%&=\overline{s}^{\ell}_{0}(\br,t) \,,
%%\langle \bs_{0}(\br) ;\rho^{\ell}(t) \rangle,
%\\
%\overline{t}_{ij}(\br |\zeta(t) 
%%\nbar(t),\ebar_{0}(t),\bu(t)
%) 
%&=\overline{t}^{\ell}_{0,ij}(\br,t) \,.
% %\langle t_{0,ij}(\br) ;\rho^{\ell}(t) \rangle  \,. 
%\label{3.5}%
%\end{align}
The simplest approximation is the ``local density approximation" whereby the
functionals are replaced by functions of the fields at the point of interest
$\br$. The local equilibrium ensemble defined as a functional of the
conjugate fields then becomes the equilibrium grand canonical ensemble,
defined as a \textit{function} of the fields at $\br$,
\begin{equation}
\rho^{\ell}_N\left[ \beta(t),\nu(t),\bw(t)
\right] \to\rho^{e}(\beta(\br,t),\nu(
\br,t),\bw(\br,t)) \,.
\label{3.5d}%
\end{equation}
Similarly the relation of these conjugate fields to the average conserved
fields, (\ref{densitieslist}), becomes%
\begin{equation}
 \langle \psi_{\kappa}(\br) ;\rho^{e}(
\beta(\br,t),\nu(\br,t),\bw(
\br,t)) \rangle  =\overline{\psi}_{\kappa}(
\br,t) . \label{eq:eq}%
\end{equation}
The local equilibrium averages on the left side then become equilibrium averages. Then (\ref{eq:eq}) expresses them as {\it functions} of the non-equilibrium conserved fields.
%\kl{Jeff's comment is not addressed here}
%with the results and 
The associated pressure tensor is
\begin{equation}
\overline{t} _{0ij}(\br |\zeta(t)
%\nbar(t),\ebar_{0}(t),\bu(t)
)   \to\delta_{ij}\overline{\pi}^{e}(\nbar(\br,t)
,\ebar_{0}(\br,t)) \,. 
\label{3.5b}%
\end{equation}
%It has been recognized in this limit that $\bw(\br,t)=\bu(\br,t) .$ 
The problem then reduces to a
determination of the equilibrium pressure for the system being considered.
This is a two step process. First calculate the pressure as a function of
$\beta(\br,t),\nu(\br,t) $ and
then use the equilibrium relations, $\beta(\br,
t) =\beta_{e}(\nbar(\br,t)
,\ebar_{0}(\br,t)),\nu(
\br,t) =\nu_{e}(\nbar(\br,t)
,\ebar_{0}(\br,t)) $ to express the
pressure as a function of $\nbar(\br,t)
,\ebar_{0}(\br,t) $. This is the usual procedure
in the derivation of Navier-Stokes hydrodynamics for states with small spatial
gradients and hence the local density approximation is justified. More
generally, it is an uncontrolled assumption that must be removed for states
with strong spatial inhomogeneities. 
%Practical forms for this case are described in Appendix \ref{ap:B}.

\subsection{Pressure tensor}
\label{sub:pressuretensor}
The pressure tensor and other local equilibrium averages simplify by transforming to a local rest frame to eliminate their dependence on the flow field.
The local equilibrium ensemble depends on the velocity field only through
$p_{\alpha j}-mw_j(\vec q_\alpha) $. A unitary
transformation can be performed to $\mathbf{p}_{\alpha}^{\prime} 
=\mathbf{p}_{\alpha} -m\mathbf{w}(\vec q_\alpha)   $ (see
Supplemental Material) 
\begin{equation}
\mathbf{p}^{\prime}_\alpha =e^{-\frac{1}{i\hbar}G}\left(  \mathbf{p}_\alpha -m\mathbf{w}(\vec q_\alpha)
   \right)  e^{\frac{1}{i\hbar}G}, \label{2.11a}%
\end{equation}
where $G([ \vec q_\alpha])$  is the generation of the transformation. Therefore
\begin{equation}
\rho^{\ell}_N\left[  \beta(t),\nu(t)  ,\mathbf{0}\right]
=e^{-\frac{1}{i\hbar}G}\rho^{\ell}_N\left[  \beta(t),\nu(t)
,\mathbf{w}(t)  \right]  e^{\frac{1}{i\hbar}G}. \label{2.11b}%
\end{equation}
In particular the average momentum density can be written
\begin{align}
\left\langle  {\mathbf{p}} \left(  \br \right)  ;\rho^{\ell}\left[
y(t)\right]  \right\rangle  &  =\left\langle e^{-\frac{1}{i\hbar}G}%
 {\mathbf{p}} \left(  \br \right)  e^{\frac{1}{i\hbar}G};e^{-\frac{1}%
{i\hbar}G}\rho^{\ell}[y(t)]  e^{\frac{1}{i\hbar}G}\right\rangle
\nonumber\\
&  =m\nbar\left(  \br,t\right)  \mathbf{w}\left(  \br%
,t\right)  . \label{2.11c}%
\end{align}
Comparison to the condition (\ref{eq:pequiv}) shows
\begin{equation}
\mathbf{w}\left(  \br,t\right)  =\mathbf{u}\left(  \br,t\right)
. \label{2.11d}%
\end{equation}

Let $A_{0}\left(  \mathbf{r;u}\left(  \br,t\right)  \right)  $ be an
operator in the local rest frame; i.e. depends on the particle momenta through
$\mathbf{p}_{\alpha}^{\prime}=\mathbf{p}_{\alpha}-m\mathbf{u}\left(
\vec q_\alpha,t\right)  .$ Then its local equilibrium average is independent of
the velocity field
\begin{align}
\overline{A}_{0}^{\ell}\left(  \br|\beta(t),\nu(t)
\mathbf{,u}\left(  \br,t\right)  \right)   &  \equiv\left\langle
A_{0}\left(  \mathbf{r;u}\left(  \br,t\right)  \right)  ;\rho_{\ell
}\left[  \beta(t),\nu(t)  ,\mathbf{u}(t)  \right]
\right\rangle \nonumber\\
&  =\overline{A}_{0}^{\ell}\left(  \br|\beta(t),\nu(t)
\mathbf{,0}\right)  \label{2.12}%
\end{align}
In all of the following the averages of interest are for operators in the
local rest frame and therefore the flow velocity is taken to be zero.

The pressure tensor is defined as the local equilibrium average of the
microscopic momentum flux in the local rest frame $t_{0ij}\left(
\br\right)  $
\begin{equation}
\overline{t}_{0ij}^{\ell}\left(  \mathbf{r}\mid y\left(  t\right)
\right)  \equiv\overline{\pi}_{ij}^{\ell}\left(  \mathbf{r}\mid\beta\left(
t\right)  ,\nu\left(  t\right)  \right)  =\overline{\pi}_{ij}\left(
\mathbf{r}\mid\overline{n}\left(  t\right)  ,\overline{e}_{0}\left(  t\right)
\right)  \label{3.0}%
\end{equation}
%above is referred to in supplementary material
with (see Supplemental Material)
\begin{align}
t_{0ij}\left(  \br\right)   &  =\frac{1}{4m}\sum_{\alpha=1}^{N}\left[
p_{i\alpha},\left[  p_{j\alpha},\Delta\left(  \mathbf{r-q}_{\alpha}\right)
\right]  _{+}\right]  _{+}\nonumber\\
&  +\frac{1}{2}\sum_{\alpha\neq\sigma=1}^{N}F_{\alpha\sigma i}\left(
\left\vert \mathbf{q}_{\alpha}-\mathbf{q}_{\sigma}\right\vert \right)
\mathcal{D}_{j}\left(  \mathbf{r,q}_{\alpha},\bq_{\sigma}\right)  ,
\label{3.0a}%
\end{align}
and%
\begin{equation}
\mathcal{D}_{i}\left(  \mathbf{r,q}_{1},\mathbf{q}_{2}\right)  \equiv
\int_{\lambda_{2}}^{\lambda_{1}}d\lambda\frac{dx_{i}\left(  \lambda\right)
}{d\lambda}\Delta\left(  \mathbf{r-x}\left(  \lambda\right)  \right)
,\hspace{0.26in}\mathbf{x}\left(  \lambda_{1}\right)  =\mathbf{q}_{1}%
,\hspace{0.26in}\mathbf{x}\left(  \lambda_{2}\right)  =\mathbf{q}_{2}.
\label{3.0b}%
\end{equation}
In all of the following the choice
\begin{equation}
\Delta\left(  \mathbf{r-q}_{\alpha}\right)  \rightarrow\delta\left(
\mathbf{r-q}_{\alpha}\right)  \label{3.0c}%
\end{equation}
is made.

Separate the pressure tensor into its diagonal and traceless parts
\begin{equation}
\overline{\pi}_{ij}^{\ell}\left(  \mathbf{r}\mid\beta\left(  t\right)
,\nu\left(  t\right)  \right)  =\delta_{ij}\overline{\pi}^{\ell}\left(
\mathbf{r}\mid\beta\left(  t\right)  ,\nu\left(  t\right)  \right)
+\widetilde{\pi}_{ij}^{\ell}\left(  \mathbf{r}\mid\beta\left(  t\right)
,\nu\left(  t\right)  \right) .
\label{3.1}%
\end{equation}
The first term on the right will be referred to as the mechanical pressure
\begin{equation}
\overline{\pi}^{\ell}\left(  \mathbf{r}\mid\beta\left(  t\right)
,\nu\left(  t\right)  \right)  =\frac{1}{3}\overline{t}_{0ii}^{\ell
}\left(  \mathbf{r}\mid y\left(  t\right)  \right) , \label{3.2}%
\end{equation}
The second term of (\ref{3.1}) is the traceless part of the pressure tensor.
Its volume integral must vanish
\begin{equation}
\int d\mathbf{r}\widetilde{\pi}_{ij}^{\ell}\left(  \mathbf{r}%
\mid\beta\left(  t\right)  ,\nu\left(  t\right)  \right)  =0, \label{3.3}%
\end{equation}
since there is no external vector from which a non-diagonal tensor can be
formed. Otherwise the off-diagonal contributions to $\widetilde{\pi}^\ell_{ij}%
\left(  \br|\beta(t),\nu(t)  \right)  $ are non-zero, in general.

\subsubsection{Local pressures}

The trace of the momentum flux operator follows from (\ref{3.0a})
\begin{align}
t_{0ii}( \br, 0)   &  =e_{0}^{K}\left(  \br\right)
+\frac{1}{2m}\sum_{\alpha=1}^{N}p_{\alpha i}\Delta\left(  \mathbf{r-q}%
_{\alpha}\right)  p_{\alpha i}\nonumber\\
&  +\frac{1}{2}\sum_{\alpha\neq\gamma=1}^{N}F_{\alpha\gamma i}\left(
\left\vert \mathbf{q}_{\alpha}-\mathbf{q}_{\gamma}\right\vert \right)
\mathcal{D}_{i}\left(  \mathbf{r,q}_{\alpha},\mathbf{q}_{\gamma}\right)
\label{3.4}%
\end{align}
where $e_{0}^{K}\left(  \br\right)  $ is the kinetic energy density
operator
\begin{equation}
e_{0}^{K}\left(  \br\right)  =\frac{1}{4m}\sum_{\alpha=1}^{N}\left[
p_{\alpha}^{2},\Delta\left(  \mathbf{r-q}_{\alpha}\right)  \right]_{+}.
\label{3.5}%
\end{equation}
The mechanical pressure is therefore

\begin{widetext}
\begin{equation}
\overline{\pi}^{\ell}\left(  \mathbf{r}\mid\beta\left(  t\right)
,\nu\left(  t\right)  \right)=\frac{2}%
{3}
\overline{
 e_{0}^{K}}^\ell(  \mathbf{r}|y(t))
  +\frac{1}{6}\left.\overline{
\sum_{\alpha\neq\gamma}\mathcal{D}_{i}\left(  \mathbf{r},\mathbf{q}_{\alpha
},\mathbf{q}_{\gamma}\right)  F_{\alpha\gamma i}\left(  \mathbf{q}_{\alpha
},\mathbf{q}_{\gamma}\right) }^{\ell}\right|_{y(  t)}   +\frac{1}{4m}\nabla_{\mathbf{r}}^{2}\overline n\left(  \mathbf{r},t\right),
 \label{eq:mechanicalpressure}
\end{equation}

where the notation introduced in Eq. (\ref{eq27}) has been extended for the local equilibrium average of products by defining
\begin{equation}
\left\langle X(\vec r)Y(\vec r');\rho^\ell[y(t)]\right\rangle\equiv \left.\overline{X(\vec r)Y(\vec r')}^\ell\right|_{y(t)}
\end{equation}
and use has been made of the identity%
\begin{equation}
\frac{1}{2m}\sum_{\alpha}p_{\alpha i}\delta\left(  \mathbf{r}-\mathbf{q}%
_{\alpha}\right)  p_{\alpha i}=e_{0}^{K}\left(  \mathbf{r}\right)  +\frac
{1}{4m}\nabla_{\mathbf{r}}^{2}n\left(  \mathbf{r}\right).
\end{equation}

To interpret the result in Eq. (\ref{eq:mechanicalpressure}) it is useful to introduce a local pressure
associated with the virial equation

\begin{equation}
\pi_{v}\left(  \br|\beta(t),\nu(t)  \right)  =\frac{2}%
{3}
\overline{ e_{0}^{K}}^\ell(  \br|y(t))  
 +\frac{1}{6}
\left.\overline{
\sum_{\alpha\neq\gamma=1}
^{N}\Delta\left(  \mathbf{r-q}_{\alpha}\right)  
\left(  \mathbf{q}_{\alpha}-\mathbf{q}_{\gamma}\right)  
\cdot\mathbf{F}_{\alpha\gamma}
\left(  \left\vert\mathbf{q}_{\alpha}-\mathbf{q}_{\gamma}\right\vert \right)}
^{\ell}\right|_{y(t)}
\label{3.7}%
\end{equation}
such that
\begin{equation}
\int d\br \pi_{v}\left(  \br|\beta(t),\nu(t)
\right)  =\frac{2}{3}
\overline{ K[y(t)]}^\ell
+\frac{1}{6}
\left.\overline{ \sum_{\alpha \neq \gamma =1}^{N}\left(
\bq_{\alpha}-\bq_{\gamma}\right)  \cdot\mathbf{F}_{\alpha \gamma} (\left\vert \bq_{\alpha}-\bq_{\gamma}\right\vert)}^\ell\right|_{y(t)}  , \label{eq:virialpressure}%
\end{equation}%

\begin{equation}
K\equiv\frac{1}{2m}\sum_{\alpha=1}^{N}p_{\alpha j}^{2} \label{eq:Kdef}%
\end{equation}
Eq.~(\ref{eq:virialpressure}) is recognized as the virial equation associated with the local equilibrium
ensemble. Then it follows that the mechanical pressure has the same volume
integral as the virial pressure
\begin{equation}
\int d\br \overline{\pi}^{\ell}\left(  \mathbf{r}\mid
\beta\left(  t\right)  ,\nu\left(  t\right)  \right)
=\int d\br \pi_{v}\left(  \br|\beta(t),\nu(t)
\right)  . \label{3.9}%
\end{equation}
To obtain this use has been made of
\begin{equation}
\int d\br\mathcal{D}( \br, \bq_\alpha, \bq_\gamma)  =\bq_{\alpha}-\bq_{\gamma}. \label{3.19}%
\end{equation}

A third local pressure, the thermodynamic pressure $\pi_{T}\left(
\br|\beta(t),\nu(t)  \right)  $, can be identified from
the partition function $Q\left[  \beta(t),\nu(t)  \right]  $ to
define the thermodynamics of a local equilibrium system
\begin{equation}
\int d\br\beta(\br,t)\pi_{T}\left(  \br|\beta
(t),\nu(t)  \right)  \equiv Q\left[  \beta(t),\nu\left(
t\right)  \right]  . \label{3.19a}%
\end{equation}
Since $Q\left[  \beta(t),\nu(t)  \right]  $ is extensive this can
also be expressed as
\begin{equation}
\frac{1}{V}\int d\br\beta(\br,t)\pi_{T}\left(  \br|%
\beta(t),\nu(t)  \right)  = \frac{\partial Q\left[
\beta(t),\nu(t)  \right]  }{\partial V}. \label{3.20}%
\end{equation}
The volume derivative in (\ref{3.20}) is evaluated in Appendix \ref{apB} with
the result that the thermodynamic local pressure is the same as the local
virial pressure
\begin{equation}
\pi_{T}\left(  \br|\beta(t),\nu(t)  \right)
=\pi_{v}\left(  \br|\beta(t),\nu(t)  \right)  .
\label{3.29}%
\end{equation}

In summary the volume integrals of all three local pressures are the same
\begin{equation}
\int d\br \pi^{\ell}\left(  \mathbf{r}\mid
\beta\left(  t\right)  ,\nu\left(  t\right)  \right)
=\int d\br \pi_{v}\left(  \br|\beta(t),\nu(t)
\right)  =\int d\br \pi_{T}\left(  \br|\beta(t),\nu\left(
t\right)  \right)%
\end{equation}
but the local mechanical pressure differs from the virial and thermodynamic
local pressures
\begin{equation}
\pi^{\ell}\left(  \mathbf{r}\mid
\beta\left(  t\right)  ,\nu\left(  t\right)  \right) \neq
\pi_{T}\left(  \br|\beta(t),\nu(t)  \right)
=\pi_{v}\left(  \br|\beta(t),\nu(t)  \right)  \,.
\end{equation}

In summary, there are several
``reasonable" definitions for a local pressure that all agree with the
thermodynamicglobal pressure but differ otherwise. In the present context the
correct local pressure for the hydrodynamic equations is that obtained by
direct evaluation of (\ref{eq:mechanicalpressure}).
A definition of the thermodynamic local pressure and its relationship to an associated pressure tensor coupling thermal and mechanical properties is given by Refs.~\onlinecite{Pozhar1993, Percus1986}.

\end{widetext}
\subsubsection{Relation of pressure gradient to free energy gradient}

It is often chosen to use a free energy density functional rather than the pressure to characterize the local equilibrium contribution to the momentum flux. To see how this is done, define the Legendre transform
\begin{align}
\mathcal{F}\left[  \beta(t),\overline n(t)  \right]   &  =-Q^\ell\left[
\beta(t),\nu(t)  \right]  +\int d\mathbf{r}\overline{n}\left(  \mathbf{r,}t\right)
\nu\left(  \mathbf{r,}t\right) \nonumber\\
&  =-\int d\br\left[  \beta(\br,t)\pi_{T}\left(  \mathbf{r\mid
}\beta(t),\nu(t)  \right)  +\overline n\left(  \br,t\right)
\nu\left(  \br,t\right)  \right] \,.
\end{align}
It follows from the definition of $Q^\ell$ that the first functional derivatives are
\begin{equation}
\left.\frac{\delta Q^\ell\left[  \beta(t),\nu(t)  \right]  }{\delta
\nu\left(  \br,t\right)  }\right|_{\beta}=\overline n\left(  \br,t\right)
\hspace{0.24in}\left.\frac{\delta\mathcal{F}\left[  \beta(t),\overline n(t)
\right] }{\delta \overline n\left(  \br,t\right)  }\right|_{\beta}=\nu\left(
\br,t\right) \,. \label{eq:fnalderivative}
\end{equation}
Here $\mathcal{F}\left[  \beta(t),n\left(  t\right)  \right]  $ is the
dimensionless free energy (in the uniform $\beta$ limit $\mathcal{F}%
\rightarrow\beta F$). Consider the special case
\begin{equation}
\delta\nu\left(  \mathbf{r},t\right)  =\nu\left(  \mathbf{r+}\delta
\mathbf{r},t\right)  -\nu\left(  \mathbf{r},t\right)  .
\end{equation}
The corresponding variation in $Q^\ell\left[  \beta(t),\nu\left(  t\right)
\right]  $ at constant $\beta$ is
\begin{align}
\left.\delta Q^\ell\left[  \beta(t),\nu\left(  t\right)  \right]   \right|_{\beta}&=\int
d\mathbf{r}\overline n\left(  \mathbf{r},t\right)  \delta\nu\left(  \mathbf{r},t\right)
\nonumber\\
&  =\int d\mathbf{r}\overline n\left(  \mathbf{r},t\right)  \left(  \nu\left(
\mathbf{r+}\delta\mathbf{r},t\right)  -\nu\left(  \mathbf{r},t\right)
\right)  \nonumber\\
&  =\int d\mathbf{r}\overline n\left(  \mathbf{r},t\right)  \delta\mathbf{\mathbf{r}%
}\cdot\nabla\left.\frac{\delta\mathcal{F}\left[  \beta(t),\overline n\left(  t\right)
\right]  }{\delta \overline n\left(  \mathbf{r},t\right)  }\right|_{\beta}%
\label{3.35}
\end{align}
Next express $\delta Q^\ell\left[  \beta(t),\nu\left(  t\right)  \right]
\mid_{\beta}$ in terms of $\delta \pi_{T}\left(  \mathbf{r\mid}\beta
(t),\nu\left(  t\right)  \right)  \mid_{\beta}$ from (\ref{3.19a})
\begin{align}
\delta Q\left[  \beta(t),\nu\left(  t\right)  \right]   \mid_{\beta}&=-\int
d\mathbf{r}\beta(\mathbf{r,}t)\delta \pi_{T}\left(  \mathbf{r\mid}\beta
(t),\nu\left(  t\right)  \right)  \mid_{\beta}\nonumber\\
&  =-\int d\mathbf{r}\beta(\mathbf{r,}t)\delta\mathbf{\mathbf{r}}\cdot\nabla
\pi_{T}\left(  \mathbf{r\mid}\beta(t),\nu\left(  t\right)  \right)  \mid_{\beta}%
\label{3.37}
\end{align}
Finally, equating (\ref{3.35}) and (\ref{3.37}) and taking $\delta
\mathbf{r\rightarrow0}$
\begin{equation}
\beta(\mathbf{r,}t)\nabla  \pi_{T}\left(  \mathbf{r\mid}\beta(t),\nu\left(
t\right)  \right)  \mid_{\beta}=\overline n\left(  \mathbf{r},t\right)  \nabla 
\left.\frac{\delta\mathcal{F}\left[  \beta(t),\overline n\left(  t\right)  \right]  }{\delta
\overline n\left(  \mathbf{r},t\right)  }\right|_{\beta}.%
\label{eq83}
\end{equation}
This is the commonly used expression by others for the case of constant
$\beta(\mathbf{r,}t)$. In our case it is important to remember there is
another contribution to $\nabla  \pi_{T}\left(  \mathbf{r\mid}\beta
(t),\nu\left(  t\right)  \right)  $ due to its variation with respect to
$\beta(\mathbf{r,}t).$ This can be treated in the same way by introducing a
double Legendre transformation from $\beta(t),\nu\left(  t\right)  $ to
$e_{0}(t),n\left(  t\right)  $. The result is
\[
\beta(\mathbf{r,}t)\nabla  \pi_{T}\left(  \mathbf{r\mid}\beta(t),\nu\left(
t\right)  \right)  =\overline n\left(  \mathbf{r},t\right)  \nabla \left.\frac
{\delta\mathcal{F}\left[  e_{0}(t),\overline n\left(  t\right)  \right]  }{\delta
\overline n\left(  \mathbf{r},t\right)  }\right|_{e_{0}}%
\]%
\begin{equation}
-e_{0}\left(  \mathbf{r},t\right)  \nabla \left.\frac{\delta\mathcal{F}\left[
 e_{0}(t),\overline n\left(  t\right)  \right]  }{\delta e_{0}\left(  \mathbf{r}%
,t\right)  }\right|_{n}.%
\end{equation}

\subsection{Relation to previous work}

As noted at the end of
the Introduction above a new review of quantum hydrodynamics for plasmas has
just appeared\cite{Bonitz2019}. It provides an overview of much of the previous work on
continuum descriptions for electrons, with extensive references. In most cases
the phenomenology does not originate with traditional hydrodynamics, but is
``reinvented". Consequently, the brief discussion here is motivated by
recovering typical examples from the present exact analysis showing how some
common limitations (weak inhomogeneities, weak coupling) can be removed in a
controlled fashion.

Most previous work is based on the continuity equation and the momentum
equation, without reference to the energy conservation law and without
dissipation (i.e., special cases of the perfect fluid equations). Exceptions
are references \cite{Eich2016,Murillo2019} and Ref.~[\onlinecite{Renziehausen2018}],
%\kl{what are they?}
although the latter is restricted to pure states. The implicit assumption in
neglecting the energy equation is that the pressure tensor is independent of
the internal energy. Then it is possible to show for the perfect fluid that
the inverse temperature $\beta$ is spatially uniform and dissipation is weak
\cite{McLennan1989}. 
%Assuming also that the flow field dependence in the pressure tensor is weak, 
The perfect fluid equations (\ref{eq:eom_density_nodissipation}) - (\ref{eq:eom_u_nodissipation}) reduce
to%
\begin{subequations}
\begin{align}
D_{t}\nbar(\br,t) +\nbar(
\br,t) \nabla\cdot\bu(\br,t) &=0,
\label{eq:weak_density}%
\\
%\begin{equation}
m\overline{n}(\br,t) D_{t}u_{i}(
\br,t) +\partial_{j}
%\overline{t}^{\ell}_{0ij}
\overline{{\pi}}_{ij}(\br |\nbar(t)) &=-n(\br,
t) \partial_{i}v^{\ext}(\br,t) \,. 
\label{eq:weak_momentum}%
%\end{equation}
\end{align}
\end{subequations}
With the condition of uniform temperature these equations are still general,
following from the underlying conservation laws for mass and momentum. In
contrast, most earlier work is restricted to pure states or phenomenological
macroscopic (coarse grained) Hamiltonian dynamics. Important restrictions to that work are
explicit at the outset: 1) there is no dissipation, 2) the velocity field is
irrotational ($\nabla\times\bu(\br,t) =0$), and 3)
there are no off-diagonal elements to the pressure tensor.
%\kl{See Jeff's comment}

The origins and applications of continuum equations such as (\ref{eq:weak_density}) and
(\ref{eq:weak_momentum}) have been reviewed extensively in references \cite{Bonitz2013,Bonitz2018}
in addition to the review noted above \cite{Bonitz2019}. Instead, the focus is limited to results
obtained from a phenomenological macroscopic Hamiltonian dynamics, due to
Bloch \cite{Bloch1934} and developed primarily by others, e.g. Refs.
\onlinecite{Jensen1937,Ying1974,Bonitz2013}. A Hamiltonian depending on the macroscopic
fields $\nbar(\br) $ and $\phi(\br) $, where the flow velocity is defined in terms of this scalar potential by $\bu(\br) =\nabla\phi (\br) $, is 
\begin{equation}
\mathcal{H}[\nbar,\phi]=F[\nbar]+\int \dr \left(\frac
{1}{2}m\left\vert \nabla\phi(\br) \right\vert
^{2}+v^{\ext}(\br,t) \right) . \label{3.30}%
\end{equation}
Here $\beta F[\nbar]$ is the dimensionless equilibrium free energy
functional defined above, and $v^{\ext}(\br) $ is the given
external single particle potential. Then the usual Hamilton's equations for
the variables $\nbar(\br) $ and $\phi(
\br) $ give the forms (\ref{eq:weak_density}) and (\ref{eq:weak_momentum}) with
\begin{equation}
%\nabla \overline{t}^{\ell}_{0ij}
\nabla {\pi}_{ij}
(\br |\nbar(t)
) \to\delta_{ij}\nbar(\br)
\nabla\frac{\delta F[\nbar]}{\delta\nbar(\br%
) }\,.
\end{equation}
(Note that the free energy here includes the Hartree contribution, whereas
that of reference \cite{Bonitz2018} has extracted it explicitly). This result is
in fact the same as the diagonal part of the pressure tensor in (\ref{3.1})  
where the thermodynamic pressure and free energies are related by (\ref{eq83})%
\begin{equation}
\nabla {\pi}(\br |\nbar(t))
=\nbar(\br) \nabla\frac{\delta F[\nbar%
]}{\delta\nbar(\br) }\,.
\end{equation}
%This is proved in Appendix \ref{ap:B}. \kl{Still not proved, right?} 
In most of the literature on
generalized hydrodynamics for electrons (\ref{eq:weak_momentum}) therefore has the form%
\begin{equation}
D_{t}\bu(\br,t) +m^{-1}\nabla\left(\frac{\delta
F[\nbar(t) ]}{\delta\nbar(\br%
,t) }+v^{\ext}(\br,t) \right)=0 \,.
\end{equation}

There is considerable analysis of $F[\nbar(t) ]$ within
recent finite temperature DFT \cite{IPAMreview}. It is typically written as
\begin{equation}
F_{e}[n]=F_{\s}[n]+F_{\H}[n]+F_{\xc}[n] \label{eq:freeenergyfnal}
\end{equation}
where $F_{\s},F_{\H},F_{\xc}$ are the non-interacting,
Hartree, and exchange-correlation contributions respectively. The
parametrization of each term by the constant temperature $\beta^{-1}$ has been
left implicit. The most common case considered is weakly coupled electrons,
for which $F_{xc}$ is neglected. Determination
of the non-interacting contribution as an explicit functional of the density
is still a formidable challenge, but for weakly inhomogeneous states it can be
calculated from a gradient expansion. To second order in the density gradient
it is \cite{Perrot1979}%
\begin{equation}
F_{\s}[\beta,n]\to F_{\TF}[\beta,n]+\int
\dr a_{2}(\beta,n(\br)) \frac{|\nabla
n(\br)|^{2}}{8n(\br)}. 
\label{eq:SGE2freeenergy}%
\end{equation}
Here $F_{\TF}[\beta,n]$ is the Thomas-Fermi free energy \cite{Feynman1949}.
%\begin{equation}
%F_{\TF}[\beta,n]=\int \dr\frac{\sqrt{2}}{\pi^{2}\beta^{5/2}%
%}\left[-\frac{2}{3}I_{\frac{3}{2}}(\beta\mu)+\beta\mu I_{\frac{1}{2}}%
%(\beta\mu)\right], \label{3.36}%
%\end{equation}
%where $I_{\alpha}$ is the Fermi-Dirac integral
%\begin{equation}
%I_{\alpha}(\eta)=\int_{0}^{\infty}\dd x\frac{x^{\alpha}}{1+\exp(x-\eta)},
%\label{3.37}%
%\end{equation}
%and the chemical potential $\mu$ $(\br)${ is a function of the density
%}$n(\br)$ {determined via
%\begin{equation}
%n(\br)=\frac{\sqrt{2}}{\pi^{2}\beta^{3/2}}I_{\frac{1}{2}}(\beta
%\mu(\br)). \label{3.38}%
%\end{equation}
The coefficient of the square gradient contribution is known exactly for the special case of the uniform electron gas.  More generally it is known in terms of the density response function for the corresponding uniform equilibrium system.

%\begin{equation}
%a_{2}(\beta,n(\br)) =-\frac{I_{1/2}(\beta\mu
%(\br))I_{-3/2}(\beta\mu(\br))}{I_{-1/2}^{2}(\beta\mu
%(\br))} \,.
%\label{3.39}%
%\end{equation}

%These results simplify considerably in the low temperature limit, using%
%\begin{equation}
%\lim\limits_{\beta \mu \to0}I_{\alpha}(\beta\mu)=(\beta\mu)^{1+\alpha
%}/(1+\alpha)\label{3.42}%
%\end{equation}

%to get {%
As a special case, the results for the uniform electron gas at zero temperature are
\begin{equation}
F_{s}[\beta,n]=\frac{3}{10}\left(  3\pi^{2}\right)  ^{2/3}\int d\mathbf{r}%
n^{5/3}\left(  \mathbf{r}\right)  +\frac{1}{9}\int d\mathbf{r}\frac{\left\vert
\nabla n\left(  \mathbf{r}\right)  \right\vert ^{2}}{8n\left(  \mathbf{r}%
\right)  }
\label{3.42a}%
\end{equation}
%\kl{May 08, 12:30, work on it later..}
where
$a_{2}\left(  \beta,\overline{n}\left(  \mathbf{r}\right)  \right)
\rightarrow1/9$ in (\ref{eq:SGE2freeenergy}). The momentum equation (\ref{eq:weak_momentum}) becomes in this case%
\begin{widetext}
\begin{align}
&D_{t}\bu(\br,t) +(m\nbar(
\br,t)) ^{-1}\nabla {\pi}_{\TF}(
\beta,\nbar(\br,t)) 
%\nonumber \\
%&
+m^{-1}\nabla\left(  v_{H}
(\mathbf{r})+v_{B}(\mathbf{r})+v_{\rm ext}(\mathbf{r},t)\right) 
 =0 \,,
\label{eq:TFvWhydrodynamics}%
\end{align}
\end{widetext}
where $v_{H}(\mathbf{r})$ is the Hartree potential, and where the Thomas-Fermi pressure ${\pi}_{\TF}(\beta,\overline
{n}(\br,t)) $ and Bohm potential $v_{\mathrm{B}%
}(\br)$ \cite{Bohm1952} are 
\begin{align}
{\pi}_{\TF}(\beta,\nbar(\br,t))
&\to
\frac{1}{5}(3\pi^2) ^{2/3}  n(\br)^{5/3}
\\
v_{\mathrm{B}}(\br)
&\to
 \frac{|\nabla n(\br)|^2}{8n(\br)^2} - \frac{\nabla^2 n(\br)}{4n(\br)} \,.
\end{align}
The appearance of the Bohm potential in this context has led to considerable confusion \cite{Bonitz2018,Stanton2015}. A result similar to (\ref{eq:TFvWhydrodynamics}) follows from an exact transformation of the Schroedinger equation (Madelung transform \cite{Madelung1927,Eich2016}) except without the Thomas-Fermi pressure and without the factor of $1/9$ for the Bohm potential. Its validity is strictly related to a pure state and does not extend to the mixed state ensembles considered here \cite{Bonitz2013}.
The approximation (\ref{eq:TFvWhydrodynamics}) is reasonable as long as the density of the underlying system stays close to homogeneity. 

\subsection{Application to strong coupling and strong inhomogeneities}

The perfect fluid equations may have important applications to the complex
electron - ion systems of warm, dense matter. Although neglecting dissipation,
they describe the dominant convective dynamics without restriction to length
and time scales. To incorporate the strong coupling all three contributions of
the free energy functional in (\ref{eq:freeenergyfnal}) must be included. Furthermore,
since electrons in the vicinity of ions will have strong inhomogeneities, the
limitation to gradient expansions must be relaxed. The range of temperatures
for the electrons should extend from near zero to well above the Fermi
temperature. This ambitious scope of applicability has been addressed successfully and two main
advances are now available. An accurate representation of the equilibrium free
energy for the uniform electron gas across the entire temperature/density
plane has been developed from recent quantum Monte Carlo simulations
\cite{Dornheim2018,Karasiev2019}. This provides the essential local density
approximation that is necessary to assure the uniform limit of any approximate
functional for real electron-ion systems \cite{Dufty2017}. Functional development for strong inhomogeneities has included a
generalized gradient approximation for both the non-interacting free energy
\cite{Luo2018} and the exchange-correlation free energy \cite{Karasiev2018}. With
these developments the perfect fluid hydrodynamics can be applied to warm dense matter conditions.

\subsection{Linear modes}

The simplest application of the perfect fluid equations is their linearization
about an equilibrium reference state,
\begin{align}
\nbar(\br,t) &=\nbar^{(0)}(
\br,t) +\nbar^{(1)}(\br,t)
+..,
\\
\ebar_{0}(\br,t) &=\overline
{e}_{0}^{(0)}(\br,t) +\ebar_{0}^{(1)}(
\br,t) +..,
\\
\bu(\br,t) &=\bu^{(0)}(\br,t) +\bu^{(1)}(\br,
t) +..,\hspace{0.25in}\label{3.44}%
\end{align}
Choose $\bu^{(0)}(\br,t) =0$ and the external
force equal to zero. The solution sought is therefore the response to an
initial perturbation. To lowest order the equations give uniform, time
independent quantities $\nbar^{(0)}(\br,t)
=\nbar^{(0)},\ebar_{0}^{(0)}(\br,t)
=\ebar_{0}^{(0)}, \overline{{\pi}}_{ij}(\br |\nbar^{(0)}(t),\ebar_{0}^{(0)}(t),\bu%
^{(0)}(t)) =\delta_{ij}  \overline{{\pi}}^{(0)}(\nbar^{(0)},\ebar_{0}^{(0)})$.
% \kl{check here}%
%\begin{subequations}
%\begin{align}
%\nbar^{(0)}(\br,t)
%&=\nbar^{(0)}%
%\\
%\ebar_{0}^{(0)}(\br,t)
%&=\ebar_{0}^{(0)},
%\\
%\overline{t}^{\ell}_{0,ij}(\br |\nbar^{(0)}(t),\ebar_{0}^{(0)}(t),\bu%
%^{(0)}(t))
%&=\delta_{ij} \mathrm{p}^{(0)}(\nbar^{(0)},\ebar_{0}^{(0)}) \,.
%\label{3.45}%
%\end{align}
%\end{subequations}
To next order they are%
\begin{subequations}
\begin{align}
\partial_{t}\nbar^{(1)}(\br,t) +\nbar%
^{(0)}\nabla\cdot\bu^{(1)}(\br,t) 
&=0 \,,
\label{3.46}%
\\
\partial_{t}\ebar_{0}^{(1)}(\br,t) +h^{(0)}(
\nbar^{(0)},\ebar_{0}^{(0)}) \nabla\cdot\bu%
^{(1)}(\br,t) 
&=0 \,,
\label{3.47}%
\\
m\nbar^{(0)}\partial_{t}u_{i}^{(1)}(\br,t)
+\partial_{j}
%\overline{t}^{L(1)}_{0ij}
\overline{{\pi}}_{ij}
(\br |\nbar(t),\ebar_{0})
&=0 \,,
\label{3.48}%
\end{align}%
\end{subequations}
where $h^{(0)}(\nbar^{(0)},\ebar_{0}^{(0)})
=\ebar_{0}^{(0)}+\overline{{\pi}}^{(0)}(\nbar^{(0)},\ebar%
_{0}^{(0)}) $ is the enthalpy density. The pressure tensor contribution is
\begin{align}
\partial_{j} 
\overline{{\pi}}_{ij}
%\overline{t}^{L(1)}_{0,ij}
(\br |\nbar(t),\ebar_{0}(t)) 
&=\partial_{j}\int \dr'
%\frac{\delta \overline{t}^{\ell}_{0,ij}(\br |\nbar(t),\ebar_{0}(t),\bu(t)) }{\delta\nbar(
%\br',t) }|_{\nbar^{(0)},\overline
%{e}_{0}^{(0)}}
c_n(|\br-\br^{\prime}|) \delta_{ij}
\nbar^{(1)}(\br',t) \nonumber
\\
&+\partial_{j}\int \dr'
%\frac{\delta \overline{t}^{\ell}_{0,ij}(\br |\nbar(t),\ebar_{0}(t),\bu(t)) }{\delta\ebar_{0}(
%\br',t) }|_{\nbar^{(0)},\overline
%{e}_{0}^{(0)}}
c_{e_0}(|\br-\br^{\prime}|) \delta_{ij}
\ebar_{0}^{(1)}(\br',t) \,.\label{3.49}%
\end{align}
The pressure derivatives are evaluated at the uniform equilibrium state and
are equilibrium response functions%
\begin{subequations}
\begin{align}
c_{n}(\left\vert \br-\br'\right\vert )
\delta_{ij}
&\equiv\frac{1}{\nbar}\frac{\delta 
%\overline{t}^{\ell}_{0,ij}
\overline{{\pi}}_{ij}
(\br |\nbar(t),\ebar_{0}(
t)) }{\delta\nbar(
\br',t) }|_{\nbar^{(0)},\overline
{e}_{0}^{(0)}}
\label{3.50}%
\\
c_{e_{0}}(\left\vert \br-\br'\right\vert )
\delta_{ij}
&\equiv\frac{1}{\nbar}
\frac{\delta 
%\overline{t}^{\ell}_{0,ij}
\overline{{\pi}}_{ij}
(\br |\nbar(t),\ebar_{0}(
t)) }{\delta\ebar%
_{0}(\br',t) }|_{\nbar%
^{(0)},\ebar_{0}^{(0)}} \,.
\label{3.50a}%
\end{align}
\end{subequations}
The linear equations (\ref{3.46}) - (\ref{3.48}) are most easily solved in a
Fourier representation, with the notation%
\begin{equation}
\widetilde{f}(\mathbf{k}) =\int \dr e^{i\mathbf{k\cdot
r}}f(\br) .\label{3.51}%
\end{equation}
The velocity field is written in terms of its longitudinal component
($\widehat{\mathbf{k}}\cdot\widetilde{\bu}^{(1)}$) and two transverse
components $(\widehat{\mathbf{e}}_{1}\cdot\widetilde{\bu}%
^{(1)},\widehat{\mathbf{e}}_{2}\cdot\widetilde{\bu}^{(1)}) $.
Transverse modes decouple from the others and have no time dependence.
The remaining three variables have coupled dynamics with three eigenvalues%
\begin{equation}
\lambda(k) =(k\sqrt{(h\widetilde{c}_{e_{0}}(
k) +n\widetilde{c}_{n}(k)) },-k\sqrt{(
h\widetilde{c}_{e_{0}}(k) +n\widetilde{c}_{n}(k)
) },0) .\label{3.51a}%
\end{equation}
The first two are the generalization of long wavelength sound modes to
arbitrary length scales, while the last is a constant
heat mode (since there is no dissipation). For Coulomb interactions they become generalizations of the plasmon modes. 
This is the simplest example of ``generalized hydrodynamics."

A more interesting example would be linearization around the stationary state
imposed by an external force, such as that of a configuration of ions or a
confining force. These will be discussed elsewhere.
All of the perfect fluid hydrodynamics above neglects dissipation. Two attempts to improve this have been described. The first \cite{Bonitz2018} makes a phenomenological modification of the coefficient $a_2$ in the gradient expansion of the free energy, Eq.(\ref{eq:SGE2freeenergy}), to make it non-local in space and time. This new dependence is then determined by the requirement that the linearized equations reproduce chosen representations for the linear response, e.g. random phase approximation.  In the present context it is seen that this is attempting to capture effects that properly are contained in the dissipative fluxes, through an ad hoc modification of the local equilibrium free energy.
A more consistent extension of the perfect fluid equations is to include an approximation to the irreversible fluxes, e.g. using those known in the Navier-Stokes limit for small space and time variations \cite{Murillo2017}. The result is a mixed representation of perfect fluid effects on all length scales but weak dissipation on long length scales. Still, it is a reasonable first attempt at including all relevant physical effects.

\section{Dissipative fluxes and generalized hydrodynamics}
\label{sec:dissipative_fluxes}

\label{sec4}Return now to the exact equations for $\overline{n}\left(
\mathbf{r},t\right)  ,\overline{e}_{0}(\mathbf{r},t),$ and $\overline
{\mathbf{u}}(\mathbf{r},t),$ (\ref{2.29}) - (\ref{2.31}), it remains to
determine the irreversible fluxes $\overline{\mathbf{s}}_{0}^{\ast}\left(
\mathbf{r},t\right)  $ and $\overline{t}_{0ij}^{\ast}\left(  \mathbf{r}%
,t\right)  $ as functionals of these fields. They are defined as averages over
the deviation of the solution to the Liouville - von Neumann equation from the
local equilibrium ensemble%
\begin{equation}
\overline{\mathbf{s}}_{0}^{\ast}\left(  \mathbf{r},t\right)  =\overline
{\mathbf{s}}_{0}\left(  \mathbf{r},t\right)  -\overline{\mathbf{s}}_{0}^{\ell
}\left(  \mathbf{r},t\right)  =\sum_{N}\Tr^{(N)}\mathbf{s}_{0}\left(
\mathbf{r}\right)  \Delta_N\left(  t\right)  ,\label{4.1}%
\end{equation}%
\begin{equation}
\overline{t}_{0ij}^{\ast}\left(  \mathbf{r},t\right)  =\overline{t}%
_{0ij}\left(  \mathbf{r},t\right)  -\overline{t}_{0ij}^{\ell}\left(
\mathbf{r},t\right)  =\sum_{N}\Tr^{(N)}t_{0ij}\left(  \mathbf{r}\right)
\Delta_N\left(  t\right)  ,\label{4.1a}%
\end{equation}
where $\Delta_N\left(  t\right)  $ is the deviation from local equibrium
\begin{equation}
\rho_{N}\left(  t\right)  =\rho_{N}^{\ell}\left[  y\left(  t\right)  \right]
+\Delta_{N}\left(  t\right)
.\label{4,2}%
\end{equation}
The formal solution for $\Delta_N\left(  t\right)  $ is obtained in Appendix
\ref{ap:B}%
\begin{widetext}
\begin{align}
\Delta_N( t)  =\int_{0}^{t}dt'e^{-\mathcal{L}(
t-t') }\int \dr(  \Psi_{\alpha}(\br|y(t') )\partial_{i}\overline{\gamma}_{i\alpha}^{\ast}(
\br,t')  -\Phi_{i\alpha}(\br|y(t'))
\partial_{i}\overline{\psi}_{\alpha}( \br,t')) 
 \rho^\ell[y( t')]  \label{4.3}%
\end{align}
where $\mathcal{L}$ denotes the Liouville operator
(\ref{eq13a}) and a time independent external force
has been chosen here for simplicity of notation. Also 
$\Phi_{i\alpha}(\br|y(t))$ is
\begin{equation}
\Phi_{i\alpha}(\mathbf{r}|y(t))=\Gamma_{i\alpha}(\mathbf{r}|y(t))-\int
d\mathbf{r}^{\prime}\Psi_{\beta}(\mathbf{r}^{\prime}|y(t))
\left.\overline{
\psi_{\beta}(\vec r')  \Gamma_{i\alpha}%
(\mathbf{r}|y(t))}^{\ell}\right|_{y(t)}  
.\label{4.4}%
\end{equation}
Recall the notation%
\begin{equation}
\left.\overline{AB}^{\ell}\right|_{ y(t)}   =\left\langle
AB;\rho^{\ell}\left[  y(t)\right]  \right\rangle.
\end{equation}
The operators $\Psi_{\beta},$ $\Gamma_{i\alpha}$ are transformations of the
those representing the local conserved densities $\psi_{\alpha}(\mathbf{r})$
and their fluxes $\gamma_{i\alpha}(\mathbf{r})$%
\begin{equation}
\Psi_{\alpha}(\mathbf{r}|y(t))=\int d\mathbf{r}^{\prime}\widetilde{\psi}_{\beta
}(\mathbf{r}^{\prime}|y(t))g_{\beta\alpha}^{-1}(\mathbf{r}^{\prime},\mathbf{r}%
|y(t)),\hspace{0.25in}\Gamma_{i\alpha}(\mathbf{r}|y(t))=\int d\mathbf{r}^{\prime
}\widetilde{\gamma}_{i\beta}(\mathbf{r}^{\prime}|y(t))g_{\beta\alpha}%
^{-1}(\mathbf{r}^{\prime},\mathbf{r}|y(t))\label{4.5}%
\end{equation}%
\begin{equation}
g_{\alpha\beta}\left(  \mathbf{r},\mathbf{r}^{\prime}|y(t)\right)  \equiv
\left.\overline{\psi_{\alpha}(\mathbf{r})\widetilde{\psi}_{\beta}(\mathbf{r}%
^{\prime}|y(t))}^{\ell}\right|_{y(t)}=-\frac{\delta\overline{\psi}_{\alpha}^{\ell}(\mathbf{r\mid
}y\left(  t\right)  )}{\delta y_{\beta}(\mathbf{r}^{\prime},t)}.\label{4.6}%
\end{equation}
%above is referred to in supplementary material
The tilde over an operator is defined by%
\begin{equation}
\widetilde{A}(\vec r|y)\equiv \int_{0}^{1}dx e^{-x \eta\left[  y  \right]  }\left(  A(\vec r)-\overline{A}^{\ell}\left( \vec r|y
\right)  \right)  e^{x \eta\left[  y  \right]
},\hspace{0.25in}\eta\left[  y  \right]  =-\ln\rho^{\ell
}_N[y ].\label{eq:xtilde}%
\end{equation}
This transformation is closely related to the Kubo transform of correlation
functions for linear response (see Appendix \ref{ap:A} for its origin here).

With these results the equation for the irreversible fluxes can be written
from (\ref{4.1}). The energy flux $\overline{s}_{0i}^{\ast}\left(
\mathbf{r},t\right)  $ \ and momentum flux $\overline{t}_{0ij}^{\ast}\left(
\mathbf{r},t\right)  $ are found to be (see Supplemental Material)%

\begin{align}
\overline{s}_{0i}^{\ast}\left(  \mathbf{r},t\mid y\right)   &  =\int_{0}%
^{t}dt\int d\mathbf{r}^{\prime}\left\{  \left.\overline{\left[  e^{\mathcal{L}%
\left(  t-t^{\prime}\right)  }s_{0i}\left(  \mathbf{r}\right)  \right]
\phi_{0j}(\mathbf{r}^{\prime}|y(t^{\prime}))}^{\ell}\right|_{y(t')}\partial
_{j}^{\prime}\beta(\mathbf{r}^{\prime},t^{\prime})\right.  \nonumber\\
&  -\left.\overline{\left[  e^{\mathcal{L}\left(  t-t^{\prime}\right)  }%
s_{0i}\left(  \mathbf{r}\right)  \right]  \sigma_{0kj}(\mathbf{r}^{\prime
}|y(t^{\prime}))}^{\ell}\right|_{y(t')}\beta(\mathbf{r}^{\prime},t^{\prime}%
)\partial_{j}^{\prime}u_{k}(\mathbf{r}^{\prime},t^{\prime})\nonumber\\
&  -\left.\overline{\left[  e^{\mathcal{L}\left(  t-t^{\prime}\right)  }%
s_{0i}\left(  \mathbf{r}\right)  \right]  \xi_0\left(  \mathbf{r}^{\prime
}|y(t^{\prime})\right)  }^{\ell}\right|_{y(t')}\partial_{k}\overline{s}_{0k}^{\ast
} (  \vec r',t'|y  )  \nonumber\\
&  \left.  +\left.\overline{\left[  e^{\mathcal{L}\left(  t-t^{\prime}\right)
}s_{0i}\left(  \mathbf{r}\right)  \right]  \widetilde{p}_{k}(\mathbf{r}%
^{\prime}|y(t^{\prime}))}^{\ell}\right|_{y(t')}\frac{\beta\left(  \mathbf{r}\right)
}{mn\left(  \mathbf{r}\right)  }\partial_{j}\overline{t}_{jk}^{\ast}(  \vec r',t'|y  )  \right\}  \label{4.8}%
\end{align}%
\begin{align}
\overline{t}_{0ij}^{\ast}\left(  \mathbf{r},t\mid y\right)     =&\int_{0}%
^{t}dt\int d\mathbf{r}^{\prime}\left\{  -\left.\overline{\left[  e^{\mathcal{L}%
\left(  t-t^{\prime}\right)  }t_{0ij}(\mathbf{r})\right]  \phi_{0k}%
(\mathbf{r}^{\prime}|y(t^{\prime}))}^{\ell}\right|_{y(t')}\partial_{k}^{\prime}%
\beta(\mathbf{r}^{\prime},t^{\prime})\right.  \nonumber\\
&  +\left.\overline{\left[  e^{\mathcal{L}\left(  t-t^{\prime}\right)  }%
t_{0ij}(\mathbf{r})\right]  \sigma_{0km}(\mathbf{r}^{\prime}|y(t^{\prime})%
)}^{\ell}\right|_{y(t')}\beta(\mathbf{r}^{\prime},t^{\prime})\partial_{m}%
^{\prime}u_{k}(\mathbf{r}^{\prime},t^{\prime})\nonumber\\
&  -\left.\overline{\left[  e^{\mathcal{L}\left(  t-t^{\prime}\right)  }%
t_{0ij}(\mathbf{r})\right]  \xi_0\left(  \mathbf{r}^{\prime}|y(t^{\prime})\right)
}^{\ell}\right|_{y(t')}\partial_{k}\overline{s}_{0k}^{\ast}(\vec r',t'| y) \nonumber\\
&  \left.  +\left.\overline{\left[  e^{\mathcal{L}\left(  t-t^{\prime}\right)
}t_{0ij}(\mathbf{r})\right]  \widetilde{p}_{m}(\mathbf{r}^{\prime}|y(t^{\prime
}))}^{\ell}\right|_{y(t')}\partial_{k}\overline{t}_{km}^{\ast}(\vec r',t'| y)  \right\}  \label{4.9}%
\end{align}
These expressions involve local equilibrium correlation functions for the flux
operators with three others%
\begin{align}
\phi_{0i}(\mathbf{r}| y(t))&=\widetilde{s}_{0i}(\vec r|y(t))-\int d\mathbf{r}%
^{\prime}\widetilde{p}_{0j}(\vec r|y(t))\frac{\beta\left(
\mathbf{r}^{\prime},t\right)  }{n\left(  \mathbf{r}^{\prime},t\right)
}
\left.\overline{ p_{0j}\left(  \mathbf{r}^{\prime}\right)  \widetilde{s}%
_{0i}(\mathbf{r}|y(t))}^{\ell}\right|_{y(t)}
 \label{4.10}\\
\sigma_{0ij}(\mathbf{r}| y(t)) &  =\widetilde{t}_{0ij}(\vec r|y(t)) -\int d\mathbf{r}^{\prime}\left(  \widetilde
{n}(\vec r|y(t))\left.\frac{\delta \pi_{0ij}(\vec r|y(t))}{\delta n\left(
\mathbf{r}^{\prime},t\right)  }\right|_{e}+\widetilde{e}_{0}(\mathbf{r}^{\prime
}|y(t))\left.\frac{\delta \pi_{0ij}(\vec r|y(t))}{\delta e\left(  \mathbf{r}^{\prime
},t^{\prime}\right)  }\right|_{n}\right)   +\beta^{-1}(\mathbf{r},t)\xi_0\left(  \mathbf{r},t\right)  \overline{t}%
_{ij}^{\ast}\left(  \mathbf{r}^{\prime},t|y\right)  \label{4.11}\\
\xi_0\left(  \mathbf{r}| y(t)\right) & =\int d\mathbf{r}^{\prime}\left(  \frac
{\delta\beta\left(  \mathbf{r,}t\right)  }{\delta\overline{n}_{0}\left(
\mathbf{r}^{\prime},t\right)  }\widetilde{n}(\vec r|y(t))+\frac
{\delta\beta\left(  \mathbf{r},t\right)  }{\delta\overline{e}_{0}\left(
\mathbf{r}^{\prime},t\right)  }\widetilde{e}_{0}(\vec r|y(t))\right)
\label{4.12}%
\end{align}

In summary, the exact generalized hydrodynamic equations obtained here are
Eqs. (\ref{2.29}) - (\ref{2.31}),
\begin{align}
& D_{t}\nbar(\br,t) +\nbar(
\br,t) \nabla\cdot\bu(\br,t) =0,
\label{eq122}
\\
%& D_{t}\ebar_{0}(\br,t) +\ebar_{0}(
%\br,t) \nabla\cdot\bu(\br,t)
%+\overline{t}_{0,ij}(\br |\nbar(t)
%,\ebar_{0}(t),\bu(t)) \partial_{i}%
%u_{j}(\br,t) +\nabla\cdot\overline{\bs}_{0}(
%\br |\nbar(t),\ebar_{0}(t)
%,\bu(t)) \nonumber
%\\
%&\qquad +t_{0,ij}^{\ast}(\br,t) \partial_{i}u_{j}(\br,t) +\nabla\cdot\bs_{0}^{\ast}(\br,t) = 0 
%\label{2.30}%
& D_{t}\ebar_{0}(\br,t) +\ebar_{0}(
\br,t) \nabla\cdot\bu(\br,t)
%+\overline{t}^{\ell}_{0,ij}(\br | \zeta(t))
+\overline{{\pi}}_{ij}(\br | \zeta(t)) \partial_{i}%
u_{j}(\br,t) 
%+\nabla\cdot\overline{\bs}^{\ell}_{0}(\br | \zeta(t)) 
\nonumber
\\
&\qquad +\overline{t}_{0ij}^{\ast}(\br,t|y) \partial_{i}u_{j}(\br,t) +\nabla\cdot \overline{\bs}_{0}^{\ast}(\br,
t|y) = -\overline n(\vec r,t)\vec u(\vec r,t)  \cdot\mathbf{\nabla}%
v^{\rm ext}\left(  \mathbf{r},t\right) 
\\
& m\nbar(\br,t) D_{t}u_{i}(\br,
t) +\partial_{j}\overline{{\pi}}_{ij}(\br | \zeta(t)) + \partial_{j} \overline{t}_{0ij}^{\ast}(\br,t|y) 
= -\nbar(
\br,t) \partial_{i}v^{\ext}(\br,t) \,,
\label{eq124}
\end{align}
with the constitutive equations%
\begin{equation}
\overline{\pi}_{ij}\left(  \mathbf{r}\mid\zeta\left(  t\right)  \right)
=\overline{\pi}_{ij}^{\ell}\left(  \mathbf{r}\mid y\left(  t\right)  \right)
=\overline{t}_{ij}^{\ell}(\mathbf{r})\mid y\left(  t\right))  ,
\end{equation}
and $\overline{\mathbf{s}}_{0}^{\ast}\left(  \mathbf{r},t|y\right)  $ and
$\overline{t}_{0ij}^{\ast}\left(  \mathbf{r},t|y\right)  $ given by Eqs.(\ref{4.8}) and
(\ref{4.9}) above.

\end{widetext}
Several remarks are appropriate at this point.

\begin{enumerate}
\item These expressions for the irreversible fluxes are exact. Although
complex, the difficult many body problem has been reduced to calculating local
equilibrium time correlation functions. This is a formidable problem,
extending that already present for linear response expressed in terms of
corresponding equilibrium time correlation functions. The fields of the local
equilibrium ensemble for the correlation functions are $\beta\left(
\mathbf{r,}t\right)  $ and $\nu\left(  \mathbf{r,}t\right)  $ only, since the latter 
are in the local rest frame.

\item They are linear integral equations for the fluxes since they appear also
on the right sides of (\ref{4.8}) and (\ref{4.9}) as well. This complication
can be eliminated by introducing an appropriate projection operator, at the
price of a more complex generator for the dynamics than the Liouville
operator. This is described briefly in Appendix C.

\item These expressions provide the desired closure of the formal hydrodynamic
equations (\ref{2.29}) - (\ref{2.31}), giving the irreversible fluxes as
functionals of the local conserved fields. This is done indirectly, with
(\ref{4.8}) and (\ref{4.9}) expressed first as functionals of the conjugate
fields of the local equilibrium ensemble. These conjugate fields are directly
related to the conserved fields through (\ref{eq:nequiv}) - (\ref{eq:pequiv}). The choice
to represent the hydrodynamic equations in terms of the conserved fields or
the conjugate fields is a matter of convenience for a given physical state.

\item The fluxes are non-local in space, just as the pressure tensor
functional, and hence can describe states with strong spatial inhomogeneity. This
is an extension beyond the usual Navier-Stokes hydrodynamics that is limited to
small spatial gradients.

\item In contrast to the perfect fluid hydrodynamics, the irreversible fluxes
are also non-local in time. Hence they describe all time scales, and also
hysteresis (memory effects).

\item The response to small initial perturbations or external forces about
equilibrium give the exact hydrodynamic response. Since the hydrodynamic
equations describe all space and time scales, the exact response functions are
obtained from this description. Of course, they are given in terms of the
corresponding limits for the local equilibrium time correlation functions.
This is the generalized linear hydrodynamics for equilibrium time correlation
functions studied some time ago \cite{Mountain1977,Boon1991}.

\item The equilibrium response functions
obtained from the hydrodynamic equations provide an interesting new exact
representation. The generalized hydrodynamic form allows a direct cross-over
to the small wavevector, low frequency limit - the Navier-Stokes limit. As
discussed by Kadanoff and Martin\cite{KadanoffMartin1963} this is a singular limit not directly
obtained by standard many body methods.

\item A linearization about local equilibrium for the pressure tensor and
irreversible fluxes yields the non-linear Navier-Stokes hydrodynamics with the
Green - Kubo expressions for transport coefficients. This is more general than
linear response about equilibrium, in that the pressure and transport
coefficients are local functions of the hydrodynamic fields at the space and
time point of interest (in density functional theory this is the ``local
density approximation", extended to all fields).
\end{enumerate}

\section{Summary and Discussion}

\label{sec5}A very general class of problems across many fields is currently
being addressed via a continuum description of a few relevant macroscopic
fields. For example, time dependent density functional theory focuses solely
on the space and time dependent average electron number density. Properties of
interest are presumed to be expressed as functionals of that density. A
broader class of macroscopic fields, as considered here, include the energy
density and momentum density which together with the number density have a
special feature: they are averages of the set of exact locally conserved
quantities. Historically, this property has provided the basic framework for
phenomenological macroscopic balance equations for number, energy, and
momentum. For special states, the phenomenology can be justified and
supplemented by materials properties from the underlying statistical mechanics.

The present context is an interest in this approach for non-traditional
systems (e.g., degenerate electrons) with the introduction of new
phenomenology and applications whose validity and context is not yet fully
explored. The objective here has been to suggest an alternative approach, that
of starting from an exact continuum formulation and building more controlled
approximations from it. For example, the Euler equations of hydrodynamics is a
well-understood and useful approximation to the full Navier-Stokes equations;
here, the perfect fluid hydrodynamics of Section
\ref{sec:perfect_fluid_hydrodynamics} is understood as the analogous approximation (no dissipation)
to the exact hydrodynamics, extended to arbitrary space and time scales.
Furthermore, the perfect fluid equations are seen to provide the \emph{exact
short time} behavior of the system.

The analysis here is quite general within the limit of a single component
non-relativistic fluid. Extensions to multi-component systems and inclusion of
relativistic effects (e.g., graphene) follow directly. The
primary results obtained are the exact balance equations (\ref{eq122}) - (\ref{eq124}) for the
average local conserved fields, and the exact ``constitutive equations" for the
fluxes (\ref{4.8}) and (\ref{4.9}). The balance equations require a pressure tensor as a
functional of the density and energy density in the local equilibrium state.
This is an extension of the corresponding problem of density functional theory
to include non-uniform energy density (or temperature) as well as number
density. This formal extension has been developed but is still awaiting
application
\cite{Eich2016,Dufty_unpublished}. The connection of the local hydrodynamic
pressure to thermodynamics and an associated pressure tensor are discussed
above and elsewhere
\cite{Percus1986,Pozhar1993,Dufty_etal_unpublished}. Recent
practical forms for strong coupling and strong spatial inhomogeneities have
been developed recently
\cite{Karasiev2018,Luo2018}. Further developments for conditions of
warm, dense matter remain a forefront computational challenge.

The constitutive equations for the dissipative energy and momentum fluxes are
non-local functionals of the fields with respect to both space and time. While
quite complex, their exact representation parallels closely that for
Navier-Stokes hydrodynamics - linear in the spatial gradients of the conjugate
fields with coefficients given by equilibrium time correlation functions that
are local in space and time functions of the fields. These are Fourier's law
and Newton's viscosity law. The many-body challenge has been compressed to
calculating or modelling the time correlation functions. The exact forms here,
(\ref{4.8}) and (\ref{4.9}), have a similar structure generalized to non-locality and
non-linearity in the conjugate fields, characterized by \emph{local
equilibrium} time correlation functions. The many-body challenge is now
calculating or modelling these new correlation functions. The advantage of all
the formal analysis behind these equations is to embed this limited (but
difficult!) computation within a structure that assures the correct physics of
the underlying conservation laws. Approximations made in the computation are
expected to have quantitative rather than qualitative consequences.

There are several directions for first exploitation of these results. As noted
above, an interesting application would be to the linear dynamics of a
strongly coupled, inhomogeneous system of electrons in an external potential
of a frozen configuration of neutralizing ions (a model for warm, dense
matter) using the ideal fluid approximation (no dissipation). This requires
calculating thermodynamic derivatives for the inhomogeneous system - time
independent two point correlation functions. Such dynamics can be studied for
comparison by existing methods of Born-Oppenheimer molecular dynamics using
DFT generated forces \cite{marx2009}.

Another direction to explore is a local density approximation for spatially
smooth states. This entail replacing all spatial functional dependence of the
conserved and conjugate fields by their values at the same external field
point. The pressure tensor and irreversible flux correlation functions then
become corresponding equilibrium properties for which methods for evaluation
are available. This can be done without a corresponding localization in time,
extending the description to all time scales. One method available for such a
calculation including strong coupling is the Kubo-Greenwood method within
standard DFT \cite{Holst2011,Dufty2018}.

A new challenge presented by the results here is a better understanding of the
local equilibrium state, both its structure and fluctuations. This is a direct extension
of the corresponding equilibrium problem to non-uniform equilibrium states. There are
many opportunities for application of systematic formal methods, models, and novel simulation
techniques.

\begin{acknowledgments}
This research was supported by the U.S. Department of Energy grant DE-SC 0002139.
\end{acknowledgments}

\bigskip

\appendix
\begin{widetext}
\section{Properties of the local equilibrium ensemble}

\label{ap:A}The objective of this Appendix is to derive identities for
$\delta\rho^{\ell}_N[y(t)] /\delta y_{\alpha}(\br%
,t),$ $\partial_{t}\rho^{\ell}_N[y(t)],$ and ${\cal L}_N \rho^{\ell}_N[y(
t)] .$ The local equilibrium ensemble is defined by%
\begin{equation}
\rho^{\ell}_N\left[ y(t) \right] =e^{-\eta\left[ y(
t) \right] } %\calS_{N}
,\quad\eta\left[ y(
t) \right] =Q^{\ell}\left[ y(t) \right] +\int
\dr y_{\kappa}(\br,t)\psi_{\kappa}(\br) \,.
\end{equation}
The functional derivative of a functional is defined from its first order
differential variation
\begin{equation}
\lim_{\delta y\to0}(e^{-\eta\left[ y+\delta y\right]
}-e^{-\eta\left[ y\right] }) \to \int \dr\frac{\delta
e^{-\eta\left[ y\right] }}{\delta y_{\kappa}(\br)}\delta y_{\kappa
}(\br). \label{a.2}%
\end{equation}
The dependence of $y$ on $t$ has been suppressed to simplify notation at this
point. Next, define
\begin{equation}
e^{-x\eta\left[ y+\delta y\right] }\equiv e^{-x\eta\left[ y\right]
}U(x), \label{a.3}%
\end{equation}
so that $U(x) $ obeys the equation
\begin{equation}
\partial_{x}U(x) =-e^{x\eta\left[ y\right] }(
\eta\left[ y+\delta y\right]-\eta\left[ y\right]) e^{-x\eta
\left[ y\right] }U(x) . \label{a.4}%
\end{equation}
Integrating from $x=0$ to $x=1$, and using the fact that $U(0)=1$ gives
\begin{equation}
e^{-\eta\left[ y+\delta y\right] }-e^{-\eta\left[ y\right] }=-\int_{0}%
^{1}dx'e^{(x'-1) \eta[ y]}(
\eta\left[ y+\delta y\right]-\eta\left[ y\right]) e^{-x^{\prime
}\eta\left[ y\right] }U(x'), \label{a.5}%
\end{equation}
and so to linear order in $\delta y$
\begin{align}
\frac{\delta\rho^{\ell}_N\left[ y\right] }{\delta y_{\kappa}(\br)} &
=-\int_{0}^{1}dx'e^{-x'\eta\left[ y\right] }\frac{\delta
\eta\left[ y\right] }{\delta y_{\kappa}(\br)}e^{x'\eta\left[
y\right] }\rho^{\ell}_N\left[ y\right] \nonumber\\
& =-\widetilde{\psi}_{\kappa}(\br|y)\rho^{\ell}_N\left[ y\right] .
\label{a.6}%
\end{align}
The tilde above an operator is defined in (\ref{eq:xtilde}). 
%
%\begin{equation}
%\widetilde{X}=\int_{0}^{1}dx'e^{-x'\eta\left[ y\right]
%}(X- \overline{X} ^{\ell}) e^{x'\eta\left[
%y\right] } \label{a.7}%
%\end{equation}
The functional derivative of any local equilibrium average is therefore%
\begin{equation}
\frac{\delta\overline{X}^{\ell}\mid y\left(  t\right)  }{\delta
y_{\kappa}(\mathbf{r},t)}=-\left.\overline{X\widetilde{\psi}_{\beta}\left(
\mathbf{r} \mid y\left(  t\right)  \right)  }^{\ell}\right|_{ y(t)}.
\label{a.7}%
\end{equation}
For example%
\begin{equation}
\frac{\delta\overline{\psi}_{\kappa}^{\ell}\left(  \mathbf{r}\mid
y\left(  t\right)  \right)  }{\delta y_{\lambda}(\mathbf{r}^{\prime}%
,t)}=-\left.\overline{\psi_{\kappa}\left(  \mathbf{r}\right)  \widetilde{\psi
}_{\lambda}\left(  \mathbf{r}^{\prime}\mid y\left(  t\right)  \right)  }%
^{\ell}\right|_{ y(t)}  =-g_{\kappa\lambda}\left(  \mathbf{r}%
,\mathbf{r}^{\prime}\mid y\left(  t\right)  \right)  .
\label{a.8a}%
\end{equation}

Now, restoring the dependence of $y$ on $t$, the time derivative of $\rho
^{\ell}\left[ y(t) \right] $ can be calculated directly%
\begin{align}
\partial_{t}\rho^{\ell}_N\left[ y(t) \right]  & =\int
\dr\frac{\delta e^{-\eta\left[ y(t) \right] }}{\delta
y_{\kappa}(\br,t)}\partial_{t}y_{\kappa}(\br,t)\nonumber\\
& =-\int \dr\widetilde{\psi}_{\kappa}(\br|y(t))\partial
_{t}y_{\kappa}(\br,t)\rho^{\ell}_N\left[ y(t) \right] \,.
\label{a.9}%
\end{align}

Finally, consider the Liouville operator acting on the local equilibrium
ensemble%
\begin{equation}
{\cal L}_N e^{-\eta}=i\left[ H,e^{-\eta}\right] =iHe^{-\eta}-ie^{-\eta}H \,.
\label{a.10}%
\end{equation}
Define%
\begin{equation}
H(x)=e^{x\eta}He^{-x\eta}, \label{a.11}%
\end{equation}
which obeys the equation%
\begin{equation}
\partial_{x}H(x)=\left[ \eta,H(x) \right] . \label{a.12}%
\end{equation}
Integrating from $0$ to $1$ gives%
\begin{equation}
e^{\eta}He^{-\eta}=H+\int_{0}^{1}\left[ \eta,H(x) \right] dx,
\label{a.13}%
\end{equation}
or%
\begin{align}
iHe^{-\eta}-ie^{-\eta}H & =\int_{0}^{1}e^{-\eta}i\left[ \eta,H(
x) \right] dx=\int_{0}^{1}e^{-\eta(1-x) }i\left[
\eta,H\right] e^{-x\eta}dx\nonumber\\
& =\left(\int_{0}^{1}e^{-z\eta}i\left[ \eta,H\right] e^{z\eta}\dd z \right)e^{-\eta} \label{a.14}%
\end{align}

and so%
\begin{align}
\mathcal{L}_Ne^{-\eta\lbrack y]} &  =-\left(  \int_{0}^{1}e^{-z\eta\lbrack
y]}\left(  \mathcal{L}_N\eta\lbrack y]\right)  e^{z\eta\lbrack y]}dz\right)
e^{-\eta\lbrack y]}\nonumber\\
&  =-\int d\mathbf{r}y_{\alpha}(\mathbf{r},t)\left(  \int_{0}^{1}%
e^{-z\eta\lbrack y]}\left(  \mathcal{L}_N\psi_{\alpha}(\mathbf{r})\right)
e^{z\eta\lbrack y]}dz\right)  e^{-\eta\lbrack y]}\nonumber\\
&  =\int d\mathbf{r}y_{\alpha}(\mathbf{r},t)\left(  \partial_{i}\left(
\widetilde{\gamma}_{i\alpha}(\mathbf{r}|y(t))+\overline{\gamma}_{i\alpha}^{\ell
}(\mathbf{r}|y(t))\right)  e^{-\eta\lbrack y]}-\left(  \widetilde{f}_{i\alpha
}(\mathbf{r}|y(t))+\overline{f}_{i\alpha}^{\ell}(\mathbf{r}|y(t))\right)  \right)
e^{-\eta\lbrack y]}\label{a.15}%
\end{align}
where the microscopic conservation law
corresponding to (\ref{3.08}) - (\ref{3.09}) has been used, and
\begin{equation}
\widetilde{\gamma}_{i\alpha}(\mathbf{r}|y(t))=\int_{0}^{1}e^{-z\eta}\left(
\gamma_{i\alpha}(\mathbf{r})-\widetilde{\gamma}_{i\alpha}^{\ell}(\mathbf{r}%
|y(t))\right)  e^{z\eta}dz\label{a.16}%
\end{equation}
One final simplification follows from the average of (\ref{a.15}),
\begin{equation}
0=
\sum_N
\Tr^{(N)}\mathcal{L}_N\rho^\ell_N[y\left(  t\right)]  =\int d\mathbf{r}y_{\alpha
}(\mathbf{r},t)\left(  \partial_{i}\overline{\gamma}_{i\alpha}^{\ell
}(\mathbf{r}|y(t))-\overline{f}_{i\alpha}^{\ell}(\mathbf{r}|y(t))\right)
,\label{a.17}%
\end{equation}
since $\overline{\widetilde{X}}^{\ell}=0$ for any operator $X$. Then
(\ref{a.15}) becomes%
\begin{equation}
\mathcal{L}_Ne^{-\eta\lbrack y]}=\int d\mathbf{r}y_{\alpha}(\mathbf{r},t)\left(
\partial_{i}\widetilde{\gamma}_{i\alpha}(\mathbf{r}|y(t))-\widetilde{f}_{i\alpha
}(\mathbf{r}|y(t))\right)  e^{-\eta\lbrack y]}\label{a.18}%
\end{equation}

\subsection{Local equilibrium averages}

Consider a local equilibrium average of an operator $A(\mathbf{r})$ in its
local rest frame (function of the relative momenta $\mathbf{p}_{\alpha
}-mu\left(  \mathbf{q}_{\alpha}\right)  $). Also note that $\rho_N^\ell$ also
depends on the flow velocity only through the local momenta. Therefore, a
unitary transformation (see (\ref{2.11a}), (\ref{2.11b}), and Supplemental Material) removes
the velocity dependence; for simplicity of notation the dependence on $y(t)$ is
suppressed in this subsection.%
\begin{equation}
\overline{ A(\mathbf{r})} ^{\ell}=%
 \sum_N
\Tr^{(N)}e^{-G}A(\mathbf{r})\rho_N^\ell e^{G}=\left.\overline{ A(\mathbf{r})} ^{\ell}\right|_{\mathbf{u}=0}\label{a.19}%
\end{equation}
Now, let $T$ denote the anti-unitary time reversal operator (see Supplemental
Material), with the property $T\mathbf{p}_{\alpha}T^{-1}=-\mathbf{p}_{\alpha}%
$,  $T\mathbf{q}_{\alpha}T^{-1}=\mathbf{q}_{\alpha}$. Suppose $A(\mathbf{r})$
has the property $TA(\mathbf{r})T^{-1}=-A(\mathbf{r})$, then its local
equilibrium average for $\mathbf{u}=0$ vanishes%

\begin{equation}
\left.\overline{ A(\mathbf{r})} ^{\ell}\right|_{\mathbf{u}=0}%
=\left.\overline{TA(\mathbf{r})T^{-1}}^{\ell}\right|_{\mathbf{u}%
=0}=-\left.\overline{ A(\mathbf{r})} ^{\ell}\right|_{\mathbf{u}%
=0}.\label{a.20}%
\end{equation}
The first equality is non-trivial since the cyclic invariance of the trace
does not hold for anti-unitary operators; it is proved in Section S3 of the Supplemental
Material. Also use has been made of $T\rho_N^\ell T^{-1}=\rho_N^\ell$. 

As an example,
consider the choice $A(\mathbf{r})\rightarrow\mathbf{s}_0\left(  \mathbf{r}%
\right)  $, the energy flux in the local rest frame. According to (\ref{a.20})
its local equilibrium average must vanish,
\begin{equation}
\overline{s}_{0}^{\ell}\left(  \mathbf{r}\right)  =\overline{Ts_{0}\left(
\mathbf{r}\right)  T^{-1}}^{\ell}=-\overline{s}_{0}^{\ell}\left(
\mathbf{r}\right)  =0
.
\end{equation} 
As a second example consider the
matrix $g_{\alpha\beta}\left(  \mathbf{r},\mathbf{r}^{\prime}\right)
=\left\langle \psi_{\alpha}(\mathbf{r})\widetilde{\psi}_{\beta}(\mathbf{r}%
^{\prime})\right\rangle ^{\ell}$ in the local rest frame. The conserved densities are odd or even
under time reversal operation. Hence the matrix elements for densities with
opposite signs must vanish.

%\section{Definitions for local pressure functional}
%
%\label{apB}Compare and contrast average virial operator; thermodynamic local
%pressure; average trace of pressure tensor; show volume average or all are equivalent.
\section{Thermodynamic pressure}

\label{apB}The objective here is to identify the local thermodynamic
pressure given in (\ref{3.20}) 
\begin{equation}
\frac{1}{V}\int d\br\beta(\br,t)\pi_{T}\left(\br|\beta(t),\nu(t)\right)\equiv\frac{\partial Q\left[\beta(t),\nu(t)\right]}{\partial V}.\label{b.1}
\end{equation}
The volume derivative is taken at constant $\beta,\nu$. To make explicit
the volume dependence of $Q\left[\beta(t),\nu(t)\right]$
assume a cubic volume with $V=L^{3}$ and define the unitary operator
$U_{L}$ with the properties\cite{DuftyTrickey2016} 
\begin{equation}
U_{L}\bq_\alpha U_{L}^{-1}=L\bq_\alpha,\hspace{0.25in}U_{L}\bp_\alpha U_{L}^{-1}=L^{-1}\bp_\alpha.\label{b.2}
\end{equation}
Then 
\begin{align}
Q\left[\beta(t),\nu(t)\right] & =\ln 
\sum_{N>0}
\Tr^{(N)}U_{L}^{-1}U_{L}e^{-\int d\br\left[\beta(\br,t)e\left(\br\right)-\nu\left(\br,t\right)n\left(\br\right)\right]}\mathcal{S}_{N}\nonumber \\
 & =\ln\sum_{N>0}\Tr^{(N)}
e^{-U_{L}\int d\br\left[\beta(\br,t)e\left(\br\right)-\nu\left(\br,t\right)n\left(\br\right)\right]U_{L}^{-1}}\mathcal{S}_{N}\label{b.3}
\end{align}
Recall that the flow velocity dependence has been transformed away.
The explicit form for the exponent is 
\begin{align}
\int d\br\left[\beta(\br,t)U_{L}e (\br)U_{L}^{-1}-\nu(\br,t)U_{L}n(\br)U_{L}^{-1}\right] & =\sum_{\alpha=1}^{N}\frac{1}{4m}\left[L^{-2}p_{\alpha j}^{2},\beta(L\bq_{\alpha}, t)\right]_{+}\nonumber \\
 & +\frac{1}{2}\sum_{\alpha\neq \gamma=1}^{N}V\left(\left\vert L\bq_\alpha-L\bq_{\gamma}\right\vert \right)\beta(L\bq_{\alpha}, t)-\sum_{\gamma=1}^{N}\nu\left(L\bq_{\gamma},t\right)\label{b.4}
\end{align}
The derivative with respect to $L$ at constant $\beta,\nu$ is 
\begin{align}
\frac{\partial}{\partial L}U_{L}\int d\br\left[\beta(\br,t)e(\br)-\nu(\br,t)n(\br)\right]U_{L}^{-1} & =\frac{1}{L}\sum_{\alpha=1}^{N}\frac{1}{4m}\left[-2L^{-2}p_{\alpha j}^{2},\beta(L\bq_{\alpha}, t)\right]_{+}\nonumber \\
 & +\frac{1}{L}\frac{1}{2}\sum_{\alpha \neq \gamma=1}^{N}\beta(L\bq_{\alpha},t)L\frac{\partial}{\partial L}V\left(\left\vert L\bq_\alpha-L\bq_{\gamma}\right\vert \right)\label{b.5}
\end{align}
The desired volume derivative is, inverting back the scale transformation
\begin{align}
\frac{\partial}{\partial V}Q\left[\beta(t),\nu(t)\right] & =\frac{\partial L}{\partial V}\frac{\partial}{\partial L}\ln\sum_{N>0}\Tr^{(N)}
e^{-U_{L}\int d\br\left[\beta(\br,t)e\left(\br\right)-\nu\left(\br,t\right)n\left(\br\right)\right]U_{L}^{-1}}\mathcal{S}_{N}\nonumber \\
 & =\frac{1}{3L^{3}}\left[
 \left.\overline{\sum_{\alpha=1}^{N}\frac{1}{2m}\left[p_{\alpha j}^{2},\beta(\bq_{\alpha},t)\right]_{+}}^{\ell}\right|_{y(t)} 
 \right.\nonumber \\
 & \qquad\left.+
  \frac{1}{2}\left.\overline{\sum_{\alpha \neq \gamma=1}^{N}\beta(\bq_{\alpha}, t)
 (\bq_{\gamma}-\bq_{\alpha})\cdot\mathbf{F}_{\alpha \gamma}(\left\vert \bq_{\alpha}-\bq_{\gamma}\right\vert )}^\ell \right|_{y(t)} 
 \right]\label{b.6}\\
 & =\frac{1}{3V}\int d\br\beta(\br,t)\left[
 \left.\overline{ \sum_{j=1}^{N}\frac{1}{2m}\left[p_{\alpha j}^{2},\Delta(\br - \bq_\alpha )\right]_{+}}^{\ell} \right|_{y(t)}
 \right.\nonumber \\
 &\qquad \left.+
 \frac{1}{2}\left.\overline{\sum_{\alpha \neq \gamma=1}^{N}\Delta(\br-\bq_\alpha)
 (\bq_\gamma-\bq_{\alpha})\cdot\mathbf{F}_{\alpha \gamma}\left(\left\vert \bq_{\alpha}-\bq_{\gamma}\right\vert \right)}^\ell \right|_{y(t)} 
 \right]\label{b.7}
\end{align}
Finally, comparison with (\ref{3.20}) gives 
\begin{align}
\pi_{T}\left(\br|\beta(t),\nu(t)\right)  =&\frac{1}{3}\left[
\left.\overline{ \sum_{\alpha=1}^{N}\frac{1}{2m}\left[p_{\alpha j}^{2},\Delta (\br-\bq_{\alpha})\right]_{+}}^{\ell} \right|_{y(t)} 
\right.\nonumber \\
 & \left.+
  \frac{1}{2}\left.\overline{\sum_{\alpha\neq \gamma=1}^{N}\Delta(\br -\bq_{\alpha})(\bq_\gamma-\bq_{\alpha})\cdot\mathbf{F}_{\alpha\gamma}\left(\left\vert \bq_{\alpha}-\bq_{\gamma}\right\vert \right)}^\ell \right|_{y(t)} 
  \right]\label{b.8}
\end{align}
\begin{align}
\pi_{T}\left(\br|\beta(t),\nu(t)\right)  =&\frac{2}{3}\overline{e_{0}^{K}}^\ell\left(\br\right|y(t)) \nonumber \\
 & +\frac{1}{6}
 \left.\overline{ \sum_{\alpha \neq \gamma=1}^{N}\Delta\left(\br -\bq_{\alpha}\right)(\bq_{\alpha}-\bq_{\gamma})\cdot\mathbf{F}_{\alpha\gamma}(\left\vert \bq_{\alpha}-\bq_{\gamma}\right\vert)}^{\ell} \right|_{y(t)} 
 \label{b.9}
\end{align}
Comparison with (\ref{3.7}) shows this is the same as the local
virial pressure 
\[
\pi_{T}\left(\br|\beta(t),\nu(t)\right)=\pi_{v}\left(\br|\beta(t),\nu(t)\right)
\]

\section{Solution to Liouville-von\ Neumann equation}

\label{ap:B}The Liouville-von Neumann equation is
\begin{equation}
(\partial_{t} + {\cal L}_N) \rho_N(t) =0. \label{d.1}%
\end{equation}
Look for solutions of the form%
\begin{equation}
\rho_N(t) =\rho^{\ell}_N[y(t)] +\Delta_N(t) .
\label{d.2}%
\end{equation}
Integrating (\ref{d.1})
\begin{equation}
\rho_N(t) =e^{-{\cal L}_N t}\rho^{\ell}_N[y(0)] =\rho^{\ell}_N[y(
t)]-\int_{0}^{t}dt'\partial_{t'}(e^{-{\cal L}_N(
t-t') }\rho^{\ell}_N[y(t')]),
\label{d.3}%
\end{equation}
gives%
\begin{equation}
\Delta_N(t) =-\int_{0}^{t}dt'e^{-{\cal L}_N (t-t^{\prime
}) }({\cal L}_N+\partial_{t'}) \rho^{\ell}_N[y(t^{\prime
})] . \label{d.4}%
\end{equation}
For simplicity of notation the case of a time
independent external force has been chosen. More generally, the solution
operator must be changed everywhere below as follows%
\begin{equation}
e^{-\mathcal{L}_N\left(  t-t^{\prime}\right)  }\rightarrow {\cal G}\left(  t,t^{\prime
}\right)
\end{equation}
with%
\begin{equation}
\left(  \partial_{t}+\mathcal{L}_N\right)   {\cal G}(t,t^{\prime})=0,\hspace
{0.28in} {\cal G}(t,t)=1
\end{equation}

Using (\ref{a.9}) and (\ref{a.18}) this becomes%
\begin{align}
\Delta_N\left(  t\right)   &  =-\int_{0}^{t}dt^{\prime}e^{-\mathcal{L}_N\left(
t-t^{\prime}\right)  }\int d\mathbf{r}\left(  -\widetilde{\psi}_{\alpha
}(\mathbf{r}|y(t'))\partial_{t^{\prime}}y_{\alpha}(\mathbf{r},t^{\prime}%
)-\widetilde{\gamma}_{i\alpha}(\mathbf{r}|y(t'))\partial_{i}y_{\alpha}%
(\mathbf{r},t^{\prime})\right.  \nonumber\\
& \qquad\qquad\qquad\qquad\qquad\qquad\qquad \left.  -y_{\alpha}(\mathbf{r},t^{\prime})\widetilde{f}_{i\alpha}%
(\mathbf{r}|y(t'))\right)  \rho^\ell_N[y\left(  t^{\prime}\right)]  \nonumber\\
&  =-\int_{0}^{t}dt^{\prime}e^{-\mathcal{L}_N\left(  t-t^{\prime}\right)  }\int
d\mathbf{r}\left(  \Psi_{\alpha}(\mathbf{r}|y(t'))\partial_{t^{\prime}}%
\overline{\psi}_{\alpha}\left(  \mathbf{r}|y(t^{\prime})\right)  +\Gamma
_{i\alpha}(\mathbf{r}|y(t'))\partial_{i}\overline{\psi}_{\alpha}\left(
\mathbf{r},t^{\prime}\right)  \right.  \nonumber\\
& \qquad\qquad\qquad\qquad\qquad\qquad\qquad  \left.  -y_{\alpha}(\mathbf{r},t^{\prime})\widetilde{f}_{i\alpha}%
(\mathbf{r}|y(t'))\right)  \rho^\ell_N[y\left(  t^{\prime}\right)]  \label{c.5}%
\end{align}
where%
\begin{equation}
\Psi_{\alpha}(\mathbf{r}|y(t'))=\int d\mathbf{r}^{\prime}\widetilde{\psi}_{\beta
}(\mathbf{r}^{\prime}|y(t'))g_{\beta\alpha}^{-1}(\mathbf{r}^{\prime},\mathbf{r}|y(t')
),\hspace{0.25in}\Gamma_{i\alpha}(\mathbf{r}|y(t'))=\int d\mathbf{r}^{\prime
}\widetilde{\gamma}_{i\beta}(\mathbf{r}^{\prime}|y(t'))g_{\beta\alpha}%
^{-1}(\mathbf{r}^{\prime},\mathbf{r}|y(t'))\label{c.6}%
\end{equation}
and $g_{\alpha\beta}^{-1}(\mathbf{r},\mathbf{r}^{\prime}|y(t'))$ is the inverse of
$g_{\alpha\beta}(\mathbf{r}^{\prime},\mathbf{r}|y(t'))$ given by (\ref{a.8a})
\begin{equation}
\int d\mathbf{r}^{\prime\prime}g_{\alpha\sigma}(\mathbf{r},\mathbf{r}%
^{\prime\prime}|y(t'))g_{\sigma\beta}^{-1}(\mathbf{r}^{\prime\prime},\mathbf{r}%
^{\prime}|y(t'))=\delta_{\alpha\beta}\delta(\mathbf{r}-\mathbf{r}^{\prime
})\label{c.7}%
\end{equation}
Next, using the macroscopic conservation law \cite{Karasiev2019}
\begin{equation}
\partial_{t}\overline{\psi}_{\alpha}\left(  \mathbf{r}|y(t)\right)  +\partial
_{i}\overline{\gamma}_{\alpha}\left(  \mathbf{r},t\right)  =\overline
{f}_{\alpha}\left(  \mathbf{r},t\right)  ,
\end{equation}
to eliminate $\partial_{t^{\prime}}\overline{\psi}_{\alpha}\left(
\mathbf{r},t^{\prime}\right)  $ leads to
\begin{align}
\Delta_N\left(  t\right)   &  =-\int_{0}^{t}dt^{\prime}e^{-\mathcal{L}_N\left(
t-t^{\prime}\right)  }\int d\mathbf{r}\left(  \Psi_{\alpha}(\mathbf{r}|y(t'))\left(
-\partial_{i}\overline{\gamma}_{i\alpha}\left(  \mathbf{r},t^{\prime}\right)
+\overline{f}_{\alpha}\left(  \mathbf{r},t\right)  \right)  +\Gamma_{i\alpha
}(\mathbf{r}|y(t'))\partial_{i}\overline{\psi}_{\alpha}\left(  \mathbf{r}|y(t^{\prime
})\right)  \right.  \nonumber\\
& \qquad\qquad\qquad\qquad\qquad\qquad\qquad \left.  -y_{\alpha}(\mathbf{r},t^{\prime})\widetilde{f}_{i\alpha}^\ell%
(\mathbf{r}|y(t'))\right)  \rho^\ell_N[y\left(  t^{\prime}\right)]  \nonumber\\
&  =-\int_{0}^{t}dt^{\prime}e^{-\mathcal{L}_N\left(  t-t^{\prime}\right)  }\int
d\mathbf{r}\left(  -\Psi_{\alpha}(\mathbf{r}|y(t'))\partial_{i}\overline{\gamma
}_{i\alpha}^{\ast}\left(  \mathbf{r},t^{\prime}\right)  -\Psi_{\alpha
}(\mathbf{r},t')\partial_{i}\overline{\gamma}_{i\alpha}^{\ell}\left(  \mathbf{r}%
,t^{\prime}\right)  +\Psi_{\alpha}(\mathbf{r},t')\overline{f}_{\alpha}\left(
\mathbf{r},t\right)  \right.  \nonumber\\
&  \qquad\qquad\qquad\qquad\qquad\qquad\qquad \left.  +\Gamma_{i\alpha}(\mathbf{r},t')\partial_{i}\overline{\psi}_{\alpha
}\left(  \mathbf{r}|y(t^{\prime})\right)  -y_{\alpha}(\mathbf{r},t^{\prime
})\widetilde{f}_{i\alpha}(\mathbf{r}|y(t'))\right)  \rho^\ell_N[y\left(  t^{\prime
}\right)]  \label{c.8a}%
\end{align}
The contribution from the local equilibrium flux can be written%
\begin{align}
\partial_{i}\overline{\gamma}_{i\alpha}^{\ell}\left(  \mathbf{r},t^{\prime
}\right)   &  =-
\overline{ \mathcal{L}_N\psi_{\alpha}\left(  \mathbf{r}%
\right) }^\ell|y(t')
+\overline{f}_{\alpha}^{\ell
}(\mathbf{r,}t^{\prime})\nonumber\\
&  =\int d\mathbf{r}^{\prime}\sum_{N}\Tr^{(N)}\psi_{\alpha}\left(
\mathbf{r}\right)  \left(  -\widetilde{\gamma}_{i\beta}\left(  \mathbf{r}%
^{\prime}|y(t')\right)  \partial_{i}y_{\beta}(\mathbf{r}^{\prime},t^{\prime
})-\widetilde{f}_{i\beta}(\mathbf{r}^{\prime}|y(t'))y_{\beta}(\mathbf{r}^{\prime
},t^{\prime})\right)  e^{-\eta\left(  y\right)  }\nonumber\\
& \qquad +\overline{f}_{\alpha}^{\ell}(\mathbf{r,}t^{\prime})\nonumber\\
&  =\int d\mathbf{r}^{\prime}\left(  -\left.\overline{\psi_{\alpha}\left(  \mathbf{r}^{\prime
}\right)  \widetilde{\gamma}_{i\beta}\left(  \mathbf{r}^{\prime}\mid y\left(
t^{\prime}\right)  \right)  }^{\ell}\right|_{y(t')}
\partial_{i}y_{\beta}(\mathbf{r}%
^{\prime},t^{\prime})-
\left.\overline{ \psi_{\alpha}\left(  \mathbf{r}\right)
\widetilde{f}_{\beta}(\mathbf{r}^{\prime},t')}^\ell\right|_{y(t')}
y_{\beta}(\mathbf{r}^{\prime},t^{\prime})\right)  \nonumber\\
& \qquad +\overline{f}_{\alpha}^{\ell}(\mathbf{r,}t^{\prime})\label{c.9}%
\end{align}
Therefore (\ref{c.8a}) becomes%
\begin{align}
\Delta_N\left(  t\right)   &  =-\int_{0}^{t}dt^{\prime}e^{-\mathcal{L}_N\left(
t-t^{\prime}\right)  }\int d\mathbf{r}\left(  -\Psi_{\alpha}(\mathbf{r}|y(t')
)\partial_{i}\overline{\gamma}_{i\alpha}^{\ast}\left(  \mathbf{r},t^{\prime
}\right)  +\Psi_{\alpha}(\mathbf{r}|y(t'))\int d\mathbf{r}^{\prime}
\left.\overline{
\psi_{\alpha}\left(  \mathbf{r}\right)  \widetilde{\gamma}_{i\beta}\left(
\mathbf{r}^{\prime},t'\right) }^\ell\right|_{y(t')}
\partial_{i}y_{\beta
}(\mathbf{r}^{\prime},t^{\prime})\right.  \nonumber\\
& \qquad \left.  +\Gamma_{i\alpha}(\mathbf{r},t')\partial_{i}\overline{\psi}_{\alpha
}\left(  \mathbf{r}|y(t^{\prime})\right)  -\left(  \widetilde{f}_{i\alpha
}(\mathbf{r}|y(t'))y_{\alpha}(\mathbf{r},t^{\prime})-\Psi_{\alpha}(\mathbf{r}|y(t')
)
\left.\overline{ \psi_{\alpha}\left(  \mathbf{r}\right)  \widetilde{f}_{i\beta
}(\mathbf{r}^{\prime}|y(t'))}^\ell\right|_{y(t')}
y_{\beta}(\mathbf{r}%
^{\prime},t^{\prime})\right)  \right)  \rho^\ell_N[y\left(  t^{\prime}\right)]
\label{c.10}\\
&  =\int_{0}^{t}dt^{\prime}e^{-{\cal L}_N\left(  t-t^{\prime}\right)  }\int
d\mathbf{r}\left(  \left(  \Psi_{\alpha}(\mathbf{r}|y(t'))\partial_{i}%
\overline{\gamma}_{i\alpha}^{\ast}\left(  \mathbf{r},t^{\prime}\right)
-\Phi_{i\alpha}(\mathbf{r}|y(t'))\partial_{i}\overline{\psi}_{\alpha}\left(
\mathbf{r},t^{\prime}\right)  \right)  \right.  \nonumber\\
&\qquad -\left.  \left(  \widetilde{f}_{i\alpha}(\mathbf{r}|y(t'))-\int d\mathbf{r}%
^{\prime}\Psi_{\beta}(\mathbf{r}^{\prime}|y(t'))
\left.\overline{ \psi_{\beta}\left(
\mathbf{r}^{\prime}\right)  \widetilde{f}_{i\alpha}(\mathbf{r}|y(t'))}^\ell\right|_{y(t')}
\right)  y_{\alpha}(\mathbf{r},t^{\prime})\right)  \rho
^\ell_N[y\left(  t^{\prime}\right)]  \label{c.11}%
\end{align}
with%
\begin{equation}
\Phi_{i\alpha}\left(  \mathbf{r}\mid y\left(  t\right)  \right)
=\Gamma_{i\alpha}\left(  \mathbf{r}\mid y\left(  t\right)  \right)  -\int
d\mathbf{r}^{\prime}\Psi_{i\beta}\left(  \mathbf{r}^{\prime}\mid y\left(
t\right)  \right) \left. \overline{\psi_{\beta}\left(  \mathbf{r}^{\prime}\right)
\Gamma_{i\alpha}\left(  \mathbf{r}\mid y\left(  t^{\prime}\right)  \right)
}^{\ell}\right|_{y(t')} 
.\label{c.12}%
\end{equation}
The last line of (\ref{c.11}) is the component of the sources $\widetilde
{f}_{\alpha}(\mathbf{r})$ orthogonal to the local conserved densities. The
energy and momentum sources are both proportional to conserved densities and
hence this term vanishes, leaving%
\begin{equation}
\Delta_N\left(  t\right)  =\int_{0}^{t}dt^{\prime}e^{-{\cal L}_N\left(  t-t^{\prime
}\right)  }\int d\mathbf{r}\left(  \left(  \Psi_{\alpha}(\mathbf{r}|y(t')
)\partial_{i}\overline{\gamma}_{i\alpha}^{\ast}\left(  \mathbf{r},t^{\prime
}\right)  -\Phi_{i\alpha}(\mathbf{r}|y(t'))\partial_{i}\overline{\psi}_{\alpha
}\left(  \mathbf{r}|y(t^{\prime})\right)  \right)  \right)  \rho^\ell_N[y\left(
t^{\prime}\right)].  \label{c.13}%
\end{equation}

\subsection{Projection operator representation}

It is useful to note some properties of the operators occurring in
(\ref{c.11}). First, it is observed that the operators $\psi_{\beta}(
\br) $ and $\Psi_{\alpha}(\br|y)$ form a biorthogonal set
in the sense%
\begin{equation}
\left.\overline{ \psi_{\alpha}(\br) \Psi_{\beta}%
(\br'|y(t))}^{\ell}\right|_{y(t)} =\int \dr''
\left.\overline{\psi_{\alpha}(\br) \widetilde{\psi}%
_{\sigma}(\br''|y(t))}^{\ell}\right|_{y(t)} g_{\sigma\beta}%
^{-1}(\br'',\br'|y(t))
=\delta(\br-\br') \delta_{\alpha\beta}.
\label{d.12}%
\end{equation}
Also $\Phi_{i\alpha}(\br|y)$ is orthogonal to this set%
\begin{align}
\left. \overline{\psi_{\alpha}(\br) \Phi_{i\beta}%
(\br'|y(t))}^{\ell}\right|_{y(t)} & = 
\left.\overline{\psi_{\beta
}(\br) \Gamma_{i\alpha}(\br'%
|y(t))}^{\ell}\right|_{y(t)} -\int \dr'' 
\left.\overline{\psi
_{\alpha}(\br') \Psi_{\sigma}(\br%
''|y(t))}^{\ell}\right|_{y(t)}\left. \overline{\psi_{\sigma}(
\br''))\Gamma_{i\alpha}(\br'|y(t))}^{\ell}\right|_{y (t)} \nonumber\\
& =0. \label{d.18}%
\end{align}
Define a related projection operator $\mathcal{P}_{t}$ for trace
operators $Y$
\begin{equation}
\mathcal{P}_{t}Y=\int \dr\Psi_{\alpha}(\br)%
\sum\limits_{N}
\Tr^{(N)}\psi_{\alpha}(\br) Y,\hspace{0.25in}%
\sum\limits_{N}
\Tr^{(N)}Y<\infty. \label{d.19}%
\end{equation}
Then it is straightforward to show%
\begin{equation}
\mathcal{P}_{t}^{2}Y=\mathcal{P}_{t}Y,\hspace{0.25in}\mathcal{P}_{t}%
\Psi_{\alpha}(\br)\rho^{\ell}_N=\Psi_{\alpha}(\br)\rho^{\ell}_N%
,\hspace{0.25in}\mathcal{P}_{t}\Phi_{\alpha}(\br)\rho^{\ell}_N=0,
\label{d.20}%
\end{equation}
and consequently, from (\ref{c.11})%
\begin{equation}
\mathcal{P}_{t}\Delta_N=0. \label{d.21}%
\end{equation}

The equation of motion for $\Delta_N$ from (\ref{c.11}) is
\begin{equation}
(\partial_{t}+\calL) \Delta_N(t) =\int \dr%
(\Psi_{\alpha}(\br,t)\partial_{i}\overline{\gamma}_{i\alpha}%
^{\ast}(\br,t)-\Phi_{i\alpha}(\br,t)\partial
_{i}\overline{\psi}_{\alpha}(\br,t)) \rho
^{\ell}_N[y(t)] \label{d.22}%
\end{equation}
Equivalently, using $\Delta_N(t) =Q_{t}\Delta_N(t) $
with $Q_{t}=1-\mathcal{P}_{t}$
\begin{align}
(\partial_{t}+Q_{t}{\cal L}_NQ_{t}) Q_{t}\Delta_N(t)  &
=Q_{t}\int \dr(\Psi_{\alpha}(\br|y(t))\partial_{i}%
\overline{\gamma}_{i\alpha}^{\ast}(\br,t)-\Phi_{i\alpha
}(\br|y(t))\partial_{i}\overline{\psi}_{\alpha}(\br,t)
) \rho^{\ell}_N[y(t)] \nonumber\\
& =-\int \dr\Phi_{i\alpha}(\br|y(t))\rho^{\ell}_N[y(t)]
\partial_{i}\overline{\psi}_{\alpha}(\br,t) \label{d.23}%
\end{align}
Finally, integrating this with the initial condition $\Delta_N(0)
=0$ gives the desired result%
\begin{equation}
\Delta_N(t) =-\int_{0}^{t}dt'U(t,t')\int
\dr\Phi_{i\alpha}(\br|y(t'))\partial_{i}\overline{\psi}_{\alpha
}(\br,t') \rho^{\ell}_N[y(t')] .
\label{d.24}%
\end{equation}
The evolution operator $U(t,t')$ gives the modified Liouville-von
Neumann dynamics%
\begin{equation}
(\partial_{t}+Q_{t}{\cal L}_N Q_{t}) U(t,t')=0,\hspace
{0.25in}U(t,t)=1. \label{d.25}%
\end{equation}
In contrast to (\ref{c.11}) there is now no longer an explicit dependence on
$\overline{\gamma}_{i\alpha}^{\ast}(\br,t) $ in
(\ref{d.24}).
\end{widetext}

\end{document}

\title{Supplemental Material\\
{\large Generalized Hydrodynamics Revisited}}
\author{James Dufty}
%\email{dufty@ufl.edu}
%\email{dufty@ufl.edu}
\affiliation{Department of Physics, University of Florida, Gainesville, FL 32611}

\author{Kai Luo}
\email{kluo@carnegiescience.edu}
%\affiliation{Quantum Theory Project, Department of Physics, University of Florida, Gainesville, FL 32611}
\affiliation{Extreme Materials Initiative, Geophysical Laboratory, Carnegie Institution for Science, Washington, DC 20015-1305, USA}

\author{Jeffrey Wrighton}
%\email{jeffwrighton@gmail.com}
\affiliation{Department of Physics, University of Florida, Gainesville, FL 32611}

\date{\today}
\maketitle

%{\color{red} Note: there are three locations with red-colored text where references are made to equation numbers in the main manuscript (this document before Eqs. (\ref{5.8}) and (\ref{5.4b}), and after Eq. (\ref{7.9}), or search for '\textbackslash red' ). Before submission, verify those numbers in the main manuscript (search for the comment "\%above is referred to in supplementary material") and then delete all red text.} 

\section{Microscopic Conservation Laws}

The quantum mechanical microscopic conservation laws for the number,
momentum, and energy density operators are given in Refs. %
\onlinecite{zubarev1996statistical,mclennan1989introduction}. However, the
details of the derivation are not given and the results do not include an
external force. For completeness the general derivation is given here.

The time dependence of an operator $A(t)$ which depends on the position and
momenta of the system particles is given by

\begin{equation}
\frac{\partial}{\partial t}A(t)=i\left[ H_{N}(t),A(t)\right] ,
\end{equation}%
where the Hamiltonian of interest is of the form 
\begin{equation}
H_{N}\left( t\right) =\sum_{\alpha =1}^{N}\left( \frac{p_{\alpha }^{2}}{2m}%
+v^{\mathrm{ext}}(\vec{q}_{\alpha },t)\right) +\frac{1}{2}\sum_{\alpha \neq
\beta =1}^{N}U\left( |\vec{q}_{\alpha }-\vec{q}_{\beta }|\right).
\label{eq2}
\end{equation}%
Associated with the continuous symmetries are the usual conservation laws
for particle number, linear momentum, energy, and angular momentum. Here
only point particles are considered so the relevant conservation laws are
those of particle number, momentum, and energy. The associated local
densities are defined by 
\begin{align}
n(\vec{r})& =\sum_{\alpha =1}^{N}\delta (\vec{r}-\vec{q}_{\alpha }), \\
\vec{p}(\vec{r})& =\sum_{\alpha =1}^{N}\frac{1}{2}\left[ \vec{p}_{\alpha
},\delta (\vec{r}-\vec{q}_{\alpha })\right] _{+},  \label{pDensity} \\
e(\vec{r})& =\sum_{\alpha =1}^{N}\frac{1}{2}\left[ \frac{p_{\alpha }^{2}}{2m}%
,\delta (\vec{r}-\vec{q}_{\alpha })\right] _{+}+\frac{1}{2}\sum_{\alpha \neq
\beta =1}^{N}U(|\vec{q}_{\alpha }-\vec{q}_{\beta }|)\delta (\vec{r}-\vec{q}%
_{\alpha }),  \label{EDensity}
\end{align}%
where $\left[ A,B\right] _{+}=AB+BA$ is the anticommutator. The symmetrized
products ensure the Hermitian nature of these local density operators. In Eq. (%
\ref{eq2}) $v^{\mathrm{ext}}\left( \vec{q}_{\alpha },t\right) $ is an
external potential. As this external potential has not been included in the
definition of $e(\vec{r})$ in Eq.~(\ref{EDensity}), $e(\vec{r})$ is referred
to as the \textquotedblleft intrinsic" energy density.

The following identities will be used below:

\begin{equation}
\left[\vec p_\alpha,f(\vec q_\alpha)\right] = -i  \frac{\partial f (\vec q_\alpha)}{\partial\vec q_\alpha} ,
\end{equation}
\begin{equation}
\left[p_\alpha^2,A(\vec q_\alpha)\right]=-i\left[p_{\alpha j},\frac{\partial%
}{\partial q_{\alpha j}}A(\vec q_\alpha)\right]_+,  \label{eq7}
\end{equation}
\begin{equation}
\left[\left[A,B\right]_+,C\right] = \left[\left[ B,C\right],A\right]_+ +%
\left[\left[ A,C\right],B\right]_+ ,  \label{eq8}
\end{equation}
\begin{equation}
\left[ \vec p_\alpha, F(\vec q_\alpha) G(\vec q_\alpha) \right]_+ = \frac{1}{2} %
\left[\left[ \vec p_\alpha, F(\vec q_\alpha)\right]_+,G(\vec q_\alpha)\right]_+.
\label{eq9}
\end{equation}

\subsection{Number conservation}

Local conservation laws follow exactly from the Hamiltonian dynamics. The
simplest is the conservation of number density. In all derivations below the
Greek letters, $\alpha, \beta, \gamma, \cdots$ index the particles while
Latin indices $i, j, k, \cdots$ denote Cartesian coordinates. Unless noted,
Einstein summation is assumed for repeated indices.

Beginning with the time derivative of the number density, 
\begin{align}
i\frac{\partial }{\partial t}n(\vec{r})& =\left[ n(\vec{r}),H_{N}\left(
t\right) \right] =\left[ \sum_{\alpha =1}^{N}\delta (\vec{r}-\vec{q}_{\alpha
}),\sum_{\beta =1}^{N}\frac{p_{\beta }^{2}}{2m}\right]   \notag \\
& =\frac{1}{2m}\sum_{\alpha =1}^{N}\left[ \delta (\vec{r}-\vec{q}_{\alpha
}),p_{\alpha }^{2}\right] .
\end{align}%
Using Eq. (\ref{eq7}), 
\begin{align}
i\frac{\partial }{\partial t}n(\vec{r})& =\frac{i}{2m}\sum_{\alpha
=1}^{N}\left( p_{\alpha j}\frac{\partial }{\partial q_{\alpha j}}\delta (%
\vec{r}-\vec{q}_{\alpha })+\frac{\partial }{\partial q_{\alpha j}}\delta (%
\vec{r}-\vec{q}_{\alpha })p_{\alpha j}\right)   \notag \\
& =-\frac{i}{2m}\sum_{\alpha =1}^{N}\left( p_{\alpha j}\frac{\partial }{%
\partial r_{j}}\delta (\vec{r}-\vec{q}_{\alpha })+\frac{\partial }{\partial
r_{j}}\delta (\vec{r}-\vec{q}_{\alpha })p_{\alpha j}\right)   \notag \\
& =-\frac{1}{m}\frac{\partial }{\partial r_{j}}\sum_{\alpha =1}^{N}\frac{1}{m%
}\left[ p_{\alpha j},\delta (\vec{r}-\vec{q}_{\alpha })\right] _{+}.
\end{align}%
Inserting the definition of the momentum density in Eq. (\ref{pDensity})
gives the final result, 
\begin{equation}
  \frac{\partial}{\partial t}n(\vec r) + \frac{1}{m}\nabla\cdot\vec
p(\vec r)=0. 
\end{equation}

\subsection{Momentum conservation}

The time derivative of $\vec p(\vec r,t)$ is

\begin{equation}
i\frac{\partial }{\partial t}\vec{p}(\vec{r})=\left[ \vec{p}(\vec{r}%
),H_{N}\left( t\right) \right],
\end{equation}%
\begin{equation}
i\frac{\partial }{\partial t}p_{j}(\vec{r})=\frac{1}{2}\sum_{\alpha =1}^{N}%
\left[ \left[ p_{\alpha j},\delta (\vec{r}-\vec{q}_{\alpha })\right]
_{+},H_{N}\left( t\right) \right].
\end{equation}

Using $\left[\left[A,B\right]_+,C\right] = \left[\left[B,C\right],A\right]_+
+\left[\left[A,C\right],B\right]_+ $ from Eq. (\ref{eq8}), this is

\begin{equation}
i\frac{\partial }{\partial t}p_{j}(\vec{r})=\frac{1}{2}\sum_{\alpha
=1}^{N}\left( \left[ \left[ \delta (\vec{r}-\vec{q}_{\alpha }),H_{N}\left(
t\right) \right] ,p_{\alpha j}\right] _{+}+\left[ \left[ p_{\alpha
j},H_{N}\left( t\right) \right] ,\delta (\vec{r}-\vec{q}_{\alpha })\right]
_{+}\right).  \label{eq27}
\end{equation}

The two commutators appearing in the above equation will be treated
separately. The former is

\begin{align}
\left[ \delta (\vec{r}-\vec{q}_{\alpha }),H_{N}\right] & =\frac{1}{2m}\left[
\delta (\vec{r}-\vec{q}_{\alpha }),p_{\alpha }^{2}\right]   \notag \\
& =\frac{i}{2m}\left[ p_{\alpha k},\frac{\partial }{\partial q_{\alpha k}}%
\delta (\vec{r}-\vec{q}_{\alpha })\right] _{+}  \notag \\
& =\frac{1}{2m}\left(-i\frac{\partial}{\partial r_k}\right)\left[ \delta (\vec{r}-\vec{q}_{\alpha
}),p_{\alpha k}\right] _{+}.  \label{eq23}
\end{align}%
The other commutator is 
\begin{align}
\left[ p_{\alpha j},H_{N}\right] & =\left[ p_{\alpha j},\left( v^{\mathrm{ext%
}}(\vec{q}_{\alpha },t)+\frac{1}{2}\sum_{\beta \neq \alpha }\left( U(|\vec{q}%
_{\alpha }-\vec{q}_{\beta }|)+U(|\vec{q}_{\beta }-\vec{q}_{\alpha }|\right)
\right) \right]   \notag \\
& =-i\frac{\partial }{\partial q_{\alpha j}}v^{\mathrm{ext}}(\vec{q}_{\alpha
},t)-i\frac{\partial }{\partial q_{\alpha j}}\sum_{\beta \neq \alpha }U(|%
\vec{q}_{\alpha }-\vec{q}_{\beta }|)  \notag \\
& =iF_{\alpha j}^{\mathrm{ext}}(\vec{q}_{\alpha },t)+i\sum_{\beta \neq
\alpha }F_{\alpha \beta j}(|\vec{q}_{\alpha }-\vec{q}_{\beta }|),
\label{eq26}
\end{align}%
where 
\begin{equation}
F_{\alpha j}^{\mathrm{ext}}(\vec{q}_{\alpha },t)=-\frac{\partial }{\partial
q_{\alpha j}}v^{\mathrm{ext}}(\vec{q}_{\alpha },t)
\end{equation}%
and 
\begin{equation}
F_{\alpha \beta j}(|\vec{q}_{\alpha }-\vec{q}_{\beta }|)=-\frac{\partial }{%
\partial q_{\alpha j}}U(|\vec{q}_{\alpha }-\vec{q}_{\beta }|).
\end{equation}

Using Eqs. (\ref{eq23}) and (\ref{eq26}) in Eq. (\ref{eq27}) yields

\begin{equation}
\frac{\partial }{\partial t}p_{j}(\vec{r})=\frac{1}{2}\sum_{\alpha
=1}^{N}\left( \left[ -\frac{1}{2m}\frac{\partial}{\partial r_k}\left[ \delta (\vec{r}-\vec{%
q}_{\alpha }),p_{\alpha k}\right] _{+},p_{\alpha j}\right] _{+}+\left[
\left( \sum_{\beta \neq \alpha }F_{\alpha \beta j}(|\vec{q}_{\alpha }-\vec{q}%
_{\beta }|)+F_{\alpha j}^{\mathrm{ext}}(\vec{q}_{\alpha },t)\right) ,\delta (%
\vec{r}-\vec{q}_{\alpha })\right] _{+}\right) .
\end{equation}%
Define the kinetic part of the total momentum flux to be 
\begin{equation}
t_{jk}^{K}(\vec{r},t)=\frac{1}{4m}\sum_{\alpha =1}^{N}\left[ p_{\alpha j},%
\left[ p_{\alpha k},\delta (\vec{r}-\vec{q}_{\alpha })\right] _{+}\right]
_{+}
\end{equation}%
so 
\begin{align}
\frac{\partial }{\partial t}p_{j}(\vec{r})& =-\frac{\partial }{\partial r_{k}%
}t_{jk}^{K}(\vec{r},t)+\frac{1}{2}\sum_{\alpha =1}^{N}\left[ \left(
\sum_{\beta \neq \alpha }F_{\alpha \beta j}(|\vec{q}_{\alpha }-\vec{q}%
_{\beta }|)+F_{\alpha j}^{\mathrm{ext}}(\vec{q}_{\alpha })\right) ,\delta (%
\vec{r}-\vec{q}_{\alpha })\right] _{+}  \notag \\
& =-\frac{\partial }{\partial r_{k}}t_{jk}^{K}(\vec{r},t)+\sum_{\alpha \neq
\beta =1}^{N}F_{\alpha \beta j}(|\vec{q}_{\alpha }-\vec{q}_{\beta }|)\delta (%
\vec{r}-\vec{q}_{\alpha })+\sum_{\alpha =1}^{N}F_{\alpha j}^{\mathrm{ext}}(%
\vec{q}_{\alpha })\delta (\vec{r}-\vec{q}_{\alpha }).
\end{align}%
The force density arising from the external
potential is defined by  
\begin{equation}
f_{j}^{\mathrm{ext}}(\vec{r},t)=\sum_{\alpha =1}^{N}F_{\alpha j}^{\mathrm{ext%
}}(\vec{q}_{\alpha })\delta (\vec{r}-\vec{q}_{\alpha })
\end{equation}%
giving 
\begin{equation}
\frac{\partial }{\partial t}p_{j}(\vec{r})=-\frac{\partial}{\partial r_k}t_{jk}^{K}(\vec{r}%
,t)+\sum_{\alpha \neq \beta =1}^{N}F_{\alpha \beta j}(|\vec{q}_{\alpha }-%
\vec{q}_{\beta }|)\delta (\vec{r}-\vec{q}_{\alpha })+f_{j}^{\mathrm{ext}}(%
\vec{r},t).  \label{eq36}
\end{equation}

The second term on the right side requires some care. First, interchanging
the dummy labels $\alpha ,\beta $ and using Newton's third law gives

\begin{align}
\sum_{\alpha \neq \beta =1}^{N}F_{\alpha \beta j}(|\vec{q}_{\alpha }-\vec{q}%
_{\beta }|)\delta (\vec{r}-\vec{q}_{\alpha })& =\sum_{\alpha \neq \beta
=1}^{N}F_{\beta \alpha j}(|\vec{q}_{\alpha }-\vec{q}_{\beta }|)\delta (\vec{r%
}-\vec{q}_{\beta })  \notag \\
& =-\sum_{\alpha \neq \beta =1}^{N}F_{\alpha \beta j}(|\vec{q}_{\alpha }-%
\vec{q}_{\beta }|)\delta (\vec{r}-\vec{q}_{\beta }).
\end{align}%
Taking half of the sum of these two equivalent expressions gives 
\begin{equation}
\sum_{\alpha \neq \beta =1}^{N}F_{\alpha \beta j}(|\vec{q}_{\alpha }-\vec{q}%
_{\beta }|)\delta (\vec{r}-\vec{q}_{\alpha })=\frac{1}{2}\sum_{\alpha \neq
\beta =1}^{N}F_{\alpha \beta j}(|\vec{q}_{\alpha }-\vec{q}_{\beta }|)\left(
\delta (\vec{r}-\vec{q}_{\alpha })-\delta (\vec{r}-\vec{q}_{\beta })\right).
\end{equation}%
Consider a parameter $\lambda $ defining an arbitrary path from $\vec{x}%
(\lambda _{\beta })=\vec{q}_{\beta }$ to $\vec{x}(\lambda _{\alpha })=\vec{q}%
_{\alpha }$. The following identity then holds,

\begin{equation}
\delta (\vec{r}-\vec{q}_{\alpha })-\delta (\vec{r}-\vec{q}_{\beta
})=\int\limits_{\lambda _{\beta }}^{\lambda _{\alpha }}d\lambda \ \frac{d}{%
d\lambda }\delta (\vec{r}-\vec{x}(\lambda ))
\end{equation}%
so that 
\begin{align}
\delta (\vec{r}-\vec{q}_{\alpha })-\delta (\vec{r}-\vec{q}_{\beta })&
=\int\limits_{\lambda _{\beta }}^{\lambda _{\alpha }}d\lambda \ \frac{dx_{k}%
}{d\lambda }\frac{d}{dx_{k}}\delta (\vec{r}-\vec{x}(\lambda ))  \notag \\
& =-\frac{d}{dr_{k}}\int\limits_{\lambda _{\beta }}^{\lambda _{\alpha
}}d\lambda \ \frac{dx_{k}}{d\lambda }\delta (\vec{r}-\vec{x}(\lambda )) 
\notag \\
& \equiv -\frac{\partial}{\partial r_{k}}\mathcal{D}_{k}(\vec{r},\vec{q}_{\alpha },\vec{q}%
_{\beta }).  \label{DiracDiff}
\end{align}%
Thus the interaction force term becomes

\begin{eqnarray}
\sum_{\alpha \neq \beta =1}^{N}F_{\alpha \beta j}(|\vec{q}_{\alpha }-\vec{q}%
_{\beta }|)\delta (\vec{r}-\vec{q}_{\alpha }) &=&-\frac{\partial}{\partial r_{k}}\frac{1}{2}%
\sum_{\alpha \neq \beta =1}^{N}F_{\alpha \beta j}(|\vec{q}_{\alpha }-\vec{q}%
_{\beta }|)\mathcal{D}_{k}(\vec{r},\vec{q}_{\alpha },\vec{q}_{\beta }) 
\notag \\
&=&-\frac{\partial }{\partial r_{k}}t_{jk}^{P}(\vec{r},t)  \label{eq47}
\end{eqnarray}%
where the potential part of the momentum flux density is 
\begin{equation}
t_{jk}^{P}(\vec{r},t)\equiv \frac{1}{2}\sum_{\alpha \neq \beta
=1}^{N}F_{\alpha \beta j}(|\vec{q}_{\alpha }-\vec{q}_{\beta }|)\mathcal{D}%
_{k}(\vec{r},\vec{q}_{\alpha },\vec{q}_{\beta }).
\end{equation}

Using Eq. (\ref{eq47}) in Eq. (\ref{eq36}) gives 
\begin{equation}
\frac{\partial }{\partial t}p_{j}(\vec{r})=-\frac{\partial}{\partial r_k}\left( t_{jk}^{K}(%
\vec{r},t)+t_{jk}^{P}(\vec{r},t)\right) +f_{j}^{\mathrm{ext}}(\vec{r},t).
\end{equation}%
Defining the total momentum flux density as the sum of its kinetic and
potential parts, $t_{jk}=t_{jk}^{K}+t_{jk}^{P}$, gives a compact form for
the conservation of momentum equation, 
\begin{equation}
  \frac{\partial}{\partial t}p_j(\vec r) + \frac{\partial}{\partial
r_k} t_{jk} (\vec r,t) = f_j^{\rm ext}(\vec r,t).  
\end{equation}

\subsection{Energy Conservation}

The conservation of energy equation also follows from the time derivative of
the local energy density, using Eqs. (\ref{eq2}) and (\ref{EDensity}),

\begin{equation}
i\frac{\partial }{\partial t}e(\vec{r},t)=\left[ e(\vec{r}),H_{N}\left(
t\right) \right]
\end{equation}%
\begin{equation}
i\frac{\partial }{\partial t}e(\vec{r},t)=\left[ \sum_{\alpha =1}^{N}\frac{1%
}{2}\left[ \frac{p_{\alpha }^{2}}{2m},\delta (\vec{r}-\vec{q}_{\alpha })%
\right] _{+},H_{N}\left( t\right) \right] +\left[ \frac{1}{2}\sum_{\alpha
\neq \beta =1}^{N}U(|\vec{q}_{\alpha }-\vec{q}_{\beta }|)\delta (\vec{r}-%
\vec{q}_{\alpha }),H_{N}\left( t\right) \right].  \label{eq52}
\end{equation}%
The following two sections deal separately with each commutator appearing in
Eq. (\ref{eq52}).

\subsubsection{First term of Eq. (\protect\ref{eq52})}

The first commutator can be written as the sum of three terms,

\begin{align}
\left[ \sum_{\alpha =1}^{N}\frac{1}{2}\right. &\left. \left[ \frac{p_{\alpha
}^{2}}{2m},\delta (\vec{r}-\vec{q}_{\alpha })\right] _{+},H_{N}\left(
t\right) \vphantom{ \sum_a^b}\right]   \notag \\
& =\frac{1}{4m}\sum_{\alpha =1}^{N}\left[ \left[ \delta (\vec{r}-\vec{q}%
_{\alpha }),H_{N}\left( t\right) \right] ,p_{\alpha }^{2}\right] _{+}+\frac{1%
}{4m}\left[ \left[ p_{\alpha }^{2},H_{N}\left( t\right) \right] ,\delta (%
\vec{r}-\vec{q}_{\alpha })\right] _{+}  \notag \\
& =\frac{1}{8m^{2}}\sum_{\alpha =1}^{N}\left[ \left[ \delta (\vec{r}-\vec{q}%
_{\alpha }),p_{\alpha }^{2}\right] ,p_{\alpha }^{2}\right] _{+}  \notag \\
& \qquad -\frac{i}{4m}\sum_{\alpha =1}^{N}\left[ \left[ p_{\alpha j},\frac{%
\partial }{\partial q_{\alpha j}}\left( v^{\mathrm{ext}}(\vec{q}_{\alpha
},t)+\frac{1}{2}\sum_{\beta \neq \alpha }\left( U(|\vec{q}_{\alpha }-\vec{q}%
_{\beta }|)+U(|\vec{q}_{\beta }-\vec{q}_{\alpha }|)\right) \right) \right]
_{+},\delta (\vec{r}-\vec{q}_{\alpha })\right] _{+}  \notag \\
& =\frac{1}{8m^{2}}\sum_{\alpha =1}^{N}\left[ p_{\alpha }^{2},\left[ \delta (%
\vec{r}-\vec{q}_{\alpha }),p_{\alpha }^{2}\right] \right] _{+}  \notag \\
& \qquad -\frac{i}{4m}\sum_{\alpha =1}^{N}\left[ \left[ p_{\alpha j},\frac{%
\partial }{\partial q_{\alpha j}}\left( v^{\mathrm{ext}}(\vec{q}_{\alpha
},t)+\sum_{\beta \neq \alpha }U(|\vec{q}_{\alpha }-\vec{q}_{\beta }|)\right) %
\right] _{+},\delta (\vec{r}-\vec{q}_{\alpha })\right] _{+}  \notag \\
& =\frac{i}{8m^{2}}\sum_{\alpha =1}^{N}\left[ p_{\alpha }^{2},\left[
p_{\alpha j},\frac{\partial }{\partial q_{\alpha j}}\delta (\vec{r}-\vec{q}%
_{\alpha })\right] _{+}\right] _{+}  \notag \\
& \qquad -\frac{i}{4m}\sum_{\alpha =1}^{N}\left[ -\left[ p_{\alpha j},\left(
\sum_{\beta \neq \alpha }F_{\alpha \beta j}(|\vec{q}_{\alpha }-\vec{q}%
_{\beta }|)+F_{\alpha j}^{\mathrm{ext}}(\vec{q}_{\alpha },t)\right) \right]
_{+},\delta (\vec{r}-\vec{q}_{\alpha })\right] _{+}  \notag \\
& =-\frac{i}{8m^{2}}\frac{\partial }{\partial r_{j}}\sum_{\alpha =1}^{N}%
\left[ p_{\alpha }^{2},\left[ p_{\alpha j},\delta (\vec{r}-\vec{q}_{\alpha })%
\right] _{+}\right] _{+}  \notag \\
& \qquad +\frac{i}{4m}\sum_{\alpha \neq \beta =1}^{N}\left[ \left[ p_{\alpha
j},F_{\alpha \beta j}(|\vec{q}_{\alpha }-\vec{q}_{\beta }|)\right]
_{+},\delta (\vec{r}-\vec{q}_{\alpha })\right] _{+}+\frac{i}{4m}\sum_{\alpha
=1}^{N}\left[ \left[ p_{\alpha j},F_{\alpha j}^{\mathrm{ext}}(\vec{q}%
_{\alpha },t)\right] _{+},\delta (\vec{r}-\vec{q}_{\alpha })\right] _{+}
\label{eq54}
\end{align}

\subsubsection{Second term of Eq. (\protect\ref{eq52})}

Returning to the final commutator of Eq.(\ref{eq52}), use the relationship $%
[p_{\alpha },A(\vec{q}_{\alpha })\delta (\vec{r}-\vec{q}_{\alpha })]_{+}=%
\frac{1}{2}\left[ [p_{\alpha },A(\vec{q}_{\alpha })]_{+},\delta (\vec{r}-%
\vec{q}_{\alpha })\right] _{+}$ from Eq. (\ref{eq9}) in several places in
the following,

\begin{align}
\left[ \frac{1}{2}\sum_{\alpha \neq \beta =1}^{N}\right.&\left.U(|\vec{q}%
_{\alpha }-\vec{q}_{\beta }|)\delta (\vec{r}-\vec{q}_{\alpha }),H_{N}\left(
t\right) \vphantom{\sum_a^b} \right]   \notag \\
& =\left[ \frac{1}{2}\sum_{\alpha \neq \beta =1}^{N}U(|\vec{q}_{\alpha }-%
\vec{q}_{\beta }|)\delta (\vec{r}-\vec{q}_{\alpha }),\frac{p_{\alpha
}^{2}+p_{\beta }^{2}}{2m}\right]   \notag \\
& =i\frac{1}{4m}\sum_{\alpha \neq \beta =1}^{N}\left( \left[ p_{\alpha
j},\frac{\partial}{\partial q_{\alpha j}}\left(U(|\vec{q}_{\alpha }-\vec{q}_{\beta }|)\delta (\vec{r%
}-\vec{q}_{\alpha })\right)\right] _{+}+\left[ p_{\beta j},\frac{\partial}{\partial q_{\beta j}}\left(U(|%
\vec{q}_{\alpha }-\vec{q}_{\beta }|)\delta (\vec{r}-\vec{q}_{\alpha })\right)\right]
_{+}\right)   \notag \\
& =-i\frac{1}{4m}\sum_{\alpha \neq \beta =1}^{N}\left( \left[ p_{\alpha
j},F_{\alpha \beta j}(\left\vert \vec{q}_{\alpha }-\vec{q}_{\beta
}\right\vert )\delta (\vec{r}-\vec{q}_{\alpha })\right] _{+}+\left[ p_{\beta
j},F_{\beta \alpha j}(\left\vert \vec{q}_{\alpha }-\vec{q}_{\beta
}\right\vert )\delta (\vec{r}-\vec{q}_{\alpha })\right] _{+}\right.   \notag
\\
& \qquad \qquad \qquad \qquad -\left. \left[ p_{\alpha j},U(|\vec{q}_{\alpha
}-\vec{q}_{\beta }|)\frac{\partial}{\partial q_{\alpha j}}\delta (\vec{r}-\vec{q}_{\alpha })%
\right] _{+}\right)   \label{eq58} \\
& =-i\frac{1}{4m}\sum_{\alpha \neq \beta =1}^{N}\left( \left[ \left(
p_{\alpha j}-p_{\beta j}\right) ,F_{\alpha \beta j}(\left\vert \vec{q}%
_{\alpha }-\vec{q}_{\beta }\right\vert )\delta (\vec{r}-\vec{q}_{\alpha })%
\right] _{+}-\frac{\partial}{\partial r_{j}}\left[ p_{\alpha j},U(|\vec{q}_{\alpha }-\vec{q}%
_{\beta }|)\delta (\vec{r}-\vec{q}_{\alpha })\right] _{+}\right) 
\label{eq59} \\
& =-\frac{i}{8m}\sum_{\alpha \neq \beta =1}^{N}\left[ \left[ p_{\alpha
j}-p_{\beta j},F_{\alpha \beta j}(\left\vert \vec{q}_{\alpha }-\vec{q}%
_{\beta }\right\vert )\right] _{+}\delta (\vec{r}-\vec{q}_{\alpha })\right]
_{+}  \notag \\
& \qquad \qquad \qquad +\frac{\partial}{\partial r_{j}}\frac{i}{4m}\sum_{\alpha \neq \beta
=1}^{N}\left[ p_{\alpha j},U(|\vec{q}_{\alpha }-\vec{q}_{\beta }|)\delta (%
\vec{r}-\vec{q}_{\alpha })\right] _{+}.  \label{eq70}
\end{align}%
In going from Eq. (\ref{eq58}) to (\ref{eq59}), Newton's third law has been
used, 
\begin{equation}
F_{\beta \alpha j}\left(\left|\vec{q}_{\alpha }-\vec{q}_{\beta }\right|\right)=-F_{\alpha \beta j}\left(\left|\vec{q}_{\alpha }-\vec{q}_{\beta }\right|\right).
\end{equation}

\subsubsection{Combining results}

Using Eqs. (\ref{eq54}) and (\ref{eq70}) in Eq. (\ref{eq52}),

\begin{align}
i\frac{\partial }{\partial t}e(\vec{r},t)& =-\frac{i}{8m^{2}}\frac{\partial 
}{\partial r_{j}}\sum_{\alpha =1}^{N}\left[ p_{\alpha }^{2},\left[ p_{\alpha
j},\delta (\vec{r}-\vec{q}_{\alpha })\right] _{+}\right] _{+}  \notag \\
& \qquad +\frac{i}{4m}\sum_{\alpha =1}^{N}\left[ \left[ p_{\alpha
j},F_{\alpha j}^{\mathrm{ext}}(\vec{q}_{\alpha },t)\right] _{+},\delta (\vec{%
r}-\vec{q}_{\alpha })\right] _{+}  \notag \\
& \qquad   +\frac{i}{4m}\sum_{\alpha \neq \beta =1}^{N}\left[ \left[
p_{\alpha j},F_{\alpha \beta j}(|\vec{q}_{\alpha }-\vec{q}_{\beta }|)\right]
_{+},\delta (\vec{r}-\vec{q}_{\alpha })\right] _{+}  \notag \\
& \qquad   -\frac{i}{8m}\sum_{\alpha \neq \beta =1}^{N}\left[ %
\left[ p_{\alpha j}-p_{\beta j},F_{\alpha \beta j}(\left\vert \vec{q}%
_{\alpha }-\vec{q}_{\beta }\right\vert )\right] _{+},\delta (\vec{r}-\vec{q}%
_{\alpha })\right] _{+}  \notag \\
& \qquad   +\frac{\partial}{\partial r_{j}}\frac{i}{4m}\sum_{\alpha \neq
\beta =1}^{N}\left[ p_{\alpha j},U(|\vec{q}_{\alpha }-\vec{q}_{\beta
}|)\delta (\vec{r}-\vec{q}_{\alpha })\right] _{+}.
\end{align}%
The terms in the third and fourth lines can be combined, 
\begin{align}
i\frac{\partial }{\partial t}e(\vec{r},t)& =-\frac{i}{8m^{2}}\frac{\partial 
}{\partial r_{j}}\sum_{\alpha =1}^{N}\left[ p_{\alpha }^{2},\left[ p_{\alpha
j},\delta (\vec{r}-\vec{q}_{\alpha })\right] _{+}\right] _{+}  \notag \\
& \qquad +\frac{i}{4m}\sum_{\alpha =1}^{N}\left[ \left[ p_{\alpha
j},F_{\alpha j}^{\mathrm{ext}}(\vec{q}_{\alpha },t)\right] _{+},\delta (\vec{%
r}-\vec{q}_{\alpha })\right] _{+}  \notag \\
& \qquad   +\frac{i}{8m}\sum_{\alpha \neq \beta =1}^{N}\left[ %
\left[ \left( p_{\alpha j}+p_{\beta j}\right) ,F_{\alpha \beta j}(\left\vert 
\vec{q}_{\alpha }-\vec{q}_{\beta }\right\vert )\right] _{+},\delta (\vec{r}-%
\vec{q}_{\alpha })\right] _{+}  \notag \\
& \qquad  +\frac{\partial}{\partial r_{j}}\frac{i}{4m}\sum_{\alpha \neq
\beta =1}^{N}\left[ p_{\alpha j},U(|\vec{q}_{\alpha }-\vec{q}_{\beta
}|)\delta (\vec{r}-\vec{q}_{\alpha })\right] _{+}.
\end{align}%
The third term can further be rewritten as

\begin{align}
\frac{i}{8m}\sum_{\alpha \neq \beta =1}^{N}&\left[ \left[
\left( p_{\alpha j}+p_{\beta j}\right) ,F_{\alpha \beta j}(\left\vert \vec{q}%
_{\alpha }-\vec{q}_{\beta }\right\vert )\right] _{+},\delta (\vec{r}-\vec{q}%
_{\alpha })\right] _{+}  \notag \\
& =\frac{i}{8m}\sum_{\alpha \neq \beta =1}^{N}\left( \left[ \left[ p_{\alpha
j},F_{\alpha \beta j}(|\vec{q}_{\alpha }-\vec{q}_{\beta }|)\right]
_{+},\delta (\vec{r}-\vec{q}_{\alpha })\right] _{+}+\left[ \left[ p_{\alpha
j},F_{\beta \alpha j}(|\vec{q}_{\alpha }-\vec{q}_{\beta }|)\right]
_{+},\delta (\vec{r}-\vec{q}_{\beta })\right] _{+}\right)   \notag \\
& =\frac{i}{8m}\sum_{\alpha \neq \beta =1}^{N}\left( \left[ \left[ p_{\alpha
j},F_{\alpha \beta j}(|\vec{q}_{\alpha }-\vec{q}_{\beta }|)\right]
_{+},\delta (\vec{r}-\vec{q}_{\alpha })\right] _{+}-\left[ \left[ p_{\alpha
j},F_{\alpha \beta j}(|\vec{q}_{\alpha }-\vec{q}_{\beta }|)\right]
_{+},\delta (\vec{r}-\vec{q}_{\beta })\right] _{+}\right)   \notag \\
& =\frac{i}{8m}\sum_{\alpha \neq \beta =1}^{N}\left[ \left[ p_{\alpha
j},F_{\alpha \beta j}(|\vec{q}_{\alpha }-\vec{q}_{\beta }|)\right]
_{+},\left( \delta (\vec{r}-\vec{q}_{\alpha })-\delta (\vec{r}-\vec{q}%
_{\beta })\right) \right] _{+}  \notag \\
& =\frac{i}{8m}\sum_{\alpha \neq \beta =1}^{N}\left[ \left[ p_{\alpha
j},F_{\alpha \beta j}(|\vec{q}_{\alpha }-\vec{q}_{\beta }|)\right]
_{+},\left( -\frac{\partial }{\partial r_{k}}\mathcal{D}_{k}(\vec{r},\vec{q}%
_{\alpha },\vec{q}_{\beta })\right) \right] _{+}  \notag \\
& =-\frac{i}{8m}\frac{\partial }{\partial r_{j}}\sum_{\alpha \neq \beta
=1}^{N}\left[ \left[ p_{\alpha k},F_{\alpha \beta k}(|\vec{q}_{\alpha }-\vec{%
q}_{\beta }|)\right] _{+},\mathcal{D}_{j}(\vec{r},\vec{q}_{\alpha },\vec{q}%
_{\beta })\right] _{+},
\end{align}%
where use has been made of Eq. (\ref{DiracDiff}) in the next to last line and
 dummy indices $r,k$  have been swapped in the last line. The energy equation is
therefore

\begin{align}
\frac{\partial}{\partial t}e(\vec r,t) &=- \frac{1}{8m^2}\frac{\partial}{%
\partial r_j}\sum_{\alpha=1}^N\left[p_\alpha^2,\left[p_{\alpha
j},\delta(\vec r-\vec q_\alpha)\right]_+\right]_+  \notag \\
&\qquad +\frac{1}{4m}\sum_{\alpha=1}^N\left[\left[p_{\alpha j},F_{\alpha j}^{%
\mathrm{ext}}(\vec q_\alpha)\right]_+,\delta(\vec r-\vec q_\alpha)\right]_+ 
\notag \\
&\qquad  -\frac{1}{8m}\frac{\partial}{\partial r_j}\sum_{\alpha\neq%
\beta=1}^N \left[ \left[ p_{\alpha k}, F_{\alpha\beta k}(|\vec q_\alpha-\vec
q_\beta|)\right]_+, \mathcal{D}_j(\vec r,\vec q_\alpha,\vec q_\beta) \right]%
_+  \notag \\
&\qquad  -\frac{\partial}{\partial r_j} \frac{1}{4m}\sum_{\alpha\neq\beta=1}^N%
\left[p_{\alpha j}, U(|\vec q_\alpha-\vec q_\beta|)\delta(\vec r-\vec
q_\alpha)\right]_+.
\end{align}
Moving the three derivative terms to the left side, this can be rewritten as

\begin{equation}
  \frac{\partial}{\partial t} e(\vec r,t) +\nabla\cdot s(\vec r,t) =
w(\vec r,t) 
\end{equation}%
with the energy flux 
\begin{align}
 s_j(\vec r,t) =  &
\frac{1}{8m^2} \sum_{\alpha=1}^N\left[p_\alpha^2,\left[p_{\alpha j},\delta(\vec r-\vec q_\alpha)\right]_+ \right]_+\nonumber\\
&  + \frac{1}{8m} \sum_{\alpha\neq\beta=1}^N  
	\left[
		\left[
			p_{\alpha k},
		F_{\alpha\beta k}(|\vec q_\alpha-\vec q_\beta|)\right]_+,  {\cal D}_j(\vec r,\vec q_\alpha,\vec q_\beta) 
		\right]_+   \nonumber\\
&   	 	  + \frac{1}{4m}\sum_{\alpha\neq\beta=1}^N\left[p_{\alpha j}, U(|\vec q_\alpha-\vec q_\beta|)\delta(\vec r-\vec q_\alpha)\right]_+
\end{align}
 and the work done by the external potential 
\begin{equation}
  w(\vec r,t) = \frac{1}{4m}\sum_{\alpha=1}^N\left[\left[p_{\alpha
j},F_{\alpha j}^{\rm ext}(\vec q_\alpha)\right]_+,\delta(\vec r-\vec
q_\alpha)\right]_+ . 
\end{equation}

This work term can be simplified as follows, 
\begin{align*}
w\left( \mathbf{r},t\right) & =\frac{1}{4m}\sum_{\alpha }\left( \mathbf{p}%
_{\alpha }\cdot \mathbf{F}^{\rm ext}\left( \mathbf{q}_{\alpha },t\right) +\mathbf{%
F}^{\rm ext}\left( \mathbf{q}_{\alpha },t\right) \cdot \mathbf{p}_{\alpha
}\right) \delta \left( \mathbf{r}-\mathbf{q}_{\alpha }\right) \\
& \qquad +\frac{1}{4m}\sum_{\alpha }\delta \left( \mathbf{r}-\mathbf{q}%
_{\alpha }\right) \left( \mathbf{p}_{\alpha }\cdot \mathbf{F}^{\rm ext}\left( 
\mathbf{q}_{\alpha },t\right) +\mathbf{F}^{\rm ext}\left( \mathbf{q}_{\alpha
},t\right) \cdot \mathbf{p}_{\alpha }\right) \\
& =\frac{1}{4m}\sum_{\alpha }\left( 2\mathbf{p}_{\alpha },\cdot \mathbf{F}%
^{\rm ext}\left( \mathbf{r},t\right) \delta \left( \mathbf{r}-\mathbf{q}_{\alpha
}\right) +\left[ \mathbf{F}^{\rm ext}\left( \mathbf{q}_{\alpha },t\right) ,\cdot 
\mathbf{p}_{\alpha }\right] \delta \left( \mathbf{r}-\mathbf{q}_{\alpha
}\right) \right) \\
& \qquad +\frac{1}{4m}\sum_{\alpha }\left( \delta \left( \mathbf{r}-\mathbf{q%
}_{\alpha }\right) \left[ \mathbf{p}_{\alpha },\cdot \mathbf{F}^{\rm ext}\left( 
\mathbf{q}_{\alpha },t\right) \right] +2\delta \left( \mathbf{r}-\mathbf{q}%
_{\alpha }\right) \mathbf{F}^{\rm ext}\left( \mathbf{r},t\right) \cdot \mathbf{p}%
_{\alpha }\right) \\
& =\frac{1}{m}\mathbf{p}\left( \mathbf{r}\right) \cdot \mathbf{F}^{\rm ext}\left( 
\mathbf{r},t\right) +\frac{1}{4m}\sum_{\alpha }\left[ \mathbf{F}^{\rm ext}\left( 
\mathbf{q}_{\alpha },t\right) ,\cdot \mathbf{p}_{\alpha }\right] \delta
\left( \mathbf{r}-\mathbf{q}_{\alpha }\right) \\
& \qquad +\frac{1}{4m}\sum_{\alpha }\delta \left( \mathbf{r}-\mathbf{q}%
_{\alpha }\right) \left[ \mathbf{p}_{\alpha },\cdot \mathbf{F}^{\rm ext}\left( 
\mathbf{q}_{\alpha },t\right) \right] \\
& =\frac{1}{m}\mathbf{p}\left( \mathbf{r}\right) \cdot \mathbf{F}^{\rm ext}\left( 
\mathbf{r},t\right) +\frac{i}{4m}\sum_{\alpha }\nabla _{\mathbf{q}_{\alpha }}%
\mathbf{F}^{\rm ext}\left( \mathbf{q}_{\alpha },t\right) \delta \left( \mathbf{r}-%
\mathbf{q}_{\alpha }\right) \\
& \qquad -\frac{i}{4m}\sum_{\alpha }\delta \left( \mathbf{r}-\mathbf{q}%
_{\alpha }\right) \nabla _{\mathbf{q}_{\alpha }}\mathbf{F}^{\rm ext}\left( 
\mathbf{q}_{\alpha },t\right).
\end{align*}%
This yields the final form for the work term as 
\begin{equation}
  w(\vec r,t) =\frac{1}{m}\mathbf{p}\left( \mathbf{r}\right)
\cdot\mathbf{F}^{\rm ext}\left( \mathbf{r},t\right). \label{eq24} 
\end{equation}

\section{Unitary transformation to rest frame momenta}

\label{UniTrans}

Operators and phase functions of interest include those in the local
reference frame. For simplicity consideration here is restricted to
single particle functions, 
\begin{equation}
A=A(\mathbf{q,p}-m\mathbf{u}\left( \mathbf{q}\right) ),  \label{1}
\end{equation}%
where $\mathbf{u}\left( \mathbf{q}\right) $ is the average flow velocity at
the particle position $\mathbf{q}$. In the classical case a change of
variables in the momentum average to 
\begin{equation}
\mathbf{p}^{\prime }=\mathbf{p}-m\mathbf{u}\left( \mathbf{q}\right) ,\hspace{%
0.25in}\mathbf{q}^{\prime }=\mathbf{q}  \label{2}
\end{equation}%
gives the simpler form 
\begin{equation}
A^{\prime }=A(\mathbf{q}^{\prime }\mathbf{,p}^{\prime }),  \label{3}
\end{equation}%
which is independent of the velocity field. The objective here is to find a
unitary operator that does the same for quantum mechanical operators, 
\begin{equation}
A^{\prime }=A(\mathbf{q}^{\prime }\mathbf{,p}^{\prime })=UA(\mathbf{q,p}-m%
\mathbf{u}\left( \mathbf{q}\right) )U^{-1}.  \label{4}
\end{equation}

The generator of the classical canonical transformation in Eq. (\ref{2})
obeys \cite{goldstein:mechanics} 
\begin{equation}
p_{i}=\frac{\partial F(\mathbf{q,p}^{\prime })}{\partial q_{i}},\hspace{%
0.25in}q_{i}^{\prime }=\frac{\partial F(\mathbf{q,p}^{\prime })}{\partial
p_{i}^{\prime }}.  \label{5}
\end{equation}%
Using the identity generator $\mathbf{q\cdot p}^{\prime }$ a new generator
for the deviation is introduced 
\begin{equation}
F(\mathbf{q,p}^{\prime })=\mathbf{q\cdot p}^{\prime }+G(\mathbf{q,p}^{\prime
}).  \label{6}
\end{equation}%
In the following $G(\mathbf{q,p}^{\prime })$ will be referred to as the
generator rather than $F(\mathbf{q,p}^{\prime })$. Note that for this
generator 
\begin{equation}
p_{i}=p_{i}^{\prime }+\frac{\partial G(\mathbf{q,p}^{\prime })}{\partial
q_{i}},\hspace{0.25in}q_{i}^{\prime }=q_{i}\mathbf{+}\frac{\partial G(%
\mathbf{q,p}^{\prime })}{\partial p_{i}^{\prime }}  \label{7}
\end{equation}%
\begin{equation}
\frac{\partial G(\mathbf{q,p}^{\prime })}{\partial q_{i}}=mu_{i}\left( 
\mathbf{q}\right) ,\hspace{0.25in}\frac{\partial G(\mathbf{q,p}^{\prime })}{%
\partial p_{i}^{\prime }}=0  \label{8}
\end{equation}%
so $G(\mathbf{q,p}^{\prime })$ is independent of $\mathbf{p}^{\prime }$ 
\begin{equation}
G(\mathbf{q,p}^{\prime })=G(\mathbf{q}),\hspace{0.25in}\frac{\partial G(%
\mathbf{q})}{\partial q_{i}}=mw_{i}\left( \mathbf{q}\right) .  \label{9}
\end{equation}%
Now consider the associated infinitesimal transformation, $mu_{i}\left( 
\mathbf{q}\right) \rightarrow \epsilon mu_{i}\left( \mathbf{q}\right) $, $G(%
\mathbf{q,p}^{\prime })\rightarrow \epsilon G(\mathbf{q,p}^{\prime })$. To
first order in $\epsilon $, 
\begin{align}
A(\mathbf{q}^{\prime }\mathbf{,p}^{\prime })-A(\mathbf{q,p})& \rightarrow 
\frac{\partial A(\mathbf{q,p})}{\partial q_{i}}\left( q_{i}^{\prime
}-q_{i}\right) +\frac{\partial A(\mathbf{q,p})}{\partial p_{i}}\left(
p_{i}^{\prime }-p_{i}\right)  \notag \\
& =\frac{\partial A(\mathbf{q,p})}{\partial q_{i}}\epsilon \frac{\partial G(%
\mathbf{q})}{\partial p_{i}}-\frac{\partial A(\mathbf{q,p})}{\partial p_{i}}%
\epsilon \frac{\partial G(\mathbf{q})}{\partial q_{i}}  \notag \\
& =\epsilon \left\{ A,G\right\}  \label{10}
\end{align}%
where $\left\{ ,\right\} $ denotes the Poisson bracket.

The Dirac - Weyl quantization of this procedure is given by 
\begin{equation}
\left\{ A,G\right\} \rightarrow \frac{1}{i\hbar }\left[ A,G\right]
\label{11}
\end{equation}%
although some limitations exist. According to this quantization procedure,
the classical infinitesimal transformation corresponds to the finite
transformation 
\begin{equation}
A(\mathbf{q}^{\prime }\mathbf{,p}^{\prime })=UA(\mathbf{q,p})U^{-1},\hspace{%
0.25in}U=e^{-\frac{1}{i\hbar }G}.  \label{12}
\end{equation}%
To demonstrate that this transformation does in fact give the desired
result, transform a quantity $A(\vec{q}^{\prime },\vec{r}^{\prime })$, 
\begin{equation}
A(\mathbf{q}^{\prime }\mathbf{,p}^{\prime })=e^{-\frac{1}{i\hbar }G}A(%
\mathbf{q,p})e^{\frac{1}{i\hbar }G}=A(e^{-\frac{1}{i\hbar }G}\mathbf{q}e^{%
\frac{1}{i\hbar }G}\mathbf{,}e^{-\frac{1}{i\hbar }G}\mathbf{p}e^{\frac{1}{%
i\hbar }G}).  \label{13}
\end{equation}%
Since $G=G\left( \mathbf{q}\right) $ it commutes with $\mathbf{q}$ and so 
\begin{equation}
e^{-\frac{1}{i\hbar }G}\mathbf{q}e^{\frac{1}{i\hbar }G}\mathbf{=q.}
\label{14}
\end{equation}%
Next define 
\begin{equation}
\mathbf{p}\left( \lambda \right) \equiv e^{-\frac{1}{i\hbar }\lambda G}%
\mathbf{p}e^{\frac{1}{i\hbar }\lambda G}  \label{15}
\end{equation}%
which obeys 
\begin{align}
\partial _{\lambda }\mathbf{p}\left( \lambda \right) & =-e^{-\frac{1}{i\hbar 
}\lambda G}\frac{1}{i\hbar }\left[ G\left( \mathbf{q}\right) ,\mathbf{p}%
\right] e^{\frac{1}{i\hbar }\lambda G}=-e^{-\frac{1}{i\hbar }\lambda
G}\nabla _{\mathbf{q}}G\left( \mathbf{q}\right) e^{\frac{1}{i\hbar }\lambda
G}  \notag \\
& =-m\mathbf{u}\left( \mathbf{q}\right).  \label{16}
\end{align}%
The following identity was used above: 
\begin{equation}
\left[ f\left( \mathbf{q}\right) ,\mathbf{p}\right] =i\hbar \nabla _{\mathbf{%
q}}f\left( \mathbf{q}\right) .  \label{17}
\end{equation}%
Integrating (\ref{16}) from $0$ to $1$ gives 
\begin{equation}
\mathbf{p}\left( 1\right) =\mathbf{p}\left( 0\right) -m\mathbf{u}\left( 
\mathbf{q}\right)  \label{18}
\end{equation}%
or 
\begin{equation}
e^{-\frac{1}{i\hbar }G}\mathbf{p}e^{\frac{1}{i\hbar }G}=\mathbf{p}-m\mathbf{u%
}\left( \mathbf{q}\right) .  \label{19}
\end{equation}%
Therefore (\ref{13}) gives the desired result 
\begin{equation}
A(\mathbf{q}^{\prime }\mathbf{,p}^{\prime })=e^{-\frac{1}{i\hbar }G}A(%
\mathbf{q,p})e^{\frac{1}{i\hbar }G}=A(\mathbf{q,p}-m\mathbf{u}\left( \mathbf{%
q}\right) ).  \label{20}
\end{equation}

In the text, most final results are expressed as local equilibrium
correlation functions in the rest frame, e.g.

\begin{equation}
C_{0}\left( \mathbf{r,r}^{\prime }\mid \mathbf{u}\right) \equiv \overline{%
A_{0}\left( \mathbf{r}\right) B_{0}\left( \mathbf{r}^{\prime }\right) }%
^{\ell }.  \label{20a}
\end{equation}%
Recall that the subscript zero denotes the chosen operator with all momenta
replaced by their local rest frame momenta $\mathbf{p}_{\alpha }\rightarrow 
\mathbf{p}_{\alpha }-m\mathbf{u}\left( \mathbf{q}_{\alpha }\right) $. The
local equilibrium ensemble also has this form. Then by cyclic invariance of
the trace%
\begin{eqnarray}
C_{0}\left( \mathbf{r,r}^{\prime }\mid \mathbf{u}\right) &=&\overline{e^{-%
\frac{1}{i\hbar }G}A_{0}\left( \mathbf{r}\right) e^{\frac{1}{i\hbar }G}e^{-%
\frac{1}{i\hbar }G}B_{0}\left( \mathbf{r}^{\prime }\right) e^{-\frac{1}{%
i\hbar }G}}^{\ell }  \notag \\
&=&C_{0}\left( \mathbf{r,r}^{\prime }\mid \mathbf{u=0}\right)  \label{20b}
\end{eqnarray}%
and such correlation functions are independent of the local flow velocity.

\section{Time reversal symmetry}

The objective of this section is to prove a property of local equilibrium
correlation functions for operators that have a definite sign with respect
to changes in the sign of the particle momentum operator. In the classical
case this is accomplished by introduction of the parity operator for the
momentum variables only. In the quantum case the parity operator changes both the
momentum and position operators and the analysis no longer holds. Instead,
an operator that changes the momentum sign but not the position operator
sign is considered.

As in the classical case the quantum time-reversal operator $T$ generates a
reversal of motion operator in the sense

\begin{equation}
\mathcal{T}\vec{p}\mathcal{T}^{-1}=-\vec{p}\qquad \qquad \mathcal{T}\vec{q}\mathcal{T}^{-1}=\vec{q}.  \label{24}
\end{equation}%
It follows (e.g., coordinate representation for $\vec{p}$) that $\mathcal{T}i=-i\mathcal{T}$,
and hence $\mathcal{T}$ is an anti-unitary operator%
\begin{equation}
\mathcal{T}\left( c_{1}|\alpha \rangle +c_{2}|\beta \rangle \right)
=c_{1}^{\ast }\mathcal{T}|\alpha \rangle +c_{2}^{\ast }\mathcal{T}|\beta
\rangle ,  \label{25}
\end{equation}%
where $c_{1}$ and $c_{2}$ are arbitrary complex numbers. In addition it is
required that $\mathcal{T}$ be norm-preserving%
\begin{equation}
\left\vert \left\langle \widetilde{\beta }|\widetilde{\alpha }\right\rangle
\right\vert =\left\vert \left\langle \beta |\alpha \right\rangle \right\vert
\label{26}
\end{equation}%
where\ $|\alpha \rangle $ and $|\beta \rangle $ are members of a complete
basis set $\left\{ \alpha \right\} $, and 
\begin{equation}
|\tilde{\alpha}\rangle =\mathcal{T}|\alpha \rangle ,\hspace{0.24in}|\tilde{%
\beta}\rangle =\mathcal{T}|\beta \rangle .  \label{26a}
\end{equation}%
It then follows that $\mathcal{T}$ is anti-unitary\cite{messiah1999quantum} 
\begin{equation}
\left\langle \widetilde{\beta }|\widetilde{\alpha }\right\rangle
=\left\langle \beta |\alpha \right\rangle ^{\ast } .  \label{26b}
\end{equation}

Consider again the correlation function of (\ref{20a}) 
\begin{equation}
C_{0}\left( \mathbf{r,r}^{\prime }\mid \mathbf{u=0}\right) \equiv \overline{%
A_{0}\left( \mathbf{r}\right) B_{0}\left( \mathbf{r}^{\prime }\right) }%
^{\ell }.  \label{25a}
\end{equation}%
It is shown here that this has the property%
\begin{equation}
C_{0}\left( \mathbf{r,r}^{\prime }\mid \mathbf{u=0}\right) \equiv \overline{%
\left( \mathcal{T}A_{0}\left( \mathbf{r}\right) \mathcal{T}^{-1}\right) \left( \mathcal{T}B_{0}\left( 
\mathbf{r}^{\prime }\right)\mathcal{T}^{-1}\right) }^{\ell }.  \label{25b}
\end{equation}%
Thus if both $A_{0}\left( \mathbf{r}\right) $ and $B_{0}\left( \mathbf{r}%
^{\prime }\right) $ have a definite sign under time reversal, the
correlation function vanishes when the signs are different. The proof of (%
\ref{25b}) is complicated by the anti-unitary nature of $\mathcal{T}$ which does not
obey the cyclic invariance of the trace.

The proof is as follows. Define 
\begin{equation}
|\gamma \rangle =A^{\dag }|\beta \rangle 
\xrightarrow{\mbox{dual
correspondence}}\langle \gamma |=\langle \beta |A
\end{equation}%
where $A$ is an arbitrary linear operator. Then%
\begin{equation}
\left\langle \gamma |\alpha \right\rangle =\left\langle \alpha |\gamma
\right\rangle ^{\ast }=\left\langle \widetilde{\alpha }|\widetilde{\gamma }%
\right\rangle
\end{equation}%
where the second equality follows from the anti-unitarity of $\mathcal{T}$, (%
\ref{26b}). Writing $\left\langle \gamma |\alpha \right\rangle $ and $%
\left\langle \widetilde{\alpha }|\widetilde{\gamma }\right\rangle $
explictly in terms of $A$%
\begin{equation}
\left\langle \beta \right\vert A\left\vert \alpha \right\rangle
=\left\langle \widetilde{\alpha }\right\vert \mathcal{T}A^{\dag }\left\vert
\beta \right\rangle =\left\langle \widetilde{\alpha }\right\vert \mathcal{T}%
A^{\dag }\mathcal{T}^{-1}\left\vert \widetilde{\beta }\right\rangle .
\end{equation}%
As a special case, the diagonal elements for a self-adjoint operator are%
\begin{equation}
\left\langle \alpha \right\vert A\left\vert \alpha \right\rangle
=\left\langle \widetilde{\alpha }\right\vert \mathcal{T}A\mathcal{T}%
^{-1}\left\vert \widetilde{\alpha }\right\rangle .  \label{27}
\end{equation}%
Now the correlation function in (\ref{25a}) can be expressed as 
\begin{eqnarray}
C_{0}\left( \mathbf{r,r}^{\prime }\mid \mathbf{u=0}\right) &\equiv
&\sum_{N}{\rm Tr}A_{0}\left( \mathbf{r}\right) B_{0}\left( \mathbf{r}^{\prime
}\right) \rho ^{\ell } \\
&=&\sum_{N}\sum_{\alpha }\left\langle \alpha \right\vert A_{0}\left( \mathbf{%
r}\right) B_{0}\left( \mathbf{r}^{\prime }\right) \rho ^{\ell }\left\vert
\alpha \right\rangle \\
&=&\sum_{N}\sum_{\alpha }\left\langle \widetilde{\alpha }\right\vert 
\mathcal{T}A_{0}\left( \mathbf{r}\right) B_{0}\left( \mathbf{r}^{\prime
}\right) \rho ^{\ell }\mathcal{T}^{-1}\left\vert \widetilde{\alpha }%
\right\rangle.  \label{28}
\end{eqnarray}%
The last line follows from (\ref{27}). Since $\left\vert \alpha
\right\rangle $ and $\left\vert \widetilde{\alpha }\right\rangle $ are in
one-to-one correspondence the latter is a complete basis set. Furthermore%
\begin{equation*}
\mathcal{T}\rho ^{\ell }\mathcal{T}^{-1}=\rho ^{\ell }
\end{equation*}%
and (\ref{28}) becomes (\ref{25b}).

\section{Transformation to local rest frame}

Operator functions of the particle positions and momenta typically refer to
a fixed laboratory frame. They define both convective motion and motion
relative to each particle's local average velocity field. Denote such an
operator by $A(\left\{ \mathbf{q,p}\right\} )$, where $\left\{ \mathbf{q,p}%
\right\} $ denotes the $N$ particle positions and momenta. It is convenient
to extract from this the corresponding operator representing only motion
relative to the average flow $A_{0}(\left\{ \mathbf{q,p}\right\} )\equiv
A(\left\{ \mathbf{q,p}-m\mathbf{u}\left( \mathbf{q}\right) \right\} )$. They
are related by the generator of eq.(\ref{20}), 
\begin{equation}
A_{0}(\left\{ \mathbf{q,p}\right\} )=e^{-\frac{1}{i\hbar }G}A(\left\{ 
\mathbf{q,p}\right\} )e^{\frac{1}{i\hbar }G}.
\end{equation}%
Operators of primary interest are the conserved densities and associated
fluxes, 
\begin{equation}
\{\psi (\vec{r})\}=\{n(\vec{r}),e(\vec{r}),\vec{p}(\vec{r})\} ,
\label{eq117}
\end{equation}%
\begin{equation}
\{\gamma _{i}\}=\left\{ p_{i},s_{i},t_{ij}\right\}.
\end{equation}%
The corresponding rest frame operators are easily found to be 
\begin{equation}
\{\psi _{0}(\vec{r})\}=%
\begin{pmatrix}
n(\vec{r}) \\ 
e(\vec{r})-\vec{u}(\vec{r})\cdot \vec{p}(\vec{r})+\frac{1}{2}u^{2}mn(\vec{r})
\\ 
p_{x}(\vec{r})-mn(\vec{r})u_{x}(\vec{r}) \\ 
p_{y}(\vec{r})-mn(\vec{r})u_{y}(\vec{r}) \\ 
p_{z}(\vec{r})-mn(\vec{r})u_{z}(\vec{r})%
\end{pmatrix}%
\end{equation}%
and the rest frame fluxes are%
\begin{equation}
\{\gamma _{0i}(\vec{r})\}=%
\begin{pmatrix}
p_{i}(\vec{r}) \\ 
s_{i}(\vec{r})-u_{j}(\vec{r})t_{ji}(\vec{r})+\frac{u^{2}}{2}p_{i}(\vec{r})
\\ 
t_{ix}^{T}(\vec{r})-p_{i}(\vec{r})u_{x}(\vec{r}) \\ 
t_{iy}^{T}(\vec{r})-p_{i}(\vec{r})u_{y}(\vec{r}) \\ 
t_{iz}^{T}(\vec{r})-p_{i}(\vec{r})u_{z}(\vec{r})%
\end{pmatrix}%
-u_{i}(\vec{r})%
\begin{pmatrix}
n(\vec{r}) \\ 
e(\vec{r})-\vec{p}(\vec{r})\cdot \vec{u}(\vec{r})+\frac{1}{2}mn(\vec{r}%
)u^{2}(\vec{r}) \\ 
p_{x}(\vec{r})-mn(\vec{r})u_{x}(\vec{r}) \\ 
p_{y}(\vec{r})-mn(\vec{r})u_{y}(\vec{r}) \\ 
p_{z}(\vec{r})-mn(\vec{r})u_{z}(\vec{r})%
\end{pmatrix}%
\end{equation}

A concise form for these transformations to the local rest frame can be
described in matrix representation, 
\begin{equation}
\psi _{0\alpha }\left( \mathbf{r}\right) =A_{\alpha \beta }(\mathbf{u})\psi
_{\beta }\left( \mathbf{r}\right) ,  \label{eq123}
\end{equation}%
and the flux vector transforms to the rest frame as 
\begin{equation}
\gamma _{0i\alpha }\left( \mathbf{r}\right) =A_{\alpha \beta }(\vec{u}%
)\left( \gamma _{i\beta }-u_{i}\left( \mathbf{r}\right) \psi _{\beta }(\vec{r%
})\right)  \label{eq124}
\end{equation}%
where the matrix $A(\mathbf{u)}$ is given by 
\begin{align}
A(\mathbf{u})& =%
\begin{pmatrix}
1 & 0 & 0 & 0 & 0 \\ 
\frac{1}{2}mu^{2} & 1 & -u_{x} & -u_{y} & -u_{z} \\ 
-mu_{x} & 0 & 1 & 0 & 0 \\ 
-mu_{y} & 0 & 0 & 1 & 0 \\ 
-mu_{z} & 0 & 0 & 0 & 1%
\end{pmatrix}
\\
& \equiv 
\begin{pmatrix}
1 & 0 & \mathbf{0}^{T} \\ 
\frac{1}{2}mu^{2} & 1 & -\mathbf{u}^{T} \\ 
-m\mathbf{u} & \vec{0} & \vec{I}%
\end{pmatrix}%
.  \label{5.7}
\end{align}%
The second equality is a simplified notation, whereby $\vec{u}$ is a
three-component column vector, $\vec{u}^{T}$ is its transpose, $\vec{0}^{T}$
is the transposed zero vector, and $\vec{I}$ is the $3\times 3$ identity
matrix. To obtain (\ref{eq124}) use has been made of the fact that 
\begin{equation}
A^{-1}(\vec{u})=A(-\vec{u}).
\end{equation}

It is clear that the results (\ref{eq123}) and (\ref{eq124}) apply as well
for the average conserved densities and fluxes%
\begin{equation}
\overline{\psi }_{0\alpha }\left( \mathbf{r}\right) =A_{\alpha \beta }(%
\mathbf{u})\overline{\psi }_{\beta }\left( \mathbf{r}\right)
\end{equation}%
and the flux vector transforms to the rest frame as

\begin{equation}
\overline{\gamma }_{0i\alpha }\left( \mathbf{r}\right) =A_{\alpha \beta }(%
\vec{u})\left( \overline{\gamma }_{i\beta }-u_{i}\left( \mathbf{r}\right) 
\overline{\psi }_{\beta }(\vec{r})\right)  \label{5.6}
\end{equation}%
since the average is a linear operation and commutes with $A$.

As an example, consider the local equilibrium ensemble given by (see Eq.
(22) in the main text) 
%{\color{red} Note: this number is hard-coded; verify that it has not changed before submission}%
\begin{equation}
\rho ^{\ell }_N\left[ y\left( t\right) \right] =e^{-Q\left[ y\left( t\right) %
\right] -\int dry_{\alpha }\left( r,t\right) \psi _{\alpha }\left( \mathbf{r}%
\right) }  \label{5.8}
\end{equation}%
where the conjugate fields $\{y(\vec{r},t)\}$ are 
\begin{equation}
\{y(\vec{r},t)\}=\left\{ \left( -\nu (\vec{r},t)+\frac{1}{2}\beta (\vec{r}%
,t)mu^{2}(\vec{r},t)\right) ,\beta (\vec{r},t),-\beta (\vec{r},t)\mathbf{u}(%
\vec{r},t)\right\}.
\end{equation}%
The local rest frame form for (\ref{5.8}) is obtained directly from (\ref%
{eq123})%
\begin{eqnarray}
\int dry_{\alpha }\left( r,t\right) \psi _{\alpha }\left( \mathbf{r}\right)
&=&\int dry_{\alpha }\left( r,t\right) A_{\alpha \beta }(\vec{u})\psi
_{0\beta }\left( \mathbf{r}\right)  \notag \\
&=&\int drA_{\beta \alpha }^{T}(\vec{u})y_{\alpha }\left( r,t\right) \psi
_{0\beta }\left( \mathbf{r}\right)  \notag \\
&=&\int dry_{0\alpha }\left( r,t\right) \psi _{0\alpha }\left( \mathbf{r}%
\right).
\end{eqnarray}%
Here $\left\{ y_0\left( \vec r,t\right) \right\} $ are the rest frame conjugate
variables

\begin{equation}
y_{0\alpha }(\vec{r})=A_{\alpha \beta }^{T}y_{\beta }(\vec{r})=\left\{ -\nu (%
\vec{r},t),\beta (\vec{r},t),0\right\}
\end{equation}%
where $y_0(\vec r,t) = \left.y(\vec r,t)\right|_{u=0}$ and $A^{T}$ is the transpose of $A$. The local equilibrium ensemble is
therefore expressed in terms of the rest frame variables%
\begin{equation}
\rho ^{\ell }\left[ y\left( t\right) \right] =\rho _{0}^{\ell }\left[
y_{0}\left( t\right) \right] =e^{-Q\left[ y_{0}\left( t\right) \right] -\int
dry_{0\alpha }\left( r,t\right) \psi _{0\alpha }\left( \mathbf{r}\right) }.
\label{5.9}
\end{equation}

As a second example consider the correlations for the derivatives of the
average conserved fields with respect to the conjugate fields (see Eq. (115)
in the main text), 
%{\color{red} Note: this number is hard-coded; verify that
%it has not changed before submission; also correct that equation in next
%iteration} 
\begin{eqnarray}
g_{\alpha \beta }\left( \mathbf{r,r}^{\prime }\mathbf{\mid }y(t)\right) &=&%
\frac{\delta ^{2}Q(t)}{\delta y_{\alpha }\left( \mathbf{r}\right) \delta
y_{\beta }\left( \mathbf{r}^{\prime }\right) }=\frac{\delta \overline{\psi }%
_{\beta }^{\ell }\left( \mathbf{r}^{\prime }\right) }{\delta y_{\alpha
}\left( \mathbf{r}\right) }=\frac{\delta \overline{\psi }_{\alpha }^{\ell
}\left( \mathbf{r}\right) }{\delta y_{\beta }\left( \mathbf{r}^{\prime
}\right) }  \label{5.4b} \\
&=&\left.\overline{\psi _{\alpha }\left( \mathbf{r}\right) \widetilde{\psi }%
_{\beta }\left( \mathbf{r}^{\prime },t\right) }^{\ell }\right|_{y( t)},
\end{eqnarray}
with the tilde representing a transform closely related to the Kubo
transform for correlation functions, 
\begin{equation}
\widetilde{\psi }_{\alpha }(\mathbf{r}|y(t))=\int_{0}^{1}e^{-z\eta \left[ y(t)%
\right] }\left( \psi _{\alpha }(\mathbf{r})-\overline{\psi }_{\alpha }^{\ell
}\left( \mathbf{r}|y(t)\right) \right) e^{z\eta \left[ y(t)\right] }dz
\label{5.4}
\end{equation}
and the local equilibrium average is defined by
\begin{equation}
\left.\overline{X(\vec r)Y(\vec r')}^\ell\right|_{y(t)} \equiv\left\langle X(\vec r)Y(\vec r');\rho^\ell[y(t)]\right\rangle
\end{equation}
It is readily seen that%
\begin{equation}
\widetilde{\psi }_{0\alpha }(\mathbf{r}|y(t))=A_{\alpha \beta }(\vec{u}\left( 
\mathbf{r}\right) )\widetilde{\psi }_{\beta }(\mathbf{r}|y(t)).  \label{5.5}
\end{equation}%
Therefore, in matrix notation, $g\left( \mathbf{r,r}^{\prime }\mathbf{\mid }%
y(t)\right) $ transforms as 
\begin{equation}
g\left( \mathbf{r,r}^{\prime }\mid y(t)\right) =A^{-1}\left( \vec{u}\left( 
\mathbf{r}\right) \right) g_{0}\left( \mathbf{r,r}^{\prime }\mid
y(t)\right) A^{T-1}\left( \vec{u}\left( \mathbf{r}^{\prime }\right) \right)
\label{5.11}
\end{equation}%
with%
\begin{equation}
g_{0}\left( \mathbf{r,r}^{\prime }\mid y_{0}\right) =\left.\overline{\psi
_{0\alpha }\left( \mathbf{r}\right) \widetilde{\psi }_{0\beta }\left( 
\mathbf{r}^{\prime },t\right) }^{\ell }\right|_{y( t)}.
\end{equation}%
Since $g_{0}\left( \mathbf{r,r}^{\prime }\mid y_{0}\right) $ is now in the
rest frame, the unitary transformation (\ref{20b}) can be used to eliminate
the dependence on $\mathbf{u}\left( \mathbf{r}\right) $. Next, the
time-reversal symmetry (\ref{25b}) implies that matrix elements of pairs
that transform oppositely under momentum reversal must vanish. Consider the
densities $\psi _{\alpha }(\vec{r})$. For $\alpha =\{1,2\}$ the density is
unchanged under time reversal, while the densities corresponding to $\alpha
=\{3,4,5\}$ change sign. The local equilibrium ensemble index function $\eta
\lbrack y_{0}(t)]$ is also unchanged. Therefore, $g_{0}\left( \mathbf{r,r}%
^{\prime }\mid y_{0}\right) $ has the form%
\begin{equation}
g_{0}\left( \mathbf{r,r}^{\prime }\mid y(t)\right) =\left( 
\begin{array}{ccc}
\frac{\delta \overline{n}^{\ell }\left( \mathbf{r}\right) }{\delta \nu
\left( \mathbf{r}^{\prime }\right) }\big|_{\beta } & -\frac{\delta \overline{%
n}^{\ell }\left( \mathbf{r}\right) }{\delta \beta \left( \mathbf{r}^{\prime
}\right) }\big|_{\nu } & \mathbf{0}^T \\ 
\frac{\delta \overline{e_{0}}^{\ell }\left( \mathbf{r}\right) }{\delta \nu
\left( \mathbf{r}^{\prime }\right) }\big|_{\beta } & -\frac{\delta \overline{%
e_{0}}^{\ell }\left( \mathbf{r}\right) }{\delta \beta \left( \mathbf{r}%
^{\prime }\right) }\big|_{\nu } & \mathbf{0}^T \\ 
\mathbf{0}  & \mathbf{0}  &\frac{n\left( \mathbf{r}%
\right) }{\beta \left( \mathbf{r}\right) }\delta \left( \mathbf{r}-\mathbf{r}%
^{\prime }\right)\vec I%
\end{array}%
\right)
\end{equation}
where use has been made of (\ref{5.4b}) in the local rest frame, and for $%
\alpha ,\beta =3,4,5$%
\begin{equation}
g_{0\alpha \beta }\left( \mathbf{r,r}^{\prime }\mid y(t)\right) =\delta
_{\alpha \beta }\frac{1}{m\beta \left( \mathbf{r}\right) }\frac{\delta 
\overline{p}_{x}^{L}\left( \mathbf{r}\right) }{\delta u_{x}\left( \mathbf{r}%
^{\prime }\right) }\mid _{\mathbf{u}=0}=\delta _{\alpha \beta }\frac{%
mn\left( \mathbf{r}\right) }{m\beta \left( \mathbf{r}^{\prime }\right) }=%
\frac{n\left( \mathbf{r}\right) }{\beta \left( \mathbf{r}\right) }\delta_{\alpha\beta}\delta
\left( \mathbf{r}-\mathbf{r}^{\prime }\right).
\end{equation}

The inverse matrix $g_{0\alpha \beta }^{-1}(\mathbf{r},\mathbf{r}^{\prime })$%
, defined by 
\begin{equation}
\int d\mathbf{r}'' g_{0\alpha \gamma }(\mathbf{r},\mathbf{r}^{\prime
\prime })g_{0\gamma \beta }^{-1}(\mathbf{r}^{\prime \prime },\mathbf{r}%
^{\prime })=\delta _{\alpha \beta }\delta \left( \mathbf{r}-\mathbf{r}%
^{\prime }\right) ,  \label{152}
\end{equation}%
is required below as well, and is found to be%
\begin{equation}
g_{0}^{-1}\left( \mathbf{r,r}^{\prime }\mid y(t)\right) =\left( 
\begin{array}{ccc}
\frac{\delta \nu \left( \mathbf{r}\right) }{\delta \overline{n}\left( 
\mathbf{r}^{\prime }\right) }\mid _{e_{0}} & \frac{\delta \nu \left( \mathbf{%
r}\right) }{\delta \overline{e_{0}}\left( \mathbf{r}^{\prime }\right) }\mid
_{n} & \mathbf{0}^T \\ 
-\frac{\delta \beta \left( \mathbf{r}\right) }{\delta \overline{n}\left( 
\mathbf{r}^{\prime }\right) }\mid _{e_{0}} & -\frac{\delta \beta \left( 
\mathbf{r}\right) }{\delta \overline{e_{0}}\left( \mathbf{r}^{\prime
}\right) }\mid _{n} & \mathbf{0}^T \\ 
\mathbf{0} & \mathbf{0} & \frac{\beta \left( \mathbf{r}\right) }{n\left( 
\mathbf{r}^{\prime }\right) }\delta \left( \mathbf{r}-\mathbf{r}^{\prime
}\right)\vec I %
\end{array}%
\right) .  \label{5.17}
\end{equation}

\section{Liouville equation solution $\Delta (t)$ in the rest frame.}

The formal solution to the Liouville equation for the deviation from local
equilibrium is given in Appendix C of the text 
\begin{equation}
\Delta \left( t\right) =\int_{0}^{t}dt^{\prime }e^{-\mathcal{L}\left(
t-t^{\prime }\right) }\int d\mathbf{r}\left( \Psi _{\alpha }(\mathbf{r}%
|y(t^{\prime }))\frac{\partial}{\partial r_{i}}\overline{\gamma }_{i\alpha }^{\ast }\left( 
\mathbf{r},t^{\prime }\right) -\Phi _{i\alpha }(\mathbf{r}|y(t^{\prime
}))\frac{\partial}{\partial r_{i}}\overline{\psi }_{\alpha }\left( \mathbf{r},t^{\prime
}\right) \right) \rho ^{\ell }_N[y\left( t^{\prime }\right)] .  \label{6.7}
\end{equation}%
\begin{align}
\Phi _{i\alpha }(\mathbf{r}|y(t))& =\Gamma _{i\alpha }(\mathbf{r}|y(t))-\int d%
\mathbf{r}^{\prime }\Psi _{\beta }(\mathbf{r}^{\prime }|y(t))\left.\overline{\psi
_{\beta }\left( \mathbf{r}^{\prime }\right) )\Gamma _{i\alpha }(\mathbf{r}|y(t))%
}^{\ell }\right|_{y(t)}.  \label{6.8} \\
\Psi _{\alpha }(\mathbf{r}|y(t))& =\int d\mathbf{r}^{\prime }\widetilde{\psi }%
_{\beta }(\mathbf{r}^{\prime }|y(t))g_{\beta \alpha }^{-1}(\mathbf{r}^{\prime },%
\mathbf{r}\mid y\left( t\right) ) \\
\Gamma _{i\alpha }(\mathbf{r}|y(t))& =\int d\mathbf{r}^{\prime }\widetilde{%
\gamma }_{i\beta }(\vec r'|y(t))g_{\beta \alpha }^{-1}(\mathbf{r}%
^{\prime },\mathbf{r}\mid y\left( t\right) ).  \label{6.9}
\end{align}%
Since $\Psi_\alpha(\vec{r}|y(t))$ and $\Gamma _{i\alpha }(\vec{r}|y(t))$ are simply
related to the local conserved density $\psi _{\alpha }(\vec{r})$ and their
fluxes $\gamma _{i\alpha }(\vec{r})$, the transformation of the latter
follow directly from the former (recall the transformations of $\widetilde{%
\psi }(\vec{r}|y(t))$ and $\widetilde{\gamma }(\vec{r}|y(t))$ are the same as
those of $\psi (\vec{r})$ and $\gamma (\vec{r}),$ see (\ref{5.5})).

Consider first $\Psi _{\alpha }(\mathbf{r}|y(t))$%
\begin{align}
\Psi _{\alpha }(\mathbf{r}|y(t)) =&\int d\mathbf{r}^{\prime }A_{\alpha \beta
}^{-1}\left( \vec{u}\left( \mathbf{r}^{\prime }\right) \right) \widetilde{%
\psi }_{0\beta }(\mathbf{r}^{\prime }|y(t))\left( A^{T}\left( \vec{u}\left( 
\mathbf{r}^{\prime }\right) \right) g_{0}^{-1}\left( \mathbf{r}^{\prime }%
\mathbf{,r}\mid y(t)\right) A\left( \vec{u}\left( \mathbf{r}\right) \right)
\right) _{\beta \alpha }  \notag \\
=&\int d\mathbf{r}^{\prime }\widetilde{\psi }_{0\beta }(\mathbf{r}^{\prime
}|y(t))\left( g_{0}^{-1}\left( \mathbf{r}^{\prime }\mathbf{,r}\mid y(t)\right)
A\left( \vec{u}\left( \mathbf{r}\right) \right) \right) _{\beta \alpha } 
\notag \\
=&\Psi _{0\gamma }(\mathbf{r}|y(t))A_{\gamma \alpha }\left( \vec{u}\left( 
\mathbf{r}\right) \right).  \label{6.10}
\end{align}%
In a similar way, using (\ref{5.6}), $\Gamma _{i\alpha }(\mathbf{r}|y(t))$
becomes%
\begin{align}
\Gamma _{i\alpha }(\mathbf{r}|y(t)) =&\int d\mathbf{r}^{\prime }A_{\beta
\sigma }^{-1}(\vec{u})\left( \widetilde{\gamma }_{0\sigma }(\vec{r}^{\prime
}|y(t))+u_{i}(\vec{r}^{\prime })\widetilde{\psi }_{0\sigma }(\vec{r}^{\prime
}|y(t))\right) \left( A^{T}\left( \vec{u}\left( \mathbf{r}^{\prime }\right)
\right) g_{0}^{-1}\left( \mathbf{r}^{\prime }\mathbf{,r}\mid y(t)\right)
A\left( \vec{u}\left( \mathbf{r}\right) \right) \right) _{\beta \alpha } 
\notag \\
=&\int d\mathbf{r}^{\prime }\left( \widetilde{\gamma }_{0\sigma }(\vec{r}%
^{\prime }|y(t))+u_{i}(\vec{r}^{\prime })\widetilde{\psi }_{0\sigma }(\vec{r}%
^{\prime }|y(t))\right) \left( g_{0}^{-1}\left( \mathbf{r}^{\prime }\mathbf{,r}%
\mid y(t)\right) A\left( \vec{u}\left( \mathbf{r}\right) \right) \right)
_{\sigma \alpha }  \notag \\
=&\Gamma _{i0\beta }(\mathbf{r}|y(t))A_{\beta \alpha }\left( \vec{u}\left( 
\mathbf{r}\right) \right) +\int d\mathbf{r}^{\prime }u_{i}(\vec{r}^{\prime })%
\widetilde{\psi }_{0\sigma }(\vec{r}^{\prime }|y(t))g_{0\sigma \nu }^{-1}\left( 
\mathbf{r}^{\prime }\mathbf{,r}\mid y(t)\right) A_{\nu \alpha }\left( \vec{u%
}\left( \mathbf{r}\right) \right).  \label{6.11}
\end{align}

Finally, with these results the transformation of $\Phi _{i}(\vec{r}|y(t))$ is
obtained,

\begin{align}
\Phi _{i\alpha }(\vec{r}|y(t)) =&\Gamma _{i\alpha }(\vec{r}|y(t))-\int d\vec{r}%
^{\prime }\ \Psi _{\beta }(\vec{r}^{\prime })\left.\overline{\psi _{\beta }(\vec{r}%
^{\prime })\Gamma _{i\alpha }(\vec{r}|y(t))}^{\ell }\right|_{y(t)}  \notag \\
=&\Gamma _{i0\beta }(\mathbf{r}|y(t))A_{\beta \alpha }\left( \vec{u}\left( 
\mathbf{r}\right) \right) +\int d\mathbf{r}^{\prime }\left( u_{i}(\vec{r}%
^{\prime })\psi _{0\sigma }(\vec{r}^{\prime })\right) g_{0\sigma \nu
}^{-1}\left( \mathbf{r}^{\prime }\mathbf{,r}\mid y(t)\right) A_{\nu \alpha
}\left( \vec{u}\left( \mathbf{r}\right) \right)  \notag \\
&-\int d\vec{r}^{\prime }\ \Psi _{0\sigma }(\vec{r}^{\prime }|y(t))\left.\overline{%
\psi _{_{0\sigma }}(\vec{r}^{\prime })\Gamma _{0i\beta }(\vec{r}|y(t))}^{\ell
}\right|_{y( t)} A_{\beta \alpha }\left( \vec{u}\left( \mathbf{r}\right)
\right)  \notag \\
&-\int d\vec{r}^{\prime }\ \Psi _{0\sigma }(\vec{r}^{\prime }|y(t))\int d\mathbf{%
r}^{\prime \prime }u_{i}(\vec{r}^{\prime \prime })\left.\overline{\psi _{_{0\beta
}}(\vec{r}^{\prime })\widetilde{\psi }_{0\sigma }(\vec{r}^{\prime \prime }|y(t))}%
^{\ell }\right|_{y(t)}g_{0\sigma \nu }^{-1}\left( \mathbf{r}^{\prime \prime }\mathbf{,r}%
\mid y(0)\right) A_{\nu \alpha }\left( \vec{u}\left( \mathbf{r}\right)
\right) ^{\ell }\left( t\right)  \notag \\
=&\left( \Gamma _{i0\beta }(\mathbf{r}|y(t))-\int d\vec{r}^{\prime }\ \Psi
_{0\sigma }(\vec{r}^{\prime }|y(t))\left.\overline{\psi _{_{0\sigma }}(\vec{r}^{\prime
})\Gamma _{0i
}(\vec{r}|y(t))}^{\ell }\right|_{y( t)} \right) A_{\beta \alpha }\left( \vec{u}%
\left( \mathbf{r}\right) \right) .
\end{align}%
\begin{equation}
\Phi _{i\alpha }(\vec{r}|y)=\Phi _{0i\beta }(\vec{r}|y)A_{\beta \alpha }\left( 
\vec{u}\left( \mathbf{r}\right) \right) .  \label{6.12}
\end{equation}%
The solution to the Liouville equation (\ref{6.7}) is therefore 
\begin{equation}
\Delta \left( t\right) =\int_{0}^{t}dt^{\prime }e^{-\mathcal{L}\left(
t-t^{\prime }\right) }\int d\mathbf{r}\left( \Psi _{0\beta }(\mathbf{r}%
|y(t^{\prime }))A_{\beta \alpha }\left( \vec{u}\left( \mathbf{r}\right) \right)
\frac{\partial}{\partial r_{i}}\overline{\gamma }_{i\alpha }^{\ast }\left( \mathbf{r}%
,t^{\prime }|y\right) -\Phi _{i0\beta }(\mathbf{r}|y(t^{\prime }))A_{\beta \alpha
}\left( \vec{u}\left( \mathbf{r}\right) \right) \frac{\partial}{\partial r_{i}}\overline{\psi }%
_{\alpha }\left( \mathbf{r},t^{\prime }\right) \right) \rho ^{\ell }_N[y\left(
t^{\prime }\right)] .  \label{6.13}
\end{equation}%
The operators $\Psi _{0\beta }(\mathbf{r}|y(t^{\prime }))$ and $\Phi _{i0\beta
}(\mathbf{r}|y(t^{\prime }))$ with subscript $0$ are the same as those in (\ref%
{6.7}) except with $\vec{u}\left( \mathbf{r}\right) =0.$

It remains to transform $\partial _{i}\overline{\gamma }_{i\alpha }^{\ast
}\left( \mathbf{r},t^{\prime }\right) $ and $\partial _{i}\overline{\psi }%
_{\alpha }\left( \mathbf{r},t\right) $ to the corresponding rest frame form.
First, note that $\Psi _{0\beta }(\mathbf{r}|y(t^{\prime }))$ and $\Phi
_{0i\beta }(\mathbf{r}|y(t^{\prime }))$ can be written as%
\begin{equation}
\Psi _{0\alpha }(\mathbf{r}|y(t))=\int d\mathbf{r}^{\prime }\widetilde{\psi }%
_{0\beta }(\mathbf{r}^{\prime }|y(t))g_{0\beta \alpha }^{-1}\left( \mathbf{r}%
^{\prime }\mathbf{,r}\mid y(t)\right) ,  \label{6.14}
\end{equation}%
\begin{equation}
\Phi _{0i\beta }(\mathbf{r}|y(t))=\int d\mathbf{r}^{\prime }\widetilde{\phi }%
_{0i\beta }(\mathbf{r}^{\prime }|y(t))g_{0\beta \alpha }^{-1}\left( \mathbf{r}%
^{\prime }\mathbf{,r}\mid y(t)\right)  \label{6.15}
\end{equation}%
where%
\begin{equation}
\widetilde{\phi }_{0i\beta }(\mathbf{r}|y(t))=\widetilde{\gamma }_{i0\beta }(%
\mathbf{r}|y(t))-\int d\mathbf{r}^{\prime }\int d\vec{r}^{\prime \prime }\ 
\widetilde{\psi }_{0\sigma }(\mathbf{r}^{\prime }|y(t))g_{0\sigma \alpha
}^{-1}\left( \mathbf{r}^{\prime }\mathbf{,r}^{\prime \prime }\mid
y(t)\right) \left.\overline{\psi _{_{0\sigma }}(\vec{r}^{\prime \prime })%
\widetilde{\gamma }_{0i\beta }(\mathbf{r}|y(t))}^{\ell }\right|_{y( t)}
\label{6.16}
\end{equation}%
Then 
\begin{align}
\int d\mathbf{r}\Phi _{i0\beta }(\mathbf{r}|y(t^{\prime }))A_{\beta \alpha
}\left( \vec{u}\left( \mathbf{r}\right) \right) \partial _{i}\overline{\psi }%
_{\alpha }\left( \mathbf{r},t\right) =&\int d\mathbf{r}\int d\mathbf{r}%
^{\prime }\widetilde{\phi }_{0i\beta }(\mathbf{r}^{\prime }|y(t))g_{0\beta
\alpha }^{-1}\left( \mathbf{r}^{\prime }\mathbf{,r}\mid y_{0}\right)
A_{\alpha \nu }\left( \vec{u}\left( \mathbf{r}\right) \right) \frac{\partial}{\partial r_{i}}%
\overline{\psi }_{\nu }\left( \mathbf{r},t\right)  \notag \\
=&\int d\mathbf{r}\int d\mathbf{r}^{\prime }\widetilde{\phi }_{0i\beta }(%
\mathbf{r}^{\prime }|y(t))A_{\beta \alpha }^{T-1}(\mathbf{r}^{\prime })\int d%
\mathbf{r}g_{\alpha \sigma }^{-1}(\mathbf{r}^{\prime },\mathbf{r}\mid
y_{0})\frac{\partial}{\partial r_{i}}\overline{\psi }_{\sigma }\left( \mathbf{r},t\right) 
\notag \\
=&-\int d\mathbf{r}^{\prime }\widetilde{\phi }_{0i\beta }(\mathbf{r}%
^{\prime }|y(t))A_{\beta \alpha }^{T-1}(\mathbf{r}^{\prime })\frac{\partial}{\partial
r_{i}'}y_{\alpha }\left( \mathbf{r}^{\prime },t\right)  \notag \\
=&-\int d\mathbf{r}^{\prime }\widetilde{\phi }_{0i\beta }(\mathbf{r}%
^{\prime }|y(t))A_{\beta \alpha }^{T-1}(\mathbf{u})\frac{\partial}{\partial r_{i}'}A_{\alpha
\sigma }^{T}(\mathbf{u})y_{0\sigma }\left( \mathbf{r}^{\prime },t\right)
\end{align}%
Finally, then%
\begin{equation}
\int d\mathbf{r}\Phi _{i0\beta }(\mathbf{r}|y(t^{\prime }))A_{\beta \alpha
}\left( \vec{u}\left( \mathbf{r}\right) \right) \frac{\partial}{\partial r_{i}}\overline{\psi }%
_{\alpha }\left( \mathbf{r},t\right) =\int d\mathbf{r}^{\prime }\left( -%
\widetilde{\phi }_{0i2}(\mathbf{r}^{\prime }|y(t))\frac{\partial}{\partial r_{i}'}\beta (\mathbf{r}%
^{\prime },t^{\prime })+\widetilde{\phi }_{0ij}(\mathbf{r}^{\prime }|y(t))\beta (%
\mathbf{r}^{\prime },t^{\prime })\frac{\partial}{\partial r_{j}'}u_{i}(\mathbf{r}^{\prime
},t^{\prime })\right)  \label{6.27}
\end{equation}%
In a similar way%
\begin{eqnarray}
\Psi _{0\beta }(\mathbf{r}|y(t))A_{\beta \alpha }\left( \vec{u}\left( \mathbf{r}%
\right) \right) \frac{\partial}{\partial r_{i}}\overline{\gamma }_{i\alpha }^{\ast }\left( 
\mathbf{r},t\right|y) &=&\Psi _{0\beta }(\mathbf{r}|y(t^{\prime }))A_{\beta
\alpha }\left( \vec{u}\left( \mathbf{r}\right) \right)  \frac{\partial}{\partial r_{i}}
\overline{\gamma }_{i\alpha }^{\ast }\left( \mathbf{r},t^{\prime }|y\right) 
\notag \\
&=&\Psi _{02}(\mathbf{r}|y(t))\left( \nabla \cdot \overline{\mathbf{s}}%
_{0}^{\ast }(\mathbf{r},t|y)+\overline{t}_{ij}^{\ast }\left( \mathbf{r}%
,t^{\prime }|y\right) \partial _{i}u_{j}\right) +\Psi _{0j}(\mathbf{r}|y(t))%
\overline{t}_{ij}^{\ast }(\mathbf{r},t|y)  \label{6.29}
\end{eqnarray}%
In the above solution to the Liouville equation $\overline{\gamma }%
_{i}^{\ast }\left( \mathbf{r},t^{\prime }|y\right) $ are the laboratory frame
fluxes, $\overline{\gamma }_{i}^{\ast }\left( \mathbf{r},t^{\prime }|y\right)
=\left\langle \gamma _{i}\left( \mathbf{r}\right) ;t\right\rangle
-\left\langle \gamma _{i}\left( \mathbf{r}\right) ;t\right\rangle _{L}$. The
corresponding rest frame fluxes $\overline{\gamma }_{0i}^{\ast }\left( 
\mathbf{r},t^{\prime }|y\right) =\left\langle \gamma _{0i}\left( \mathbf{r}%
\right) ;t\right\rangle -\left\langle \gamma _{0i}\left( \mathbf{r}\right)
;t\right\rangle _{L}$ differ for the energy flux $\overline{\gamma }%
_{02i}^{\ast }\left( \mathbf{r},t^{\prime }|y\right) =\overline{\gamma }%
_{2i}^{\ast }\left( \mathbf{r},t^{\prime }|y\right) -u_{j}\left( \mathbf{r}%
,t^{\prime }\right) \overline{t}_{ij}^{\ast }\left( \mathbf{r},t^{\prime }|y\right) 
$ $\equiv \overline{s}_{0i}^{\ast }(\mathbf{r},t^{\prime }|y)$.

The form for the solution to the Liouville equation (\ref{6.13}) becomes 
\begin{align}
\Delta \left( t\right)  =&\int_{0}^{t}dt^{\prime }e^{-\mathcal{L}\left(
t-t^{\prime }\right) }\int d\mathbf{r}\left( \widetilde{\phi }_{0i2}(\mathbf{%
r}|y(t)) \frac{\partial}{\partial r_{i}}\beta (\mathbf{r},t^{\prime })-\widetilde{\sigma }_{0ij}(%
\mathbf{r}|y(t^{\prime }))\beta (\mathbf{r},t^{\prime }) \frac{\partial}{\partial r_{j}}u_{i}(%
\mathbf{r},t^{\prime })\right.   \notag \\
&\left. +\Psi _{02}(\mathbf{r}|y(t^{\prime }))\nabla \cdot \overline{\mathbf{s}%
}_{0}^{\ast }(\mathbf{r},t^{\prime }|y)+\Psi _{0j}(\mathbf{r}|y(t^{\prime
}))\partial _{i}\overline{t}_{ij}^{\ast }(\mathbf{r},t'|y)\right) \rho ^{\ell
}_N[y\left( t^{\prime }\right)] .  \label{6.30}
\end{align}%
More explicitly%
\begin{equation}
\widetilde{\phi }_{0i2}(\mathbf{r}|y(t))=\widetilde{s}_{0i}(\mathbf{r}|y(t))-\int d%
\mathbf{r}^{\prime }\int d\vec{r}^{\prime \prime }\ \widetilde{\psi }%
_{0\sigma }(\mathbf{r}^{\prime }|y(t))g_{0\sigma \alpha }^{-1}\left( \mathbf{r}%
^{\prime }\mathbf{,r}^{\prime \prime }\mid y(t)\right) \left.\overline{\psi
_{_{0\sigma }}(\vec{r}^{\prime \prime })\widetilde{s}_{0i}(\mathbf{r}|y(t))}%
^{\ell }\right|_{y( t)} .  \label{6.31}
\end{equation}%
By time reversal symmetry $\left.\overline{\psi _{_{0\sigma }}(\vec{r}^{\prime
\prime })\widetilde{s}_{0i}(\mathbf{r}|y(t))}^{\ell }\right|_{y\left( t\right)} $ is
non-vanishing only for $\psi _{_{0\sigma }}(\vec{r}^{\prime \prime })=p_{i}(%
\vec{r}^{\prime \prime })$, and similarly $g_{0\sigma \alpha }^{-1}\left( 
\mathbf{r}^{\prime }\mathbf{,r}^{\prime \prime }\mid y(t)\right) $ is
non-zero only for the diagonal momentum components, so%
\begin{eqnarray}
\widetilde{\phi }_{0i2}(\mathbf{r}|y(t)) &=&\widetilde{s}_{0i}(\mathbf{r}%
|y(t))-\int d\mathbf{r}^{\prime }\int d\vec{r}^{\prime \prime }\ \widetilde{p}%
_{0j}(\mathbf{r}^{\prime }|y(t))\delta \left( \mathbf{r}^{\prime }-\mathbf{r}%
^{\prime \prime }\right) \frac{\beta (\mathbf{r}^{\prime },t^{\prime })}{mn(%
\mathbf{r}^{\prime },t^{\prime })}\left.\overline{p_{_{0j}}(\vec{r}^{\prime \prime
})\widetilde{s}_{0i}(\mathbf{r}|y(t))}^{\ell }\right|_{y\left( t\right)}   \notag \\
&=&\widetilde{s}_{0i}(\mathbf{r}|y(t))-\int d\mathbf{r}^{\prime }\ \widetilde{p}%
_{0j}(\mathbf{r}^{\prime }|y(t))\frac{\beta (\mathbf{r}^{\prime },t^{\prime })}{%
mn(\mathbf{r}^{\prime },t^{\prime })}\left.\overline{p_{_{0j}}(\vec{r}^{\prime })%
\widetilde{s}_{0i}(\mathbf{r}|y(t))}^{\ell }\right|_{y\left( t\right)}   \label{6.33}
\end{eqnarray}%
Also 
\begin{align}
\Psi _{02}(\mathbf{r}|y(t))& =\int d\mathbf{r}^{\prime }\widetilde{\psi }_{0\beta
}(\mathbf{r}^{\prime }|y(t))g_{0\beta 2}^{-1}\left( \mathbf{r}^{\prime }\mathbf{,r%
}|y(t)\right)   \notag \\
& =\int d\mathbf{r}^{\prime }\left( \widetilde{\psi }_{01}(\mathbf{r}%
^{\prime }|y(t))g_{012}^{-1}\left( \mathbf{r}^{\prime }\mathbf{,r}|y(t)\right) +%
\widetilde{\psi }_{02}(\mathbf{r}^{\prime }|y(t))g_{022}^{-1}\left( \mathbf{r}%
^{\prime }\mathbf{,r}|y(t)\right) \right)   \notag \\
& =-\int d\mathbf{r}^{\prime }\left( \frac{\delta \beta \left( \mathbf{r}%
\right) }{\delta \overline{n}_{0}^{L}\left( \mathbf{r}^{\prime }\right) }%
\widetilde{n}(\mathbf{r}^{\prime }|y(t))+\frac{\delta \beta \left( \mathbf{r}%
^{\prime }\right) }{\delta \overline{e}_{0}^{L}\left( \mathbf{r}\right) }%
\widetilde{e}_{0}(\mathbf{r}^{\prime }|y(t))\right)   \label{7.7}
\end{align}%
and%
\begin{align}
\widetilde{\sigma }_{0ij}(\mathbf{r}|y(t)) =&\widetilde{\phi }_{0ij}(\mathbf{r}%
|y(t))-\beta ^{-1}(\mathbf{r},t)\Psi _{02}(\mathbf{r}|y(t))\overline{t}_{ij}^{\ast
}\left( \mathbf{r},t|y\right)   \notag \\
=&\widetilde{t}_{0ij}(\mathbf{r}|y(t))-\int d\mathbf{r}^{\prime }\int d\vec{r}%
^{\prime \prime }\ \widetilde{\psi }_{0\sigma }(\mathbf{r}^{\prime
}|y(t))g_{0\sigma \alpha }^{-1}\left( \mathbf{r}^{\prime }\mathbf{,r}^{\prime
\prime }\mid y(t)\right) \left.\overline{\psi _{_{0\alpha }}(\vec{r}^{\prime
\prime })\widetilde{t}_{0ij}(\mathbf{r}|y(t))}^{\ell }\right|_{y\left( t\right)}   \notag \\
&-\beta ^{-1}(\mathbf{r},t)\Psi _{02}(\mathbf{r}|y(t))\overline{t}_{ij}^{\ast
}\left( \mathbf{r},t|y\right)   \label{7.8}
\end{align}%
To evaluate the second term on the right note that%
\begin{align}
\left.\overline{\psi _{_{0\alpha }}(\vec{r}^{\prime \prime })\widetilde{t}_{0ij}(%
\mathbf{r}|y(t))}^{\ell }\right|_{y\left( t\right)}  &=\left.\overline{t_{0ij}(\mathbf{r})%
\widetilde{\psi }_{0\alpha }\left( \mathbf{r}^{\prime \prime }|y(t)\right) }%
^{\ell }\right|_{y\left( t\right)} =-\frac{\delta \left.\overline{t_{0ij}(\mathbf{r})}^{\ell
}\right|_{y\left( t\right)} }{\delta y_{0\alpha }\left( \mathbf{r}^{\prime \prime
}\right) }  \notag \\
&=-\frac{\delta \pi_{ij}(\mathbf{r}|y(t))}{\delta y_{0\alpha }\left( \mathbf{r}%
^{\prime \prime },t\right) } , \label{7.9}
\end{align}%
where $\pi_{ij}$ is the pressure (see Eq. (60) in the main text), so that 
%{\color{red} Note: this number is hard-coded; verify that
%it has not changed before submission; also correct that equation in next
%iteration}
%
\begin{align}
\widetilde{\sigma }_{0ij}(\mathbf{r}|y(t)) =&\widetilde{t}_{0ij}(\mathbf{r}%
|y(t))+\int d\mathbf{r}^{\prime }\int d\vec{r}^{\prime \prime }\ \widetilde{%
\psi }_{0\sigma }(\mathbf{r}^{\prime }|y(t))g_{0\sigma \alpha }^{-1}\left( 
\mathbf{r}^{\prime }\mathbf{,r}^{\prime \prime }\mid y(t)\right) \frac{%
\delta \pi_{ij}(\mathbf{r}|y(t))}{\delta y_{0}(\mathbf{r}^{\prime \prime 
},t)}  \notag \\
&-\beta ^{-1}(\mathbf{r},t)\Psi _{02}(\mathbf{r}|y(t))\overline{t}_{ij}^{\ast
}\left( \mathbf{r},t|y\right) .  \label{7.10}
\end{align}%
The second term simplifies%
\begin{align}
\int d\vec{r}^{\prime \prime }\widetilde{\psi }_{0\sigma }(\mathbf{r}%
^{\prime }|y(t))&g_{0\sigma \alpha }^{-1}\left( \mathbf{r}^{\prime }\mathbf{,r}%
^{\prime \prime }\mid y(t)\right) \frac{\delta \pi_{ij}(\mathbf{r}|y(t))}{\delta
y_{0}(\mathbf{r}^{\prime \prime \prime },t)} \notag\\
 =&-\widetilde{n}(\mathbf{r}^{\prime }|y(t))\int d\vec{r}^{\prime \prime }\left(
-g_{011}^{-1}\left( \mathbf{r}^{\prime }\mathbf{,r}^{\prime \prime }\mid
y(t)\right) \frac{\delta \pi_{ij}(\mathbf{r}|y(t))}{\delta \nu \left( \mathbf{r}%
^{\prime \prime }|y(t)\right) }\big|_{\beta }+g_{012}^{-1}\left( \mathbf{r}%
^{\prime }\mathbf{,r}^{\prime \prime }\mid y(t)\right) \frac{\delta \pi_{ij}(%
\mathbf{r}|y(t))}{\delta \beta \left( \mathbf{r}^{\prime \prime }|y(t)\right) }%
\big|_{\nu }\right) \notag \\
& -\widetilde{e}_{0}(\mathbf{r}^{\prime }|y(t))\int d\vec{r}^{\prime \prime
}\left( -g_{021}^{-1}\left( \mathbf{r}^{\prime }\mathbf{,r}^{\prime \prime
}\mid y(t)\right) \frac{\delta \pi_{ij}(\mathbf{r}|y(t))}{\delta \nu \left( 
\mathbf{r}^{\prime \prime }|y(t)\right) }\big|_{\beta }+g_{022}^{-1}\left( 
\mathbf{r}^{\prime }\mathbf{,r}^{\prime \prime }\mid y(t)\right) \frac{%
\delta \pi_{ij}(\mathbf{r}|y(t))}{\delta \beta \left( \mathbf{r}^{\prime \prime
}|y(t)\right) }\big|_{\nu }\right)  \notag \\
 =&-\widetilde{n}(\mathbf{r}^{\prime }|y(t))\int d\vec{r}^{\prime \prime }\left(
-g_{011}^{-1}\left( \mathbf{r}^{\prime \prime },\mathbf{r}^{\prime }\mid
y(t)\right) \frac{\delta \pi_{ij}(\mathbf{r}|y(t))}{\delta \nu \left( \mathbf{r}%
^{\prime \prime }|y(t)\right) }\big|_{\beta }+g_{021}^{-1}\left( \mathbf{r}%
^{\prime \prime },\mathbf{r}^{\prime }\mid y(t)\right) \frac{\delta \pi_{ij}(%
\mathbf{r}|y(t))}{\delta \beta \left( \mathbf{r}^{\prime \prime }|y(t)\right) }%
\big|_{\nu }\right)   \notag \\
& -\widetilde{e}_{0}(\mathbf{r}^{\prime }|y(t))\int d\vec{r}^{\prime \prime
}\left( -g_{012}^{-1}\left( \mathbf{r}^{\prime \prime }\mathbf{,r}^{\prime
}\mid y(t)\right) \frac{\delta \pi_{ij}(\mathbf{r}|y(t))}{\delta \nu \left( 
\mathbf{r}^{\prime \prime }|y(t)\right) }\big|_{\beta }+g_{022}^{-1}\left( 
\mathbf{r}^{\prime \prime }\mathbf{,r}^{\prime }\mid y(t)\right) \frac{%
\delta \pi_{ij}(\mathbf{r}|y(t))}{\delta \beta \left( \mathbf{r}^{\prime \prime
}|y(t)\right) }\big|_{\nu }\right)   \label{7.11}
\end{align}%
In this last equality use has been made of $g_{0\alpha \beta }^{-1}\left( 
\mathbf{r,r}^{\prime }\mid y(t)\right) =g_{0\beta \alpha }^{-1}\left( 
\mathbf{r}^{\prime }\mathbf{,r}\mid y(t)\right) $. Finally, then 
\begin{align*}
& \widetilde{\psi }_{0\sigma }(\mathbf{r}^{\prime }|y(t))\int d\vec{r}^{\prime
\prime }g_{0\sigma \alpha }^{-1}\left( \mathbf{r}^{\prime }\mathbf{,r}%
^{\prime \prime }\mid y(t)\right) \frac{\delta \pi_{ij}(\mathbf{r}|y(t))}{\delta
y_{0}(\mathbf{r}^{\prime \prime },t)} \\
& =-\widetilde{n}(\mathbf{r}^{\prime }|y(t))\int d\mathbf{r}^{\prime \prime
}\left( -\frac{\delta \pi_{ij}(\mathbf{r}|y(t))}{\delta \nu \left( \mathbf{r}%
^{\prime \prime }|y(t)\right) }\big|_{\beta }\frac{\delta \nu \left( \mathbf{r}%
^{\prime \prime }|y(t)\right) }{\delta \overline{n}\left( \mathbf{r}^{\prime
}|y(t)\right) }\mid _{e}-\frac{\delta \pi_{ij}(\mathbf{r}|y(t))}{\delta \beta \left( 
\mathbf{r}^{\prime \prime }|y(t)\right) }\big|_{\nu }\frac{\delta \beta \left( 
\mathbf{r}^{\prime \prime }|y(t)\right) }{\delta \overline{n}\left( \mathbf{r}%
^{\prime }|y(t)\right) }\mid _{e}\right)  \\
&\qquad -\widetilde{e}_{0}(\mathbf{r}^{\prime }|y(t))\int d\mathbf{r}^{\prime \prime
}\left( -\frac{\delta \pi_{ij}(\mathbf{r}|y(t))}{\delta \nu \left( \mathbf{r}%
^{\prime \prime }|y(t)\right) }\big|_{\beta }\frac{\delta \nu \left( \mathbf{r}%
^{\prime \prime }|y(t)\right) }{\delta \overline{e}\left( \mathbf{r}^{\prime
}|y(t)\right) }\mid _{n}-\frac{\delta \pi_{ij}(\mathbf{r}|y(t))}{\delta \beta \left( 
\mathbf{r}^{\prime \prime }|y(t)\right) }\big|_{\nu }\frac{\delta \beta \left( 
\mathbf{r}^{\prime \prime }|y(t)\right) }{\delta \overline{e}\left( \mathbf{r}%
^{\prime }|y(t)\right) }\mid _{n}\right) \nonumber\\
&=\widetilde{n}(\mathbf{r}^{\prime }|y(t))\frac{\delta \pi_{ij}(\mathbf{r}|y(t))}{%
\delta n\left( \mathbf{r}^{\prime }|y(t)\right) }\mid _{e}+\widetilde{e}_{0}(%
\mathbf{r}^{\prime }|y(t))\frac{\delta \pi_{ij}(\mathbf{r}|y(t))}{\delta e\left( 
\mathbf{r}^{\prime }|y(t)\right) }\mid _{n}  \label{7.12}
\end{align*}%
With this result (\ref{7.10}) becomes 
\begin{eqnarray}
\widetilde{\sigma }_{0ij}(\mathbf{r}|y(t)) &=&\widetilde{t}_{0ij}(\mathbf{r}%
|y(t))+\int d\mathbf{r}^{\prime }\left( \widetilde{n}(\mathbf{r}^{\prime }|y(t))%
\frac{\delta \pi_{ij}(\mathbf{r}|y(t))}{\delta n\left( \mathbf{r}^{\prime
}|y(t)\right) }\mid _{e}+\widetilde{e}_{0}(\mathbf{r}^{\prime }|y(t))\frac{\delta
\pi_{ij}(\mathbf{r}|y(t))}{\delta e\left( \mathbf{r}^{\prime }|y(t)\right) }\mid
_{n}\right)   \notag \\
&&-\beta ^{-1}(\mathbf{r},t)\Psi _{02}(\mathbf{r}|y(t))\overline{t}_{ij}^{\ast
}\left( \mathbf{r},t\right|y) .  \label{7.13}
\end{eqnarray}

\section{Irreversible fluxes in the rest frame}

The irreversible contributions to the energy and momentum fluxes are defined
by%
\begin{equation}
\overline{\mathbf{s}}_{0}^{\ast }(\mathbf{r},t|y)\equiv \sum_{N}{\rm Tr}^{(N)}%
\mathbf{s}_{0}(\mathbf{r})\Delta_N \left( t\right) ,\hspace{0.25in}\overline{t}%
_{0ij}^{\ast }(\mathbf{r},t|y)\equiv \sum_{N}{\rm Tr}^{(N)}t_{0ij}(\mathbf{r})\Delta_N
\left( t\right)   \label{7.1}
\end{equation}%
where $\mathbf{s}_{0}(\mathbf{r})$ and $t_{0ij}(\mathbf{r})$ are the
operators for the energy and momentum fluxes in the local rest frame. With (%
\ref{6.30}) these are%
\begin{align}
\overline{\mathbf{s}}_{0}^{\ast }(\mathbf{r},t|y) =&\int_{0}^{t}dt^{\prime
}\int d\mathbf{r}^{\prime }\left[ \left.\overline{\left( e^{\mathcal{L}\left(
t-t^{\prime }\right) }\mathbf{s}_{0}(\mathbf{r})\right) \widetilde{\phi }%
_{0i2}(\mathbf{r}^{\prime }|y(t^{\prime }))}^{\ell }\right|_{y\left( t^{\prime }\right)}
\frac{\partial}{\partial r_{i}}\beta (\mathbf{r}^{\prime },t^{\prime })\right.  \nonumber\\
&-\left.\overline{\left( e^{\mathcal{L}\left( t-t^{\prime }\right) }\mathbf{s}%
_{0}(\mathbf{r})\right) \widetilde{\sigma }_{0ij}(\mathbf{r}|y(t^{\prime }))}%
^{\ell }\right|_{y\left( t^{\prime }\right)} \beta (\mathbf{r}^{\prime },t^{\prime
})\frac{\partial}{\partial r_{j}}u_{i}(\mathbf{r}^{\prime },t^{\prime })\nonumber \\
&+\left.\overline{\left( e^{\mathcal{L}\left( t-t^{\prime }\right) }\mathbf{s}%
_{0}(\mathbf{r})\right) \Psi _{02}(\mathbf{r},t^{\prime })}^{\ell }\right|_{y\left(
t^{\prime }\right)} \nabla \cdot \overline{\mathbf{s}}_{0}^{\ast }(\mathbf{r}%
,t^{\prime }|y) \nonumber\\
&\left. +\left.\overline{\left( e^{\mathcal{L}\left( t-t^{\prime }\right) }%
\mathbf{s}_{0}(\mathbf{r})\right) \Psi _{0j}(\mathbf{r},t^{\prime })}^{\ell
}\right|_{y\left( t^{\prime }\right)} \right] \overline{t}_{ij}^{\ast }(\mathbf{r},t|y)
\end{align}%
and%
\begin{align}
\overline{t}_{0ij}^{\ast }(\mathbf{r},t|y)  =&\int_{0}^{t}dt^{\prime }\int d%
\mathbf{r}^{\prime }\left[ \left.\overline{\left( e^{\mathcal{L}\left( t-t^{\prime
}\right) }t_{0ij}(\mathbf{r})\right) \widetilde{\phi }_{0k2}(\mathbf{r}%
^{\prime }|y(t^{\prime }))}^{\ell }\right|_{y\left( t^{\prime }\right)} \frac{\partial}{\partial r_{k}}\beta
(\mathbf{r}^{\prime },t^{\prime })\right. \nonumber \\
& -\left.\overline{\left( e^{\mathcal{L}\left( t-t^{\prime }\right) }t_{0ij}(%
\mathbf{r})\right) \widetilde{\sigma }_{0kl}(\mathbf{r}|y(t^{\prime }))}^{\ell
}\right|_{y\left( t^{\prime }\right)} \beta (\mathbf{r}^{\prime },t^{\prime })\frac{\partial}{\partial
r_{k}}u_{l}(\mathbf{r}^{\prime },t^{\prime })\nonumber \\
& +\left.\overline{\left( e^{\mathcal{L}\left( t-t^{\prime }\right) }t_{0ij}(%
\mathbf{r})\right) \Psi _{02}(\mathbf{r}|y(t^{\prime }))}^{\ell }\right|_{y\left(
t^{\prime }\right)} \nabla \cdot \overline{\mathbf{s}}_{0}^{\ast }(\mathbf{r}%
,t^{\prime }|y)\nonumber \\
& \left. +\left.\overline{\left( e^{\mathcal{L}\left( t-t^{\prime }\right)
}t_{0ij}(\mathbf{r})\right) \Psi _{0l}(\mathbf{r}|y(t^{\prime }))}^{\ell
}\right|_{y\left( t^{\prime }\right)} \right] \frac{\partial}{\partial r_{k}}\overline{t}_{kl}^{\ast }(%
\mathbf{r},t|y)
\end{align}